\documentclass[
    aps,
    physrev,
    reprint,
    superscriptaddress,
    amsmath,
    amssymb,
    longbibliography,
    nofootinbib
]{revtex4-2}

\usepackage{graphicx}
\usepackage{mathtools}
\usepackage{braket}
\usepackage{slashed}
\usepackage{maksmath}
\usepackage{xcolor}
\usepackage{tabularx,array}
\usepackage{hyperref}
\usepackage{booktabs}

\usepackage[T5,T1]{fontenc}

\hypersetup{
    colorlinks,
    linkcolor={red!50!black},
    citecolor={blue!50!black},
    urlcolor={blue!80!black}
}

\newlength{\outerparindent}
\begin{document}

\title{Many-body topology in parity-preserving tensor networks}

\author{Maksimilian Usoltcev}
\affiliation{Institut f\"ur Theoretische Physik, Universit\"at zu K\"oln, 50937 Cologne, Germany}

\author{{\fontencoding{T5}\selectfont Nguyễn Hạnh Dung}}
\affiliation{Institut f\"ur Theoretische Physik, Universit\"at zu K\"oln, 50937 Cologne, Germany}

\author{Carolin Wille}
\affiliation{London Centre for Nanotechnology, University College London, London WC1H 0AH, United Kingdom}

\author{Matteo Rizzi}
\affiliation{Institut f\"ur Theoretische Physik, Universit\"at zu K\"oln, 50937 Cologne, Germany}
\affiliation{Institute of Quantum Control, Peter Grünberg Institut (PGI-8), Forschungszentrum Jülich GmbH, 52425 Jülich, Germany}

\author{Alexander Altland}
\affiliation{Institut f\"ur Theoretische Physik, Universit\"at zu K\"oln, 50937 Cologne, Germany}

\begin{abstract}
Planar parity-preserving tensor networks (ppTNs) admit a fermionic formulation in which the Gaussian, efficiently contractible limit is deformed by local interactions. We investigate how many-body methods can be used to analyze such contractions in a minimal two-parameter ppTN. Its contraction is simultaneously an interacting fermionic partition function, a loop gas, a deformed toric-code norm, and a quartic Ising model, allowing the same phase diagram to be approached with complementary tools. At its center lies a `topological island', characterized beyond the Gaussian limit by a boundary-twist $\Zbb_2$ indicator that reduces to Chern-number parity and admits a loop-winding interpretation under duality. Fermionic perturbation theory predicts the interacting phase boundaries, tensor-network numerics establish their critical behavior, and the loop and spin descriptions reveal a self-dual line with a $c=1$ four-state-Potts multicritical fixed point.
\end{abstract}

\maketitle

\section{Introduction}

Exact tensor network contraction in more than one dimension is computationally hard~\cite{SchuchWolfVerstraeteCirac2007,Markov_2008}. This motivates the search for classes of tensor networks that are expressive enough to describe nontrivial phenomena, while still providing sufficient structure for analytic or controlled numerical analysis. Planar parity-preserving tensor networks (ppTNs) are such a family. The parity constraint is physically natural and recurs across tensor-network formulations of fermionic systems~\cite{BarthelPinedaEisert2009,Bravyi_2008}, $\mathbb Z_2$ spin and gauge models, including topologically ordered examples~\cite{LevinNave2007,BuerschaperAguadoVidal2009,Schuch2013,TagliacozzoCeliLewenstein2014,ZhangXuWangZhang2020}, statistical-mechanical formulations of quantum error correction~\cite{Dennis_2002,FerrisPoulin2014,BravyiSucharaVargo2014,ChubbFlammia2019,BehrendsBeri2025}, and parity-preserving quantum circuits~\cite{Brod2011,Oszmaniec2017,Mocherla2024,ReardonSmith2024,WilleStrelchuk2025}.
Planar ppTNs also stand out because the trade-off between simplicity and expressivity has a particularly intuitive interpretation: they correspond to fermionic Gaussian states plus non-Gaussian deformations and their contraction corresponds to the evaluation of an interacting fermionic partition function. This feature makes it possible to understand complexity as a direct consequence of going beyond the free-fermion limit~\cite{TerhalDiVincenzo2002,JozsaMiyake2008,Bravyi_2008,Brod2011,Oszmaniec2017,HebenstreitJosza2019,Dias_2024,ReardonSmith2024,WilleStrelchuk2025}.

This fermionic formulation also provides us with access to the extensive toolbox of quantum many-body theory --- including band structure analysis, mean-field approaches, and topological considerations. This toolbox, combined with tensor-network-intrinsic analytical concepts, such as virtual symmetries, and numerical methods, provides a strong set of complementary methods.
%


In this work, we make this general perspective concrete through a representative case study of a two-parameter ppTN~\cite{Wille_quarticTN_2025}, in which one parameter explicitly controls the strength of the deformation away from the free-fermion limit. 
The phase diagram features a `topological island' --- a robust topological phase, characterized by a many-body fermionic topological invariant, that extends deep into the interacting regime. We show that the phase boundaries of this island can be obtained analytically using a mean-field-inspired approach: the `interacting' fermionic tensor network is mapped to an effective free-fermion theory and the phase transition lines are detected via the gap closing of an effective Hamiltonian.

Apart from the topological island, the phase diagram has several other marked features, which are accessible with complementary methods. Among them are a high-symmetry line and a multicritical point, where two second-order transition lines meet a first-order transition line. We present a summary of the phase diagram and the medley of tools used to derive it in the following section.

\subsection{Overview of Results}

\begin{figure*}
    \centering
    \includegraphics[width=1\linewidth]{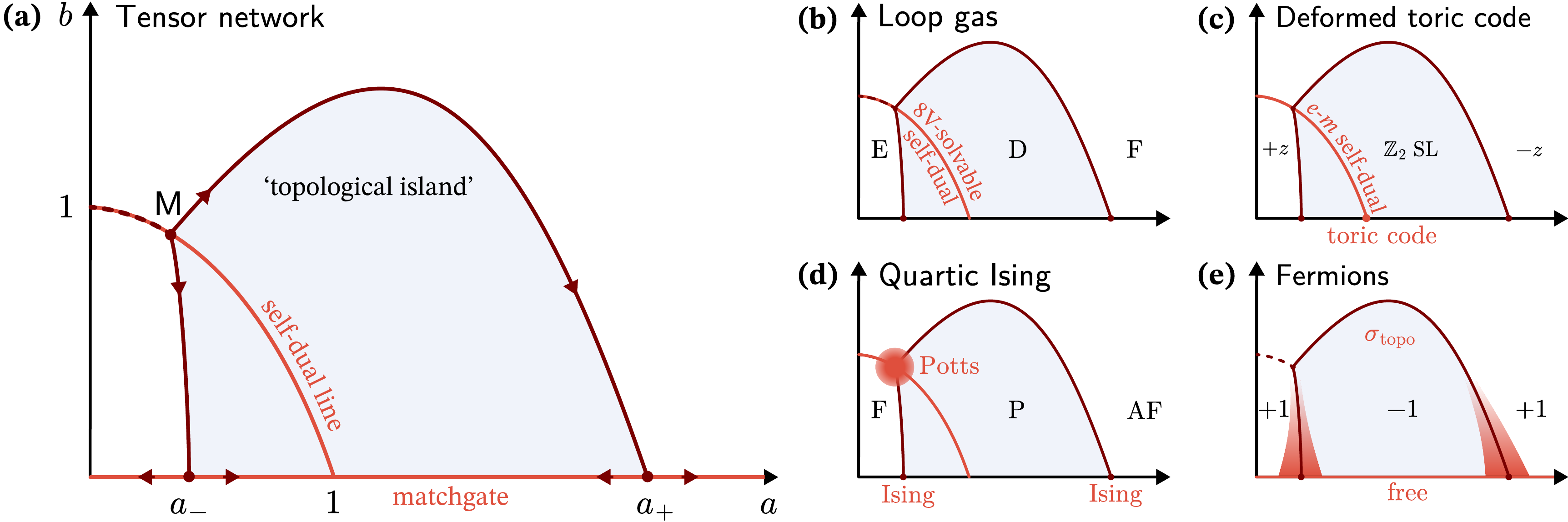} 
     \caption{%
Overview of the common phase diagram and its complementary
representations.
(a) Phase diagram of the two-parameter tensor network. The shaded
region is the topological island, bounded by two continuous transition
lines that meet the first-order line at the
multicritical point \textsf{M}. Along $b=0$, the transition points are
$a_\pm=\sqrt{2}\pm1$. The line $b=1-a^2$ is self-dual; its segment with $0\leq a<1/3$ coincides with the coexistence line, while the segment with $1/3 < a \leq 1$ lies inside the topological phase.
(b)--(e) The same three phases expressed in the loop gas,
deformed toric code, quartic Ising, and fermionic languages, respectively.
The red annotations indicate the structures made especially
transparent by each representation: eight-vertex solvability and
self-duality, the toric-code point and its $e\leftrightarrow m$
interpretation, Ising criticality and the four-state Potts point, and
the free-fermion line together with the boundary-twist invariant
$\sigma_{\tx{topo}}$.
     }
    \label{fig:dualities}
\end{figure*}

As illustrated in Fig.~\ref{fig:dualities}, the model~\cite{Wille_quarticTN_2025} is defined as a physically agnostic
two-dimensional planar tensor network --- panel (a),
but admits several complementary
physical interpretations:
(b) a classical loop-gas (eight-vertex) model,
(c) a deformed toric-code state,%
\footnote{More precisely, the contraction of this tensor-network model equals
the norm of a deformed toric-code PEPS (projected entangled-pair state); see
Ref.~\cite{Wille_2024}. The precise mapping is explained in
Sec.~\ref{subsec:recap} around Eq.~\eqref{eq:toric-code-norm}.}
(d) a classical Ising model with quartic interactions, and
(e) a class-D topological superconductor with an additional local quartic deformation.
The equivalence of these formulations establishes a web of dualities: they encode the same phase diagram, while emphasizing different aspects of its physics. 
Taken together, they extend our understanding beyond any single description and reveal how seemingly distinct structures are connected across the parameter space.

The central organizing feature of this phase diagram is the topological island [Fig.~\ref{fig:dualities}(a)]. Along the line of vanishing fermionic interactions,
$b=0$, also referred to as the Gaussian or matchgate line, this region is characterized in the fermionic description by Chern number $\mathcal C=1$ and lies between two phases with $\mathcal C=0$. Its two boundaries extend into the interacting regime and meet the first-order coexistence line of the two non-topological phases at the multicritical point \textsf{M}. 
Away from the Gaussian limit, the single-particle Chern number is no longer available as an exact diagnostic. Instead, the region is characterized by an interacting $\Zbb_2$ indicator $\sigma_{\tx{topo}}$, which reduces at $b=0$ to Chern-number parity, $(-1)^{\mathcal C}$.
This indicator remains fixed throughout the phase and motivates its designation as the topological island.

Across these formulations, the topological island appears as a deconfined loop
regime (b), the $\Zbb_2$ spin-liquid region of the deformed toric-code PEPS (c),
a paramagnet in the quartic Ising model (d) and a fermionic topological phase
(e). 
The topological distinction is defined by stability under local fermionic
perturbations that preserve fermion parity. Under the duality, these map to a
restricted class of perturbations in the loop and spin descriptions. In
particular, the Ising paramagnet is a dual encoding of the fermionic topological
region, but a generic local Ising perturbation need not belong to this inherited
class. A longitudinal magnetic field, for example, is local in the unconstrained
Ising model but lies outside the inherited class.

The different formulations emphasize complementary aspects of this common phase structure
(see also Table~\ref{tab:model-overview}). In the tensor-network formulation, $a$ and
$b$ simply specify the local tensor entries, without assigning them an a priori
physical meaning. This representation provides direct numerical access to the
full phase diagram and the universality classes of the transitions via
established tensor-network techniques.

In the loop-gas formulation, $a$ controls the string tension and $b$ modifies
the weight of loop crossings. The three phases are the empty (E), deconfined
(D), and fully covered (F) regimes. This description provides an intuitive
picture of the phases and connects the model to the well-known eight-vertex
models. Along the special line $b=1-a^2$, exchanging occupied and unoccupied links
leaves the local weights unchanged. This symmetry gives rise to a self-duality
mapping whose fixed point locates the multicritical point. At the same time,
the line satisfies the zero-field condition of the eight-vertex model and
therefore belongs to Baxter's exactly solvable Bethe-ansatz family.

In the deformed toric-code PEPS, the same parameters control string tension and
a local crossing deformation. The three regions become the $+z$-polarized,
$\Zbb_2$ spin-liquid (SL), and $-z$-polarized phases. This formulation connects
the topological island to many-body topological order and gives another
interpretation of the self-dual line as the interchange
$e\leftrightarrow m$ of the toric-code anyonic excitations.

In the quartic Ising model, $a$ determines the nearest-neighbor coupling while
$b$ introduces a four-spin interaction favoring domain-wall crossings. The
phases are ferromagnetic (F), paramagnetic (P), and antiferromagnetic (AF).
This description is most useful for identifying conventional order, Ising
criticality, and the Ashkin--Teller/four-state-Potts character of the
multicritical point.

The fermionic formulation plays a central role in our analysis. The parameter $a$ controls the quadratic Majorana couplings, while $b$ introduces a local quartic interaction and hence measures the departure from the Gaussian limit. 
At $b=0$, the resulting four-band model realizes the phase sequence
trivial--topological--trivial, with transitions at zero-energy gap closings.
For finite $b$, perturbatively integrating out the gapped high-energy bands yields an effective quadratic low-energy theory with interaction-renormalized masses. 
Their zeros predict the continuous phase boundaries, which we benchmark against direct tensor-network contractions.
Furthermore, the fermionic formulation supplies the twisted partition functions used to construct the interacting $\Zbb_2$ indicator. Finally, it gives the self-dual line a distinct interpretation as a closing of the higher-band gap, rather
than the zero-energy gap that delimits the topological phase.


\begin{table*}[t]
    \centering
    \small
    \renewcommand{\arraystretch}{1.35}
    \setlength{\tabcolsep}{6pt}

    \begin{tabular}{@{}l|l|l|l@{}}
        \parbox[t]{0.13\linewidth}{\raggedright Formulation}
        &
        \parbox[t]{0.21\linewidth}{\raggedright Basic objects}
        &
        \parbox[t]{0.33\linewidth}{\raggedright Phases}
        &
        \parbox[t]{0.25\linewidth}{\raggedright Best for}
        \\
        \hline

        \parbox[t]{0.13\linewidth}{\raggedright tensor network}
        &
        \parbox[t]{0.21\linewidth}{\raggedright local tensors $T$}
        &
        \parbox[t]{0.33\linewidth}{\raggedright --}
        &
        \parbox[t]{0.25\linewidth}{\raggedright contraction numerics}
        \\

        \parbox[t]{0.13\linewidth}{\raggedright loop gas}
        &
        \parbox[t]{0.21\linewidth}{\raggedright link occupations $n_e=0,1$}
        &
        \parbox[t]{0.33\linewidth}{\raggedright empty\slash deconfined\slash full}
        &
        \parbox[t]{0.25\linewidth}{\raggedright self-duality \& solvability}
        \\

        \parbox[t]{0.13\linewidth}{\raggedright toric-code PEPS}
        &
        \parbox[t]{0.21\linewidth}{\raggedright edge qubits $\sigma_e$}
        &
        \parbox[t]{0.33\linewidth}{\raggedright
            $+z$ polarized\slash $\Zbb_2$ spin liquid\slash $-z$ polarized}
        &
        \parbox[t]{0.25\linewidth}{\raggedright topological interpretation}
        \\

        \parbox[t]{0.13\linewidth}{\raggedright quartic Ising}
        &
        \parbox[t]{0.21\linewidth}{\raggedright site spins $s_i=\pm1$}
        &
        \parbox[t]{0.33\linewidth}{\raggedright ferro\slash para\slash antiferro}
        &
        \parbox[t]{0.25\linewidth}{\raggedright criticality \& universality}
        \\

        \parbox[t]{0.13\linewidth}{\raggedright fermionic TN}
        &
        \parbox[t]{0.21\linewidth}{\raggedright Grassmann modes $\theta_i$}
        &
        \parbox[t]{0.33\linewidth}{\raggedright trivial\slash topological\slash trivial}
        &
        \parbox[t]{0.25\linewidth}{\raggedright topology \& perturbation theory}
    \end{tabular}

    \caption{
        Conceptual dictionary for the representations shown in
        Fig.~\ref{fig:dualities}. The tensor network itself is treated as an
        agnostic contraction object, so no intrinsic phase labels are assigned
        in the first row. In each dual description, the three phases are listed
        in their left-to-right order as in Fig.~\ref{fig:dualities}. The final
        column indicates the structures for which each representation is used
        most directly in this work.
    }
    \label{tab:model-overview}
\end{table*}

The remainder of the paper is organized as follows. 
Sec.~\ref{sec:pd} introduces the two-parameter model and its dual formulations, and determines its phase diagram and critical behavior by combining tensor-network numerics with an effective fermionic theory.
Sec.~\ref{sec:self-dual} establishes the self-dual line and its fixed point, identifies the multicritical theory through the statistical-mechanical descriptions, and develops the corresponding higher-band interpretation in the fermionic picture.
Sec.~\ref{sec:topo} constructs the interacting $\Zbb_2$ indicator from fermionic partition functions
with twisted boundary conditions, derives its loop and spin representations using the torus Jordan--Wigner mapping, and verifies it numerically throughout the phase diagram. 
Sec.~\ref{sec:con} concludes with a discussion of the broader implications and open directions.

%
%

\section{Model and phase diagram} 
\label{sec:pd}

This section reviews the equivalent interpretations of our two-parameter
model~\cite{Wille_quarticTN_2025} and fixes notation. We then present its phase
diagram together with the thermodynamic-limit numerical and analytical results
supporting it. Here, we focus on the continuous transitions and their critical exponents. 
The special line $b=1-a^2$, the multicritical theory, and the
interpretation of the topological phase are discussed in later
Secs.~\ref{sec:self-dual} and \ref{sec:topo}, respectively.

\subsection{Dual models}
\label{subsec:recap}
The model is a planar, parity-preserving, translation-invariant
tensor network of bond dimension two, with local tensors determined by
two parameters, $a$ and $b$. 
For $a,b\geq0$, the contracted TN admits an interpretation as a statistical-mechanical partition function. 
The parity constraint restricts the
allowed local configurations to closed loops, which leads to the loop-gas description.

From this starting point, several equivalent formulations follow, as visualized in Fig.~\ref{fig:dualities}. 
Promoting the loop variables to physical indices defines a projected
entangled-pair state (PEPS), i.e. a two-dimensional quantum state, obtained by deforming the toric-code ground-state PEPS through string tension and local crossing weights.
A Kramers--Wannier-type duality allows us to interpret the same tensor-network
contraction as the partition function of an interacting Ising model, thereby
connecting it to the extensive literature on two-dimensional spin systems.
Finally, a Jordan--Wigner transformation yields a
fermionic tensor network that realizes a class-D topological
superconductor with additional local quartic terms. 

We discuss the individual descriptions below.

%

\subsubsection{Tensor network, loop gas, and toric code PEPS}
\paragraph{Loop gas.}
An intuitive way to present the model is a loop gas~\cite{BaxterExactly,Nienhuis1984}. Consider a two-dimensional square lattice where each link can be either `unoccupied' or `occupied'. Constraining to even numbers of occupied links adjacent to each vertex results in closed loop configurations, where loops are allowed to cross.

\begin{figure*}[t]
    \centering

    \begin{minipage}{0.35\textwidth}\centering \label{fig:vertex}
        \includegraphics[width=\textwidth]{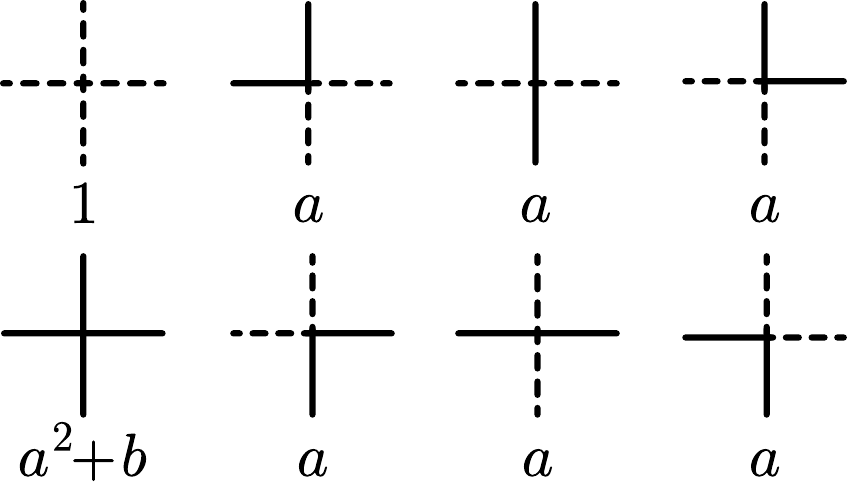}
        \par\smallskip\textbf{(a)}
    \end{minipage}%
    \hspace{8mm}
    \begin{minipage}{0.255\textwidth}\centering \label{fig:tensor-vertex}
        \includegraphics[width=\textwidth]{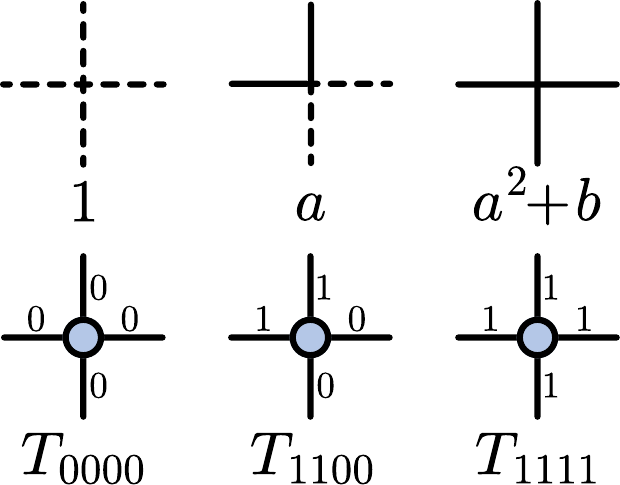}
        \par\smallskip\textbf{(b)}
    \end{minipage} 
    \hspace{8mm}   
    \begin{minipage}{0.163\textwidth}\centering \label{fig:ising-vertex}
                \includegraphics[width=\textwidth]{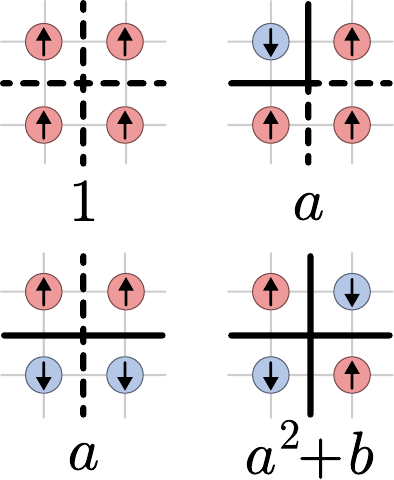}
        \par\smallskip\textbf{(c)}
    \end{minipage}%
    \caption{Vertex configurations in different languages.
    (a) Weights of the eight allowed loop-gas vertex configurations.
    (b) Interpretation of vertex weights as components of a four-legged, bond-dimension-two tensor.
    (c) Spin configurations around the vertex leading to interpretation as a (generalized) Ising model.
    }%
    \label{fig:weights}
\end{figure*}

We then use two parameters $a,b>0$ to assign a weight to each of the eight allowed vertex configurations as in Fig.~\ref{fig:weights}(a).
These eight even-parity local configurations are precisely the vertices of an eight-vertex model (an equivalent encoding is the conventional arrow representation). Our weights therefore define a restricted two-parameter subclass of the general eight-vertex model~\cite{BaxterExactly}.
The parameter $a$ controls the relative weight of unoccupied and occupied links and hence the effective string tension, while $b$ shifts the weight of a crossing configuration. The sum of the total weights of all allowed loop configurations yields the partition function
\begin{equation}
    Z = \sum_{\ell:\;\text{loops}} w(\ell), 
    \qquad 
    w(\ell) \coloneqq a^{|\ell|} (1 + b/a^2)^{n_\text{c} (\ell)}
    \label{eq:loop-gas-Z}
\end{equation}
where $|\ell|$ denotes the total loop length of a given configuration $\ell$ (i.e.\ the total number of occupied edges) and $n_\text{c}(\ell)$ the number of loop crossings in this configuration.
 
Along the Gaussian line $b=0$, crossings carry no additional weight and the model reduces to the domain-wall representation of the nearest-neighbor Ising model, which is free-fermion solvable. 
By contrast, along the line $b=1-a^2$, the weights of the vertex configurations are invariant under exchanging occupied and unoccupied links simultaneously for all four links.
This symmetry corresponds to the zero-field condition of the eight-vertex model, which is exactly solvable by Baxter's commuting-transfer-matrix method and the associated Bethe ansatz~\cite{BaxterExactly}.

\paragraph{Bosonic tensor network model.} 

The partition function above corresponds to a planar tensor network on a square lattice, where the vertex configurations encode the eight entries of a four-legged, parity-preserving tensor of bond dimension two.%
\footnote{Here and below, ``bosonic'' refers to the ordinary commuting
tensor-network contraction, in contrast to the fermionic/Grassmann
formulation. It does not refer to bosonic local degrees of freedom or imply
an infinite-dimensional local Hilbert space.}
Concretely, we define a tensor with entries $T_{i_1,i_2,i_3,i_4}$ where each index $i_a$ is $i_a=0$ for an unoccupied adjacent link and $i_a=1$ for an occupied one, cf.~Fig.~\ref{fig:weights}(b), and assign to it the corresponding vertex weights, e.g., $T_{0000} = 1$, $T_{1100} = a$, $\dots$, $T_{1111} = a^2 + b$. Crucially, $T_{i_1,i_2,i_3,i_4}=0$ if the sum of the entries is odd. This enforces the closed loop conditions and we refer to tensors with this property as parity preserving.

\paragraph{Deformed toric-code PEPS.}
Following Ref.~\cite{Wille_2024}, the same local tensor can be read as a toric-code-like PEPS by forwarding the virtual $\mathbb Z_2$ indices to physical legs and choosing local amplitudes $t_{i_1i_2i_3i_4}=\sqrt{T_{i_1i_2i_3i_4}}$, cf.\ Fig.~\ref{fig:square-lat}~(right). In the loop basis this defines
\[
|\Psi(a,b)\rangle \propto \sum_{\ell: \text{loops}}
a^{|\ell|/2}(1+b/a^2)^{n_c(\ell)/2}\,|\ell\rangle,
\]
whose norm reduces to the original single-layer tensor network,
\begin{equation}
\label{eq:toric-code-norm}
 	\langle\Psi(a,b)|\Psi(a,b)\rangle \eqqcolon Z
\end{equation}

For $b=0$ this is the string-tension deformation of the toric-code PEPS discussed e.g.\ in Refs.~\cite{Wille_2024,Castelnovo2008}, while finite $b$ introduces a local nonlinear deformation that changes only the weight of loop crossings. Along the special line $b=1-a^2$ one furthermore has $t_{0000}=t_{1111}$, so the PEPS treats electric and magnetic strings on the same footing; in this sense, the line realizes the virtual $e\leftrightarrow m$ self-duality~\cite{Kitaev2003,Bombin2010,Bridgeman2017} of the toric-code description. We return to this point in Sec.~\ref{subsec:loop-gas}. 

\subsubsection{Quartic Ising model} 
The loop gas model above is dual to an Ising model with two- and four-spin
interactions. This correspondence is established by interpreting strings as
domain walls between spins on the dual lattice as in Fig.~\ref{fig:weights}(c)
and yields what we call the quartic Ising (QI) model: a nearest-neighbor (NN)
Ising model with an additional plaquette term
\begin{equation}
    H_{\text{QI}} =  - J_2 \sum_{\av{ij}} s_i s_j - J_{4}\sum_{\square} P(\square)\,,
\end{equation}
where $P(\square)$ is one for the two 4-spin checkerboard configurations (bottom right diagram of Fig.~\ref{fig:weights}(c)) and zero otherwise, and $s_i=\pm1$ are spin variables. The couplings are given by $J_2 = - \frac12 \ln a$ and $J_4 = \ln(1 + b/a^2)$. For $b = 0$, the model reduces to the standard Ising model and is ferromagnetic for $a<1$ ($J_2>0$) and antiferromagnetic for $a>1$ ($J_2<0$), with phase transitions at $a_{\pm} = \sqrt2\pm1$~\cite{Wille_2024,Wille_quarticTN_2025}. For $b>0$, the $J_4$-term favors domain wall intersections and thus enforces antiferromagnetic order for large enough $b$ regardless of the sign of $J_2$.

The special parameter line $b = 1 - a^2$ affords a more intuitive interpretation when we reformulate the checkerboard projector as $
P(\square) =\frac{1}{16} \prod_{i \in \square} (1 - s_i s_{i+1})$, 
and write the Hamiltonian in terms of nearest-neighbor (NN), diagonal next-nearest neighbor (NNN) and four-spin couplings,
\begin{equation} \label{eq:spinHam-Kform}
	H_{\tx{QI}} = -K_1 \sum_{\av{ij}} s_i s_j - K_2 \sum_{\dis{ij}} s_i s_j - K_3 \sum_{\square} s_1 s_2 s_3 s_4 \,,
\end{equation}
with coupling constants\footnote{For the interested reader, we remark that this Hamiltonian with general $K_1,K_2,K_3$ has been studied extensively, see e.g.\ Ref.~\cite{van_Leeuwen_1975}. However, our focus is on a different parameter regime than the ones considered in the literature. We will return to a more extensive contextualization of our result within various established statmech models in Sec.~\ref{sec:self-dual}.} 
$K_1 = J_2 - J_4/4$ and $K_2 = K_3 = J_4/8$.
The special parameter line $b = 1 - a^2$ then corresponds to a vanishing of the net NN interaction, $K_1=0$.

\subsubsection{Fermionic tensor network model and class-D topological superconductor} 

 \begin{figure}[t]
     \centering
	\includegraphics[width=\linewidth]{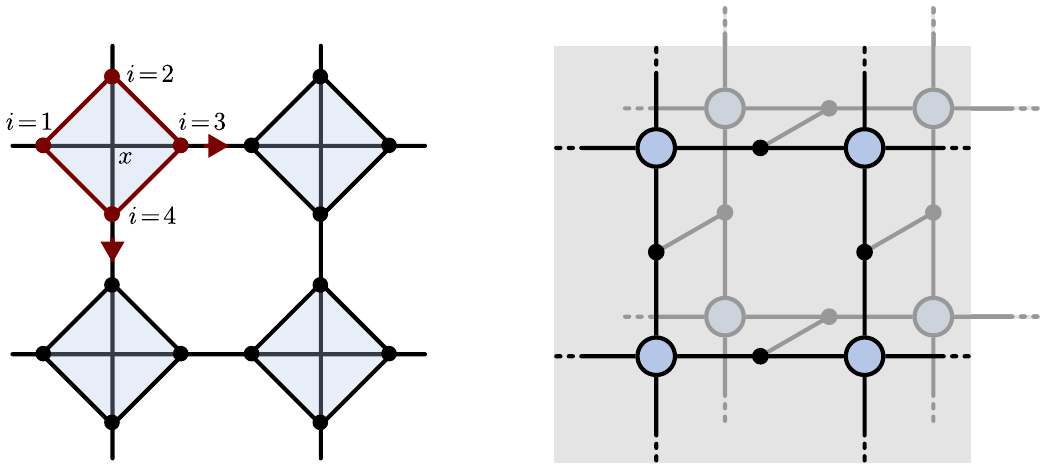}
     \caption{Left: The fermionic tensor network on the square lattice. Right: The toric code PEPS construction.}
     \label{fig:square-lat}
 \end{figure}
%

The planarity and parity preservation of the tensor network allow for a Jordan--Wigner transformation to a \emph{fermionic tensor network} (cf.\ Ref.~\cite{Wille_2024,Wille_quarticTN_2025}), specified by the fermionic partition sum $Z^{\tx f} = Z$ (the equality provided by a consistent ordering of the fermionic modes).
To each tensor index $i_a = 0,1$, we associate a fermionic mode represented by a (real) Grassmann variable $\theta^{i_a}$ with $\theta^0 = 1$, $\theta^1 = \theta$.
This defines the fermionic tensors 
\begin{equation}
\begin{aligned}
T_{\tx f} 
&= T_{i_1,i_2,i_3,i_4} \theta_1^{i_1} \theta_2^{i_2} \theta_3^{i_3} \theta_4^{i_4} 
\\
&= \exp\left(\frac12\theta^\T A \theta + b \theta_1 \theta_2 \theta_3 \theta_4\right)\,,
\label{eq:fermionic-tensor-def}
\end{aligned}
\end{equation} 
where $A$ is a $4 \times 4$ antisymmetric matrix with elements $A_{ij} = a \sgn (j-i)$.

The contraction of the fermionic TN can then be represented as a Grassmann integral 
\begin{equation}
\begin{aligned}
    Z^{\tx f}
    &=
    \int \dd \theta \, \e^{-S},
    \\
    S 
    &=
    -\frac 1 2 \theta^\T
    (C + \textstyle{\bigoplus_x} A ) \theta
    - b\sum_x
    \theta^1_x\, \theta^2_x \,\theta^3_x \,\theta^4_x \,,
\end{aligned}
\end{equation}
where $\theta = \crl{\theta^i_x}$ denotes the vector of fermionic modes $\theta^i$, $i = 1,\dots,4$ at sites $x$, and $C$ is the antisymmetric connectivity matrix of the square lattice, specifying which inter-cell links are present, cf.\ Fig.~\ref{fig:square-lat}. Importantly, for a square lattice with empty boundary as in Ref.~\cite{Wille_2024}, the statement $Z=Z^{\tx{f}}$ holds exactly. For other geometries, in particular those with non-trivial topology, the relationship between the statmech partition sum and the fermionic one can still be established upon accounting for an additional global sign factor that emerges from the non-locality of the Jordan--Wigner map (cf.\ Sec.~\ref{subsec:JW_torus} for the torus construction).

\paragraph{Class~D topological superconductor.}
From the fermionic TN model, the duality to a class D topological superconductor follows straightforwardly~\cite{Wille_2024,Wille_quarticTN_2025,Usoltcev2026}.
Concretely, in the free limit, $b=0$, we interpret the fermionic TN as a Majorana tight-binding model with BdG Hamiltonian $H = -\imagunit (C + \textstyle{\bigoplus_x} A)$. Fourier-transforming from site space to momentum space, the model is described by a four-band Bloch Hamiltonian $H(k) = -\imagunit(C(k) + A)$, with $A = a \sgn(j-i)$ as above. 

\begin{figure*}[t]
	\centering
	\includegraphics[width=0.32\textwidth]{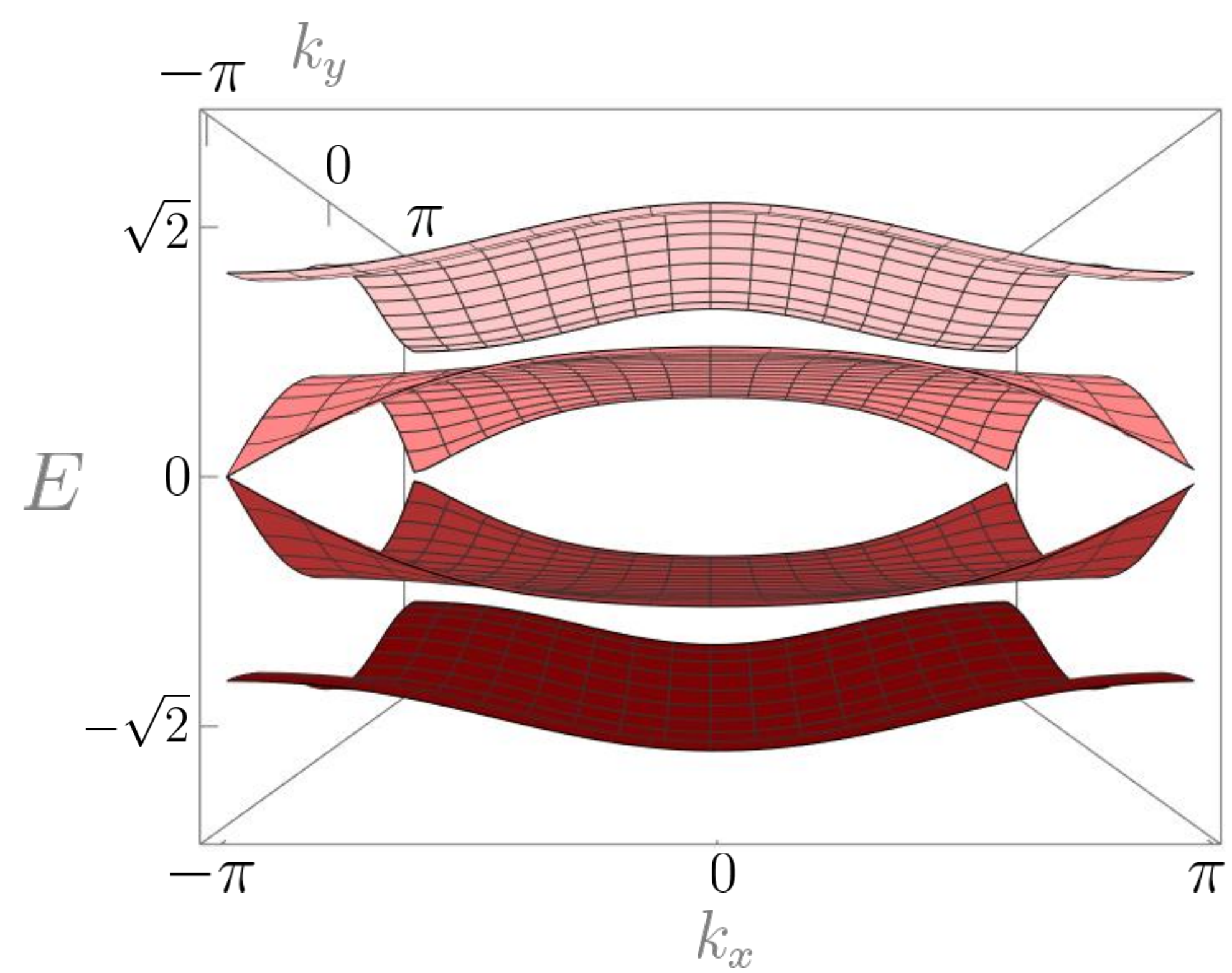}
	\hfill
	\includegraphics[width=0.32\textwidth]{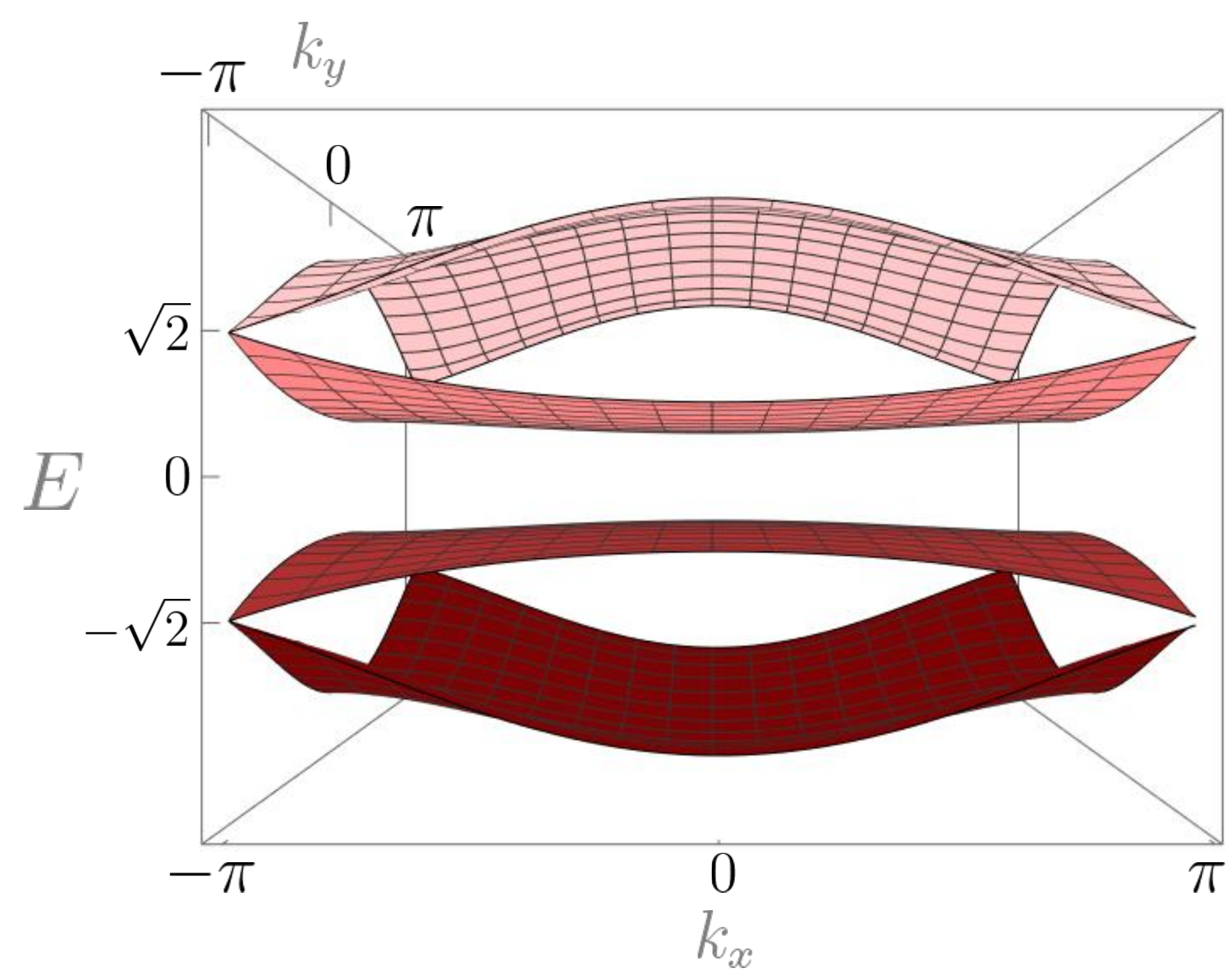}
	\hfill
	\includegraphics[width=0.32\textwidth]{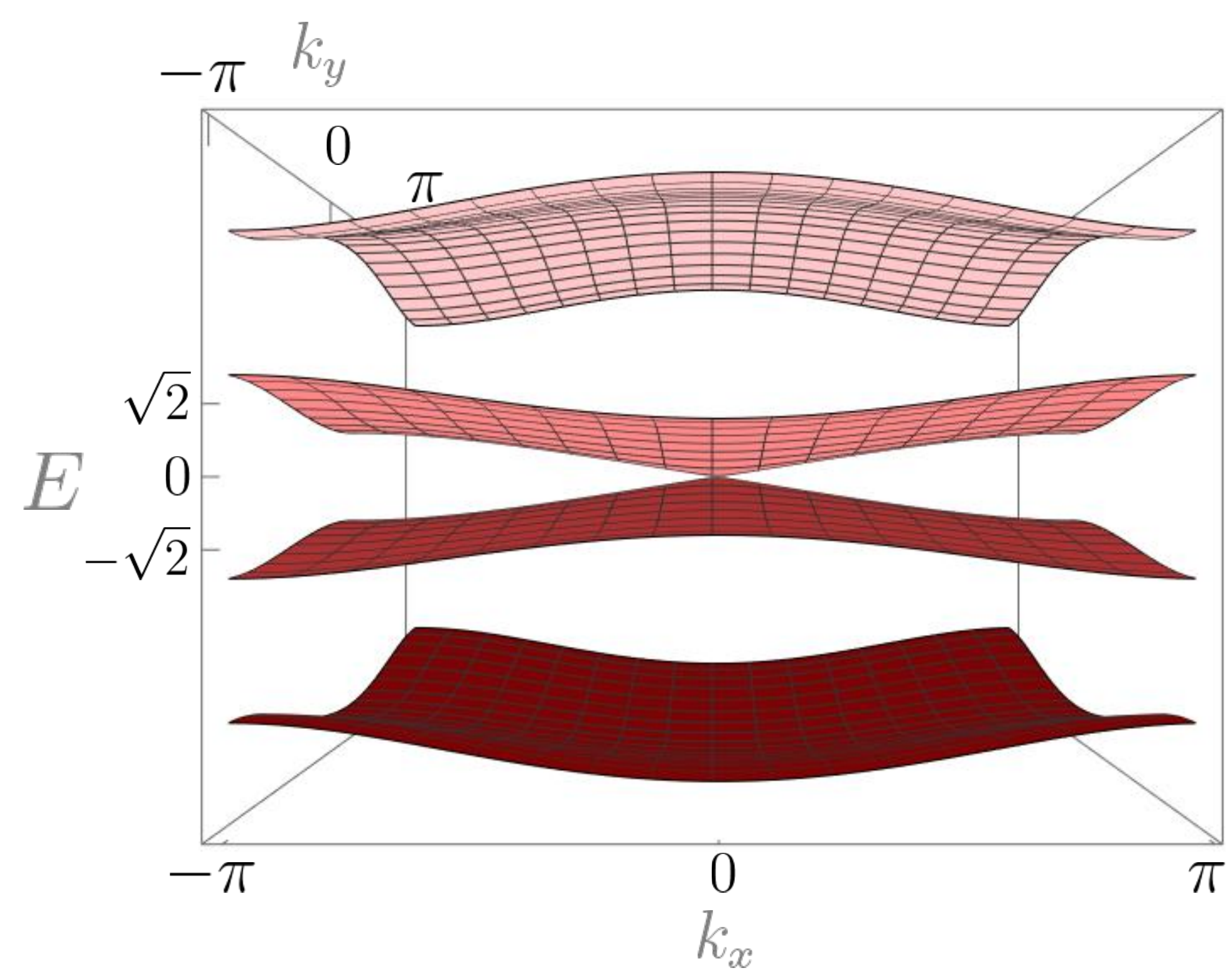}
	\caption{Band structure of the (four-band) Hamiltonian $H= -\imagunit (C(k)+A)$ for $A_{i,j}=a \operatorname{sgn}(j-i)$ at $a=a_-$ (left), $a=1$ (center) and $a=a_+$ (right).}
	\label{fig:bands}
\end{figure*}

A two-dimensional superconductor in class~D carries a $\Zbb$-valued topological
index $\mathcal C$ (total Chern
number)~\cite{AltlandZirnbauer,QiZhang2011review}. Evaluating this invariant
along the free line gives the progression $\mathcal C: 0 \to 1 \to 0$ for $a \in
(0,a_-)$, $a \in (a_-, a_+)$, and $a \in (a_+, \infty)$, respectively, with gap
closings at $a_\pm = \sqrt2 \pm 1$ and momenta $k_\pm$, where $k_+=(0,0)$,
$k_-=(\pi,\pi)$ (cf.\ Sec.~\ref{subsec:fermionic-topo} for details). It follows
that $H(k)$ exhibits topological phase transitions at values of $a = a_{\pm}$,
with the topological phase, $a_- < a < a_+$, characterized by the total Chern
number\footnote{Only the total Chern number of the occupied subspace is a stable
topological invariant of the gapped phase; the individual band Chern numbers may
rearrange when bands touch away from the Fermi level: cf.\ discussion in
Sec.~\ref{subsec:fermionic-topo}.} $\mathcal C=1$. Projecting onto the two bands
closest to zero energy in the vicinity of one of these phase transitions leads
to a convenient description by the two-band low-energy Hamiltonian $H(q) =
\sum_{i=1}^3 h_{i}(q) \sigma_i$ (cf.\ Ref.~\cite{QiWuZhang_2006,Usoltcev2026})
at relative momentum $q=k-k_\pm$.

Away from the free-fermion limit, sufficiently weak quartic terms shift the
above transitions only perturbatively and do not immediately destroy the
topological phase~\cite{Wille_quarticTN_2025}. This shift is presented
quantitatively in Sec.~\ref{subsec:phase-diag}.

The superconductor formulation is the natural one for topology: it makes the
free-line invariant explicit and motivates the interacting parity diagnostic of
Sec.~\ref{sec:topo}.

\subsection{Phase diagram} 
\label{subsec:phase-diag}

We now summarize the phase diagram of the model. Throughout this subsection, we use `empty', `topological', and `full' as convenient labels for the three parameter regions, following their loop-gas and fermionic
interpretations; the tensor network itself remains an agnostic contraction object. In the quartic-Ising description, the same regions correspond to the ferromagnetic, paramagnetic, and antiferromagnetic phases, respectively.

\paragraph{Phase structure.}

The phase diagram (cf.\ Fig.~\ref{fig:dualities}) in the $(a,b)$ plane contains three phases. 
For small to moderate $b$, we find an empty phase at small $a$, a full phase at large $a$, and a topological phase in between. The topological phase is separated from the other two phases by continuous transition lines.
At large $b\simeq1$, a first-order coexistence line between the empty and full phases meets these two critical lines at a multicritical point. This coexistence line forms part of, but is not identical to, the high-symmetry line $b=1-a^2$, discussed in detail in Sec.~\ref{sec:self-dual}. There we also derive the multicritical point $M=(1/3,8/9)$.

The three phases admit the following equivalent descriptions:
\begin{enumerate}
    \item[i)]
    \emph{Empty phase:} loops are suppressed in the loop-gas picture. In the
    quartic-Ising formulation this is the ferromagnetically ordered phase,
    while along the free-fermion line it is a trivial phase with
    $\mathcal C=0$.

    \item[ii)]
    \emph{Topological phase:} loops are deconfined and fluctuate on all scales.
    In the quartic-Ising formulation this is the disordered (paramagnetic)
    phase, while along the free-fermion line it is the $\mathcal C=1$
    topological phase.

    \item[iii)]
    \emph{Full phase:} loop occupation is favored, leading to densely packed
    configurations. In the quartic-Ising formulation this is the
    antiferromagnetically ordered phase, while along the free-fermion line it
    is again a trivial phase with $\mathcal C=0$, not adiabatically connected
    to the empty phase within the present two-parameter description.
\end{enumerate}

Two exactly solvable lines anchor the phase diagram:
\begin{itemize}
    \item
    \emph{Free-fermion line $b=0$:} the model reduces to a quadratic Majorana
    problem, equivalently the nearest-neighbor Ising model, and can therefore
    be analyzed by standard free-fermion methods.

    \item
    \emph{Self-dual line $b=1-a^2$:} the model maps to the zero-field
    eight-vertex model and is exactly solvable by Baxter's
    commuting-transfer-matrix construction and the associated Bethe
    ansatz~\cite{BaxterExactly}. For small $a$, this line coincides with the
    coexistence line between the empty and full phases; beyond the
    multicritical point \textsf{M}, it continues through the topological phase.
\end{itemize}

Along the free-fermion line $b=0$, the transitions at $a_\pm$ reduce to the
standard nearest-neighbor Ising critical points and therefore lie in the Ising
universality class with central charge $c=1/2$~\cite{DiFrancesco1997}. The
thermodynamic-limit numerics strongly support the continuation of this
universality class along both continuous transition lines away from $b=0$.
By contrast, the multicritical point \textsf{M} is governed by a $c=1$ theory
and is identified with the four-state Potts point of the Ashkin--Teller critical
line. The analytical interpretation of this multicritical structure is
developed in Sec.~\ref{subsec:spins-AT}.

\begin{figure*}[t]
	\centering
	\raisebox{2.5mm}{\includegraphics[width=0.45\linewidth]{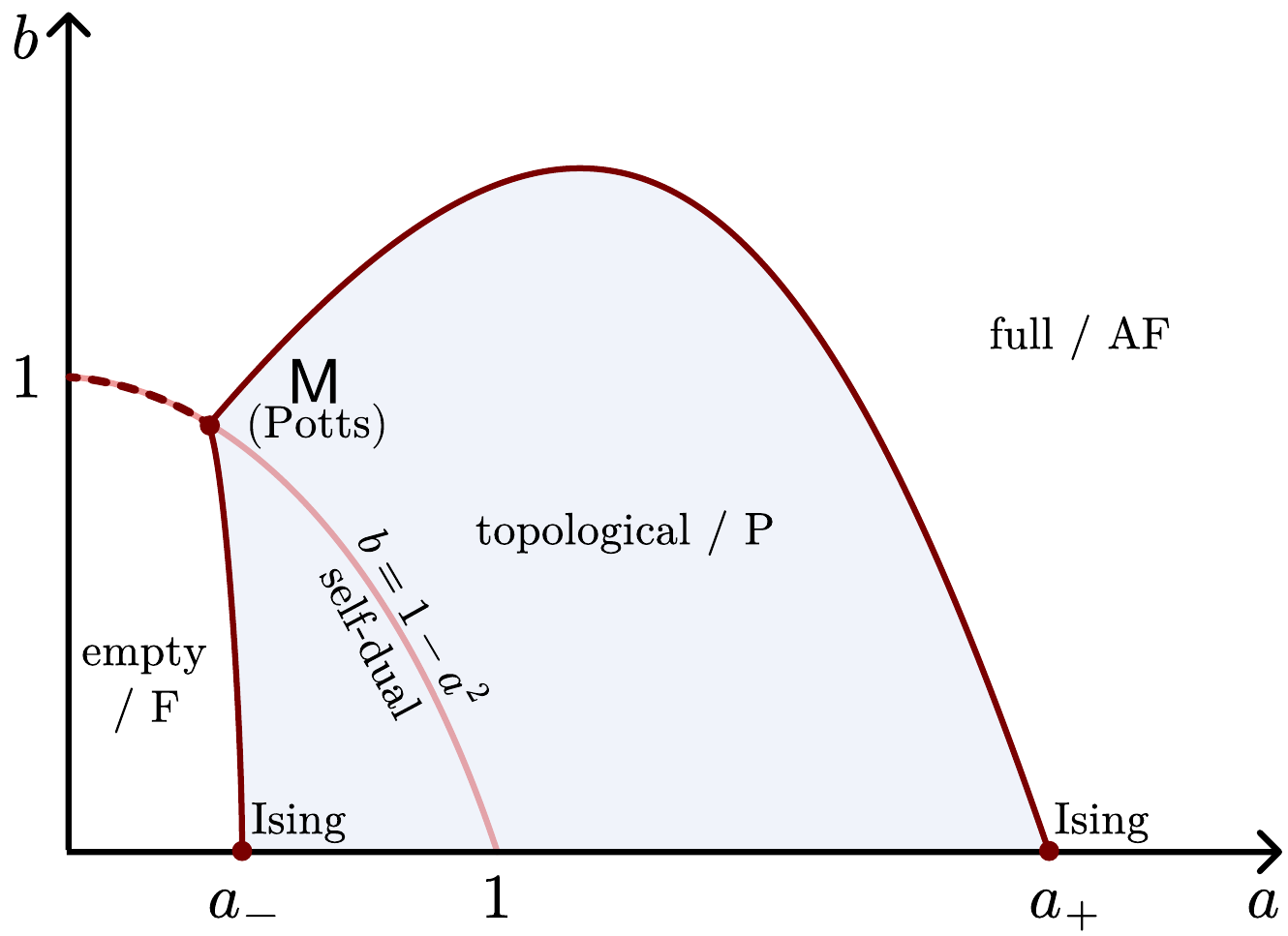}}
	\hspace{1mm}
	\includegraphics[width=0.49\linewidth]{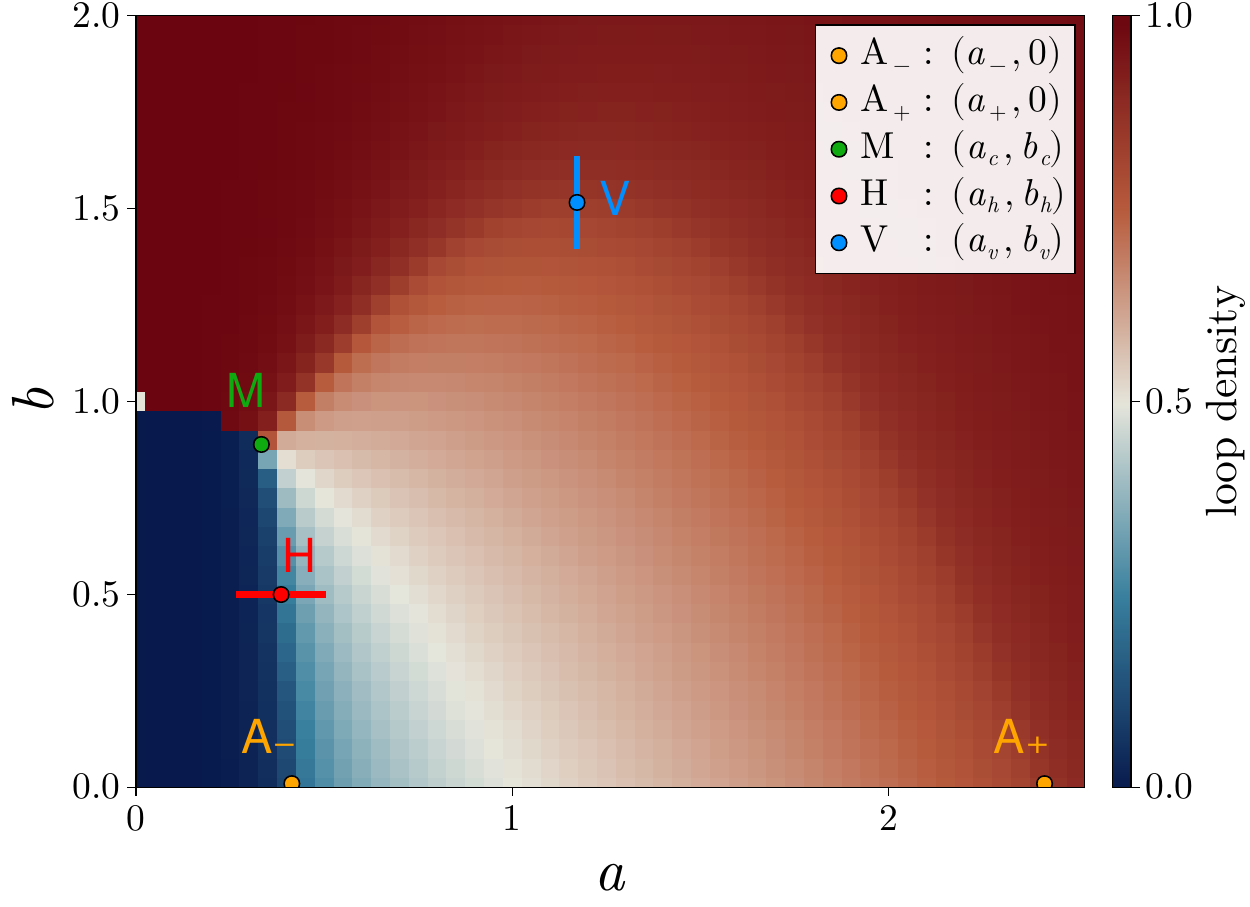}
	\caption{
	The numerical phase diagram obtained from the loop-density observable $\rho_{\tx{loop}} = \frac12(1-\av{\sigma_z})$, with $\rho_{\tx{loop}} = 0$ deep in the empty phase and $\rho_{\tx{loop}} = 1$ deep in the full phase. The points $\mathsf{A}_{\pm}$ mark the free-fermion (Ising) critical points and \textsf{M} marks the multicritical (Potts) point.
}%
	\label{fig:pd-num}
\end{figure*}

\paragraph{Numerical analysis.}

\begin{figure*}[t]
    \centering

    \begin{minipage}{0.43\textwidth}\centering
        \includegraphics[width=\textwidth]{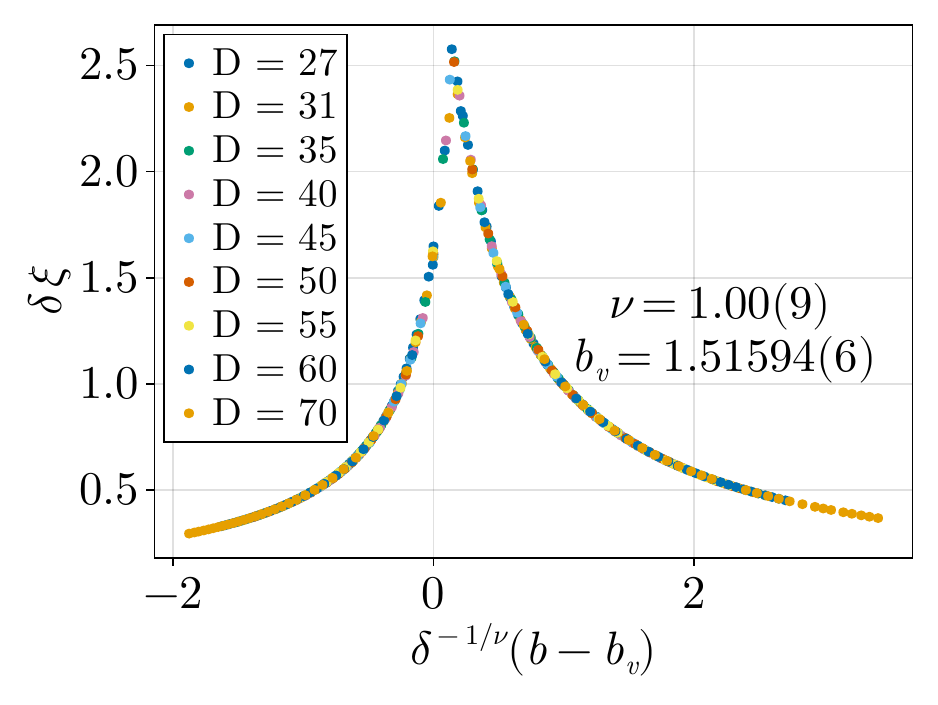}
        \par\smallskip\textbf{(a)}
    \end{minipage}%
    \hspace{1em}
    \begin{minipage}{0.43\textwidth}\centering
        \includegraphics[width=\textwidth]{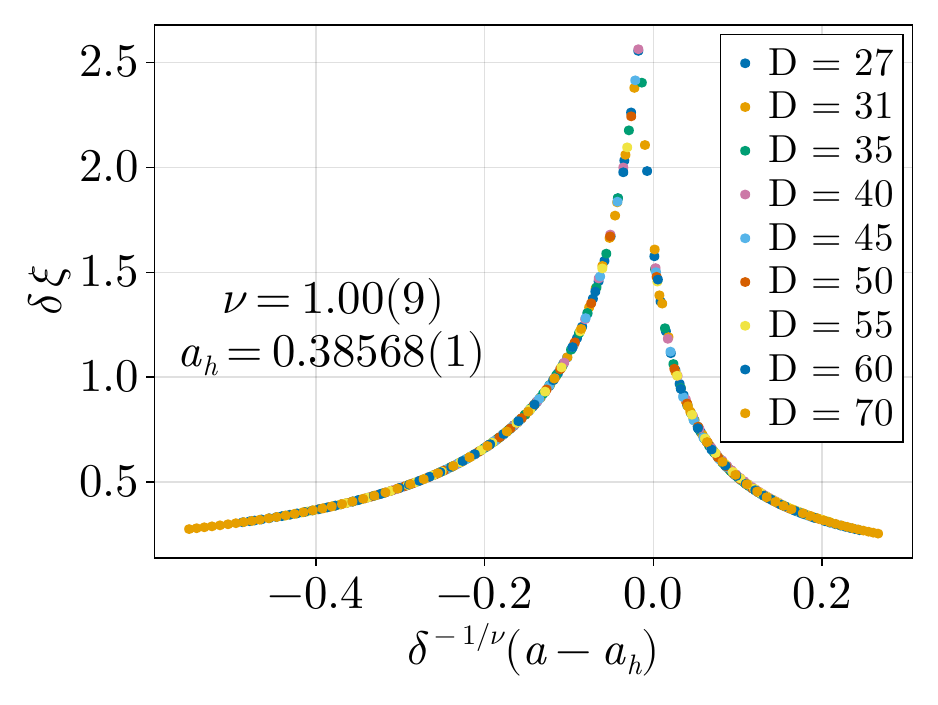}
        \par\smallskip\textbf{(b)}
    \end{minipage}
    
        \par\medskip
    
    \begin{minipage}{0.43\textwidth}\centering
                \includegraphics[width=\textwidth]{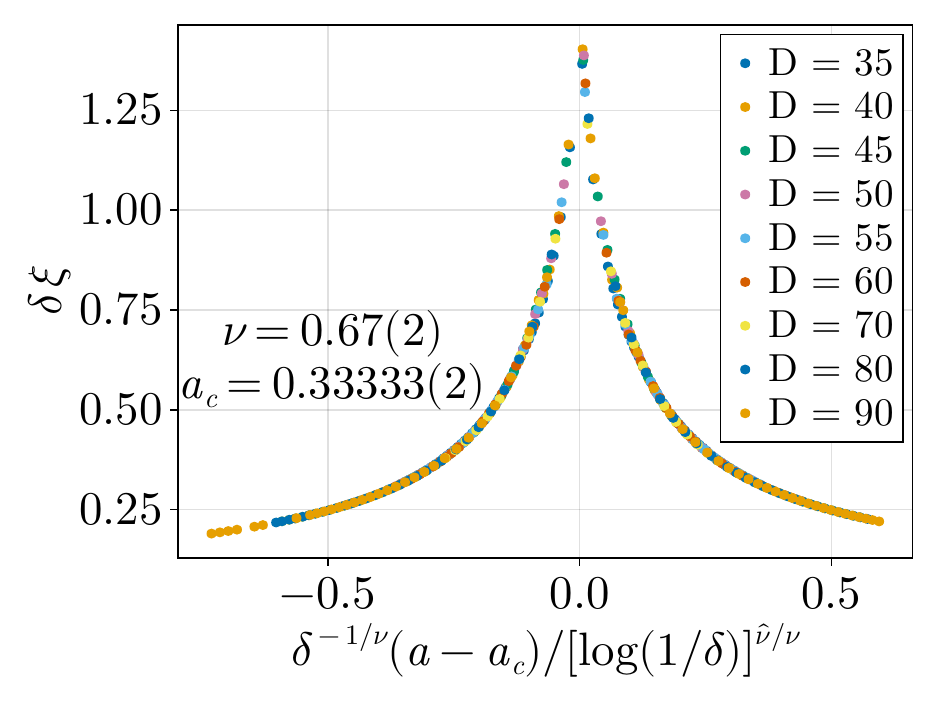}
        \par\smallskip\textbf{(c)}
    \end{minipage}%
    \hspace{1em}
    \begin{minipage}{0.43\textwidth}\centering
        \raisebox{2.5mm}{\includegraphics[width=\textwidth]{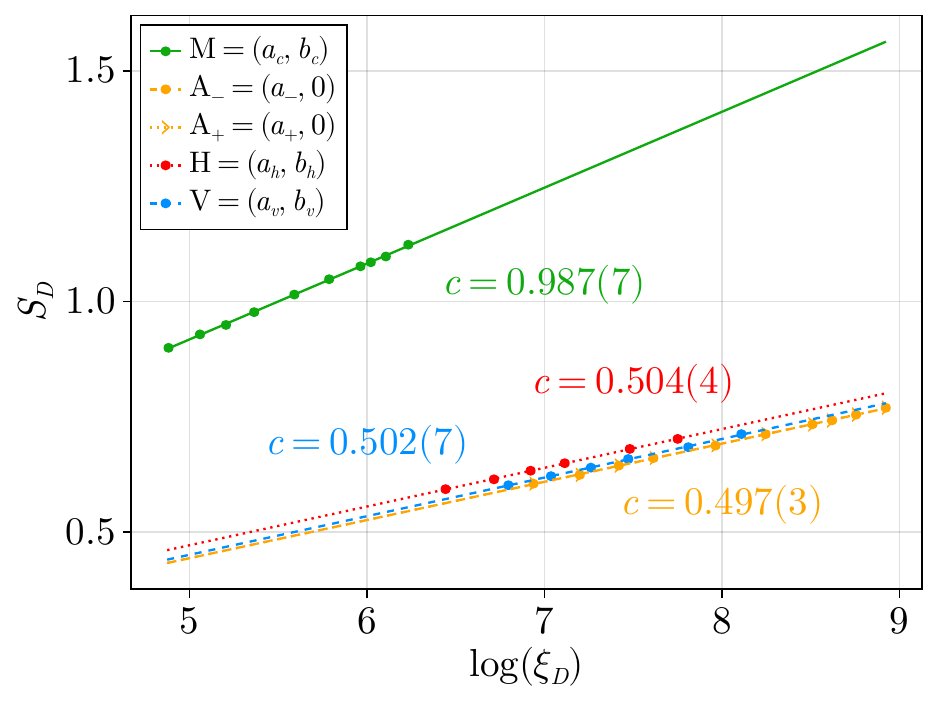}}
        \par\smallskip\textbf{(d)}
    \end{minipage}
     \caption{
     Determination of critical exponents and central charges. (a--c) Finite-entanglement scaling collapse of the correlation length $\xi_D$ along different cuts of the $(a,b)$-plane (see Fig.~\ref{fig:pd-num}). (a) Vertical cut at $a=1.172$, (b) horizontal cut at $b=0.5$, both collapsed according to Eq.~\eqref{eq:delta-based-app} and (c) cut along the self-dual line collapsed according to Eq.~\eqref{eq:delta-log-corrected-app} including logarithmic corrections.
     (d) Scaling of the entanglement entropy as a function of correlation length for bond dimensions $D \in \{35, 40,\dots, 70 \}$ fitted to Eq.~\eqref{eq:central-charge-app} at the four points $\mathsf A_\pm$, $\mathsf H$, $\mathsf V$ on the critical lines and at the multicritical point $\mathsf M$.
    }%
    \label{fig:data-collapse}
\end{figure*}

To map out the phase diagram directly in the thermodynamic limit, we contract the bosonic tensor network using boundary-MPS-based methods, CTMRG and VUMPS~\cite{Naumann_2024,Zauner2018,Fishman_2018}. 
Full details of the numerical procedure and scaling analysis are given in App.~\ref{app:numerics}.

Using the expectation value of the loop density at a single site,
\[
    \rho_{\tx{loop}}=\frac12(1-\av{\sigma_z}),
\]
as a diagnostic of the local loop occupation, we obtain a direct visualization of the phase-diagram structure discussed above, as shown in Fig.~\ref{fig:pd-num}. 
In particular, the empty and full
phases correspond to low- and high-density regimes, with
$\rho_{\tx{loop}}\to0$ and $\rho_{\tx{loop}}\to1$, respectively, deep inside the two phases. The topological phase lies between
them. Along the high-symmetry line $b=1-a^2$, the exchange symmetry between occupied and unoccupied links fixes $\rho_{\tx{loop}}=1/2$.

We also identify the two distinct parts of this high-symmetry line. For
$0\leq a<1/3$, it coincides with the first-order coexistence line between the empty and full phases, marked by a discontinuity in $\rho_{\tx{loop}}$. At $a=1/3$, this discontinuity terminates at the multicritical point, while for $1/3 < a \leq 1$ the line continues inside the topological phase with
$\rho_{\tx{loop}}=1/2$. This provides direct numerical support for the location of the multicritical point.

Turning towards a more detailed analysis of the critical lines, we monitor the transfer-matrix correlation length across the $(a,b)$ plane to resolve their location and characterize their universality class by performing finite-entanglement scaling of the MPS's correlation length and entanglement entropy. Representative scaling collapses are shown in Fig.~\ref{fig:data-collapse}(a) and (b). Along both continuous transition lines, the extracted correlation-length critical exponent is consistent with the Ising value $\nu=1$. 

We moreover perform a finite-entanglement scaling of the correlation length at the multicritical point \textsf{M}. Anticipating the identification of \textsf{M} with the four-state Potts critical point as argued in Sec.~\ref{subsec:spins-AT}, we include the known logarithmic corrections to the correlation-length scaling ansatz (cf.\ Eq.~\eqref{eq:delta-log-corrected-app}). The resulting scaling collapse
(cf.\ Fig.~\ref{fig:data-collapse}(c)) yields a correlation-length exponent consistent with the four-state Potts value,
$\nu=\nu_{\tx{4P}}=2/3$.

Entropy scaling, shown in Fig.~\ref{fig:data-collapse}(d), yields the expected central charge $c\simeq1/2$ at all four points considered along the continuous transition lines: the two points $a_\pm$ at $b=0$ and one point on each of the two finite-$b$ branches. At the multicritical point \textsf{M},
however, we find $c\simeq1$, again consistent with its identification as the four-state Potts point.

\paragraph{Analytical analysis.}
The numerical phase diagram can be complemented by a perturbative analysis in the
fermionic formulation. Along the free line $b=0$, the tensor-network contraction reduces to a Gaussian Majorana problem with Bloch Hamiltonian $H(k)$, whose gap closes at $a_\pm=\sqrt2\pm1$. Following the strategy of our prior work~\cite{Wille_quarticTN_2025}, we treat the quartic term at small $b$ as a perturbation and track the position of the gap closings as we move from $(a=a_\pm,b=0)$ into the interacting region. To do so, we construct a low-energy effective action by integrating out the high-bands and retain an (approximate) quadratic theory in the low-energy variables.

After Fourier transformation and decomposition into low- and high-band modes $\theta_\ell, \theta_h$, at first order in $b$, the quartic interaction separates into four classes. Terms with an odd number
of high-band fields vanish under the Gaussian integration at this order. 
Terms containing four high-band fields generate only a field-independent constant and can therefore be discarded from the low-energy effective action.
Similarly, the four-low-band terms vanish at zeroth order in momentum (a product of four fields constructed from only two distinct low-band variables necessarily contains a repeated Grassmann variable), and the higher order contributions in momentum contain at least two gradients, thus being irrelevant at large length scales~\cite{Wille_quarticTN_2025}.
The leading correction to the quadratic
low-energy theory therefore comes from terms containing two low- and
two high-band fields.

We label the two free transition points by $s=\pm$, with $(a_-,k_-)=(\sqrt2-1,(\pi,\pi))$ and $(a_+,k_+)=(\sqrt2+1,(0,0))$,
and write $q=k-k_s$ for the momentum relative to the corresponding
gap-closing point.
We then contract the high-band pair using its free Gaussian covariance,
\begin{equation}
    \left\langle
        \theta_{h,i}^{(s)}(p)
        \theta_{h,j}^{(s)}(p')
    \right\rangle_{0,h}
    =
    \Gamma_{h,s}^{ij}(p)\,
    \delta_{p,-p'},
    \label{eq:high-band-contraction}
\end{equation}
where $p$ is the internal momentum carried by the contracted
high-band modes. 
After summing over $p$, the remaining contribution is quadratic in the low-energy fields and diagonal in their external
momentum $q$. Denoting the two low-band Grassmann modes on branch $s$
by $\eta_\pm^{(s)}(q)$, we obtain
\begin{equation}
\begin{aligned}
    \delta(-S_{\ell,s})_{\mathrm{quad}}
    &=
    b\sum_q
    \mu_s(q)\,
    \eta_+^{(s)}(-q)
    \eta_-^{(s)}(q),
    \\
    \mu_s(q)
    &=
    \mu_s+\mathcal O(q^2),
\end{aligned}
    \label{eq:main-text-tadpole}
\end{equation}
where $\mu_s\equiv\mu_s(0)$. The coefficient $\mu_s(q)$ contains the
internal momentum sum of the high-band covariance, weighted by the
low- and high-band eigenvectors. This momentum-dependent construction
was derived in Ref.~\cite{Wille_quarticTN_2025}; the corresponding
lattice expressions used here are summarized in
App.~\ref{app:tadpole-slopes}.

The interaction thus shifts the zero-momentum mass of the two-band
Hamiltonian on branch $s$ by $b\mu_s$. We define
\begin{equation}
    m_s(a,b)
    :=
    h_{2,s}(0,a)
    +
    b\mu_s,
    \label{eq:effective-mass-condition}
\end{equation}
where $h_{2,s}$ is the coefficient of $\sigma_2$ in the bare
two-band Hamiltonian expanded around $(a_s,k_s)$. The continuation of
the corresponding critical line is obtained by setting
$m_s(a,b)=0$ and solving for $b_s(a)$.

\begin{figure*}[t]
    \centering
    \begin{minipage}{0.51\textwidth}\centering \label{fig:pd-slopes-global}
        \includegraphics[width=\textwidth]{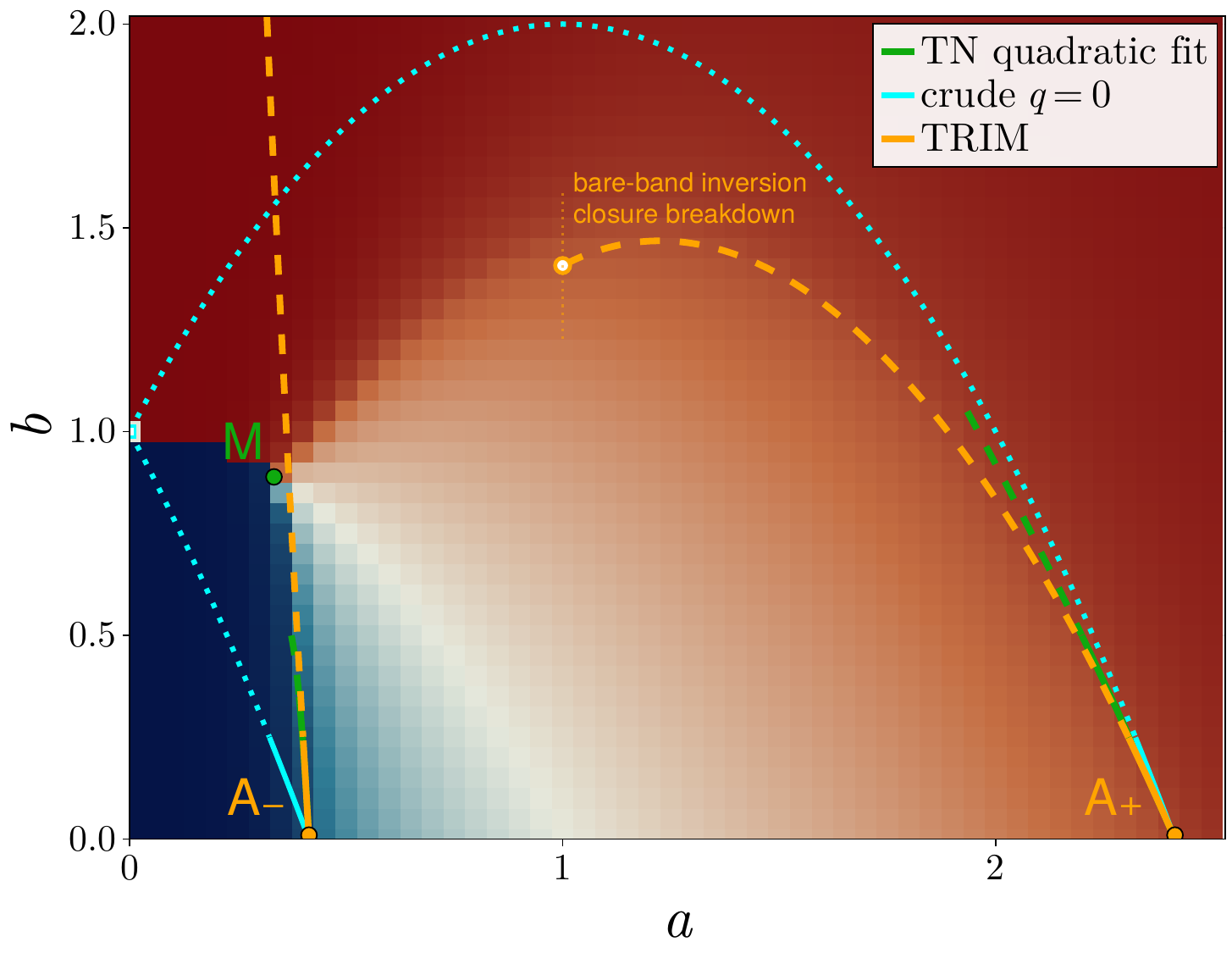}
        \par\smallskip\textbf{(a)}
    \end{minipage}%
    \hspace{2mm}
    \begin{minipage}{0.205\textwidth}\centering \label{fig:pd-slopes-zoom-min}
        \raisebox{-0.5mm}{
        \includegraphics[width=\textwidth]{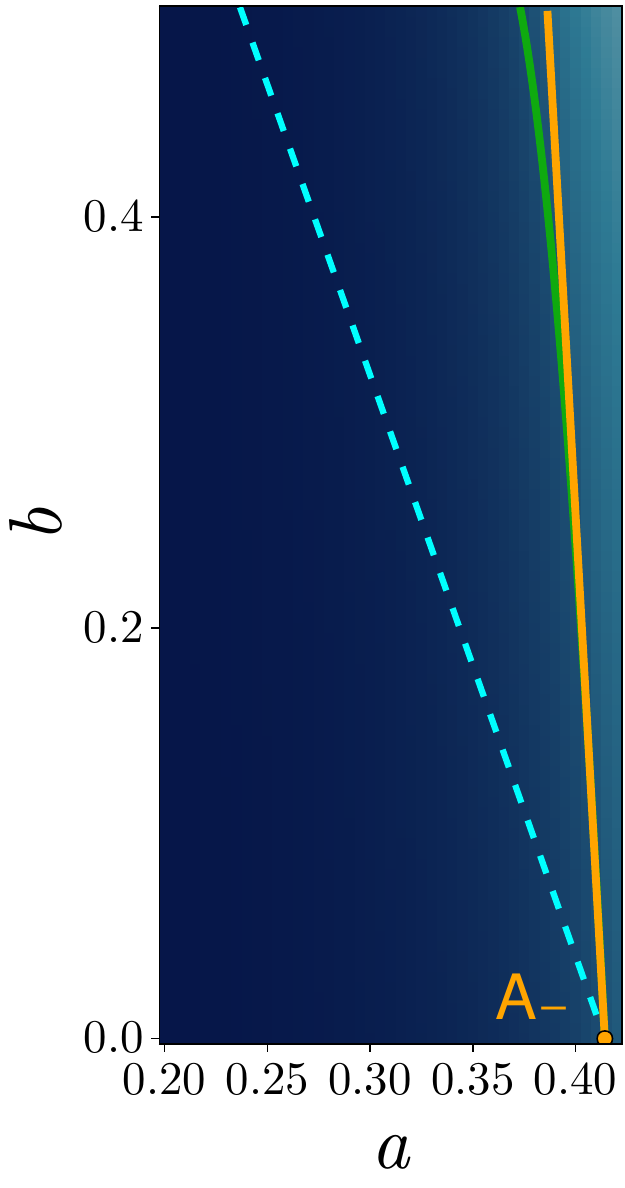}}
        \par\smallskip\textbf{(b)}
    \end{minipage} 
    \hspace{1mm}   
    \begin{minipage}{0.205\textwidth}\centering \label{fig:pd-slopes-zoom-plus}
		\raisebox{-0.5mm}{                
        \includegraphics[width=\textwidth]{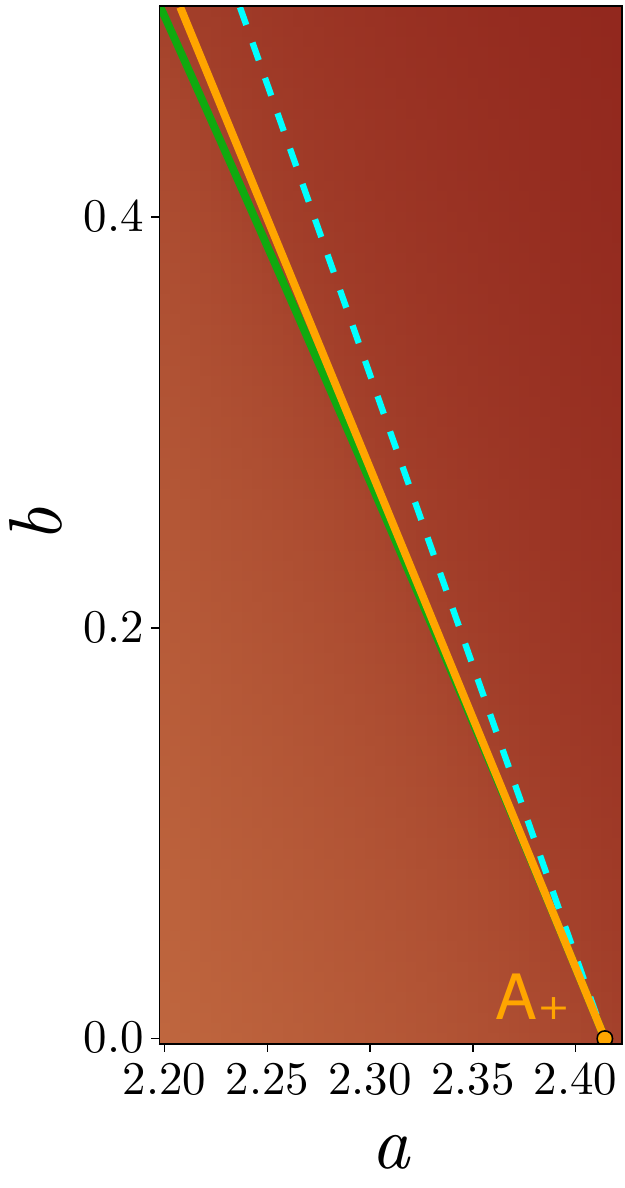}}
        \par\smallskip\textbf{(c)}
    \end{minipage}%
    \caption{
 Estimates of the phase boundaries emanating
from the free-fermion points
$\mathsf A_\pm=(\sqrt2\pm1,0)$. The background shows the loop-density calculated via tensor network numerics as in
Fig.~\ref{fig:pd-num}.
(a) Estimates obtained by the simple single-momentum calculation (cyan), cf.\ Eq.~\eqref{eq:single-point-domes-main}, and the 4-momenta (TRIM) calculation (orange) compared to quadratic fits of the boundaries obtained from the TN calculation (green).
The exact multicritical point
$\mathsf M=(1/3,8/9)$ is included as a reference.
(b,c) Close-ups near $\mathsf A_\pm$ comparing the slopes of the critical lines.
    }%
    \label{fig:pd-slopes}
\end{figure*}

While the technical details are provided in App.~\ref{app:tadpole-slopes},
we now summarize the three levels of approximation we use to compute the
first-order mass shift on each branch. The crudest estimate evaluates
$\Gamma_{h,s}$ only at the critical momentum. 
A more symmetric four-point estimate averages over the self-inverse momenta $(0,0),(\pi,0),(0,\pi),(\pi,\pi)$; we refer to this as the TRIM\footnote{In condensed matter literature, TRIM stands for ``time-reversal invariant momenta'', i.e., such momenta $k$ that $k = - k \mod 2\pi$. Here, we use this as a shorthand for such self-inverse momenta; this terminology does not imply the presence of
time-reversal symmetry in our class~D setting.} average. Finally, the
most reliable version integrates $\Gamma_{h,s}(p)$ over the full Brillouin zone. Since $b$ is the vertical axis in Fig.~\ref{fig:pd-num}, we state the resulting slopes as $\frac{\dd b}{\dd a}$:
\begin{center}
    \renewcommand{\arraystretch}{1.2}
    \setlength{\tabcolsep}{7pt}
    \begin{tabular}{c|cc}
        & $a_-=\sqrt2-1$ & $a_+=\sqrt2+1$
        \\
        \hline
        crude $q=0$ estimate & $-2.83$ & $-2.83$ \\
        TRIM average         & $-17.91$ & $-2.43$ \\
        full BZ estimate     & $-18.06$ & $-2.49$ \\
        TN linear fit        & $-15.13$ & $-2.42$ \\
        TN quadratic fit     & $-19.22$ & $-2.43$
    \end{tabular}
\end{center}
The labels `TN linear fit' and `TN quadratic fit' refer to fits of the TN data that we use as comparison. The local $q=0$ estimate captures the sign of the shift, but misses the strong asymmetry between the two sides of the phase diagram.
Including the momentum dependence of the high-band covariance repairs
this: the full-BZ contraction gives a steep tangent near
$a_-$ and a much gentler one near $a_+$, in good agreement with the numerical quadratic fits. The quadratic fit is the better comparison close to $b=0$, since it is constrained to pass through the exact free point while allowing for the visible curvature of the numerical line; the linear fit is more sensitive to the
chosen fitting window. The corresponding local
comparison is shown in Figs.~\ref{fig:pd-slopes}(b)
and~\ref{fig:pd-slopes}(c).

Beyond the strictly perturbative neighborhood of $b=0$, the same
fermionic reductions remain surprisingly instructive for the
qualitative behavior of the phase boundaries; the two finite-$b$
continuations are shown in Fig.~\ref{fig:pd-slopes}. First, retaining
only the low- and high-energy pairs at the corresponding critical
momentum $k_s$ --- the `crude $q\!=\!0$' approximation --- gives a simple expression for the mass shift, 
\begin{equation}
\begin{aligned}
    m_s^{(0)}(a,b)
    &=
    E_{\ell,s}(a)
    +
    \frac{b}{E_{h,s}(a)}
    \\[0.5ex]
    &\Rightarrow
    \quad
    b_s^{(0)}(a)
    =
    -E_{\ell,s}(a)E_{h,s}(a),
\end{aligned}
    \label{eq:single-point-mass-main}
\end{equation}
where both pair energies are evaluated at $k=k_s$, i.e., at $q=0$ (cf.\ Eq.~\eqref{eq:app-single-point-pair-energies} for explicit energy expressions and sign conventions).
This leads to two quadratic curves for the critical lines,
\begin{equation}
    b_-^{(0)}(a)=1-2a-a^2,
    \qquad
    b_+^{(0)}(a)=1+2a-a^2.
    \label{eq:single-point-domes-main}
\end{equation}
These curves are exact within the single-momentum problem and form a
closed dome terminating at $(a,b)=(0,1)$. This back-of-the-envelope
calculation already captures the bounded shape of the topological
region, the direction in which both transition lines move, and their
eventual merger, although it misses both the strong asymmetry of their
initial slopes and the location of the multicritical point. (It nevertheless hints at the existence of a multicritical point for small $a$ and large $b\approx 1$).

A more accurate finite-$b$ continuation is obtained from the TRIM
average as follows. For each branch $s$, we keep the low-mode coefficients fixed at the corresponding free critical point $(a_s,k_s)$, but reevaluate the bare high-band covariance at the four self-inverse momenta as $a$ is varied. The resulting coefficient $\mu_s^{\rm TRIM}(a)$ defines the critical lines
\begin{equation}
    b_s^{\rm TRIM}(a)
    =
    -\frac{h_{2,s}(0,a)}
    {\mu_s^{\rm TRIM}(a)}.
    \label{eq:running-trim-main}
\end{equation}
The expression for the mass shift $\mu_s^{\rm TRIM}(a)$ is less tractable analytically, but can be easily tracked numerically. 
The resulting continuation gives a considerably better account of the
quantitative phase boundaries. However, in this approach, the branch emanating from $\mathsf A_+$ terminates at $a=1$: at the opposite self-inverse momentum $k=(\pi,\pi)$, the two positive-energy Gaussian band pairs become degenerate, so that their separation into retained low modes and integrated high modes is no longer defined. 
(The single-momentum $\mathsf A_+$ approximation above remains smooth because it samples only $k_+=(0,0)$ and therefore does not resolve this higher-band crossing.)

As discussed in the next section, the Gaussian band crossing at
$a=1$ continues into the interacting problem along the self-dual
line. Thus, it is natural to expect the breakdown of one low-energy theory following one of the two $\mathsf{A}_\pm$ branches to be located at the self-dual line, rather than at a fixed, static $a=1$. However, our approximation of the mass shift so far uses the bare band structure. A more refined approach would renormalize the entire band structure instead, corresponding to a self-consistent self-energy calculation that can be solved iteratively.
We leave this extension for future work.

\section{Self-dual line and multicritical point} 
\label{sec:self-dual}

In this section, we discuss the significance of the special line
$b=1-a^2$ across the different interpretations of our model and use these
complementary descriptions to clarify the location and nature of the
multicritical point. Along this line, the vertex weights are invariant under
simultaneous inversion of all four adjacent links. This additional symmetry
induces a self-duality transformation, closely related to Kramers--Wannier
duality and to the $e\leftrightarrow m$ duality of the toric code.

We show that the fixed point of this self-duality is the multicritical point
\textsf{M} and determine its location, $a=1/3$. We then analyze the criticality
and universality class of \textsf{M} using the statistical-mechanical
descriptions and compare the resulting picture with tensor-network numerics.
Finally, we turn to the fermionic formulation, where the self-dual line is
associated with a higher-band closing that can be tracked into the interacting
regime, $b>0$.

\subsection{Loop gas picture: symmetries and self-duality}
\label{subsec:loop-gas}
Along the line $b=1-a^2$, vertex weights are invariant under the simultaneous exchange of occupied and unoccupied states on all four adjacent links. This implies that the expectation value of the link occupancy --- the loop density --- is exactly $1/2$ along this line. This is consistent with either of two scenarios: coexistence of the empty and full phases or, alternatively, the topological, loop-deconfined phase. We have already seen in the numerical phase diagram that both scenarios are realized along the line and are separated by a multicritical point. To track down its location, we invoke a self-duality map.

As a prerequisite, we need to establish an understanding of two special points along the $b=1-a^2$ line. While we have already established the topological phase at $a=1$, $b=0$, we now show the phase coexistence (first-order transition) at the extreme point $a=0$, $b=1$. In the loop gas picture, we observe that for \(a=0\) we can only have no strings at all or a fully occupied lattice. In a finite system with $N$ sites, the partition function (for periodic boundary conditions) reduces to $Z(0,b)=1+b^N$ and we find the loop density
\begin{equation}
    \rho_\text{loop}=\frac{b^N}{1+b^N}
    \xrightarrow[N\to\infty]{}
    \begin{cases}
        0, & b<1,\\
        1, & b>1 
    \end{cases} \;.
\end{equation}
This implies a first-order transition in $b$ with coexistence of the empty and full phases at $b=1$. 

Having established that there are only two possible scenarios\footnote{We
refrain from using the word `phase' to denote a coexistence of phases.} along
the $b=1-a^2$ line and understanding their two respective anchor points, it is
reasonable to assume that there is only one transition between the two scenarios
along this line. This assumption will be relevant momentarily.

We now identify a self-duality map along $b=1-a^2$ which relates small and large
$a$. To motivate this, let us revisit the notion of duality in the classical
two-dimensional Ising model. There, both the low-temperature and
high-temperature expansions admit a formulation in terms of a loop gas: in the
former case the loops are physical domain walls, while in the latter they arise
after a local change of variables and can be understood  particularly naturally
in the tensor-network language as a Hadamard gauge transformation on all virtual
bonds~\cite{Wille_2024,KramersWannier1941,Savit1980}. This suggests applying the
same transformation directly to the present loop-gas tensor network.

For a generic point in parameter space, the Hadamard transformation maps the
model to a different spin-like description. Along the line $b=1-a^2$, however,
the transformed tensor again admits a loop-gas interpretation. Concretely, on
this line the local tensor $T$ enjoys, in addition to the loop-gas parity
constraint $Z^{\otimes 4}T=T$, an additional inversion symmetry $X^{\otimes
4}T=T$. Since the Hadamard matrix $H$ interchanges $X$ and $Z$, the transformed
tensor
\[
T' = H^{\otimes 4} T 
\]
satisfies the same two virtual symmetries. Together with the full leg-permutation symmetry of the tensor, this implies that $T'$ has the same one-parameter loop-gas form as $T$. 
Thus, along $b=1-a^2$, the model is dual to itself: the Hadamard transformation preserves the loop-gas form and acts only by changing the string-tension parameter $a \mapsto a'$.
Imposing the normalization $T'_{0000}=1$, we find 
\begin{equation}
a'=\frac{1-a}{1+3a}.
\label{eq:selfdual_map_a}
\end{equation}

This map exchanges the two extreme limits, $(a=1,b=0) \,\leftrightarrow \, (a'=0,b=1)$, that is, a point deep in the topological phase is mapped onto a point of phase coexistence and vice-versa. Assuming that there is only one transition between the topological phase and the coexistence along this line, we can identify the fixed point of the self-duality map $a=a'=1/3$ with the multicritical point.

\paragraph{Toric code.}
As a side remark, let us note that the two-fold virtual symmetry ($X^{\otimes 4}$, $Z^{\otimes 4}$) admits a natural interpretation in the (deformed) toric-code PEPS language. In the toric code, $e$ charges and $m$ fluxes are created by $Z$- and $X$-string operators, respectively~\cite{Kitaev2003,Castelnovo2008}. Since the Hadamard gate exchanges $X$ and $Z$, it realizes the anyon-permuting symmetry $e \leftrightarrow m$~\cite{Bombin2010,Bridgeman2017}. Thus, the extra virtual symmetry $X^{\otimes 4}T=T$ present on the line $b=1-a^2$ means that, in addition to the usual $\mathbb Z_2$ loop constraint, electric and magnetic strings are treated on equal footing. In this sense, the loop-gas self-duality derived above is related to the toric-code $e\leftrightarrow m$ self-duality which, along a special parameter line, extends beyond the known self-duality of the unperturbed toric code (which lives at the fine-tuned point ($a=1, b=0$)).


\subsection{Spin picture: Ashkin--Teller and four-state Potts model}
\label{subsec:spins-AT}
We now investigate the ramifications of the self-dual line and the multicritical
point in the spin model language. As shown in Sec.~\ref{subsec:recap} and
Eq.~\eqref{eq:spinHam-Kform}, for $b=1-a^2$ our model reduces to an Ising model
without nearest-neighbor interactions ($K_1=0$), but with next-nearest neighbor
and 4-spin-exchange terms of equal strength $K_2=K_3$. 
In this regime, our model is equivalent to a particular submanifold of the
symmetric Ashkin--Teller (AT) model that is in one-to-one correspondence with
the four-state Potts (4P) model. The multicritical point is therefore identified
with the standard 4P critical point.

\vspace{0.5ex}
\noindent
\begin{minipage}[c]{0.62\linewidth}
\setlength{\parindent}{\outerparindent}
\indent
To arrive at this identification it is instructive to divide the square lattice
into two sublattices $A$ and $B$ and to denote the spins on $A$ ($B$) by
$\sigma_i$ ($\tau_i$), as illustrated schematically on the right. The diagonal (NNN)
\end{minipage}
\hfill
\begin{minipage}[c]{0.27\linewidth}
    \centering
    \includegraphics[width=\linewidth]{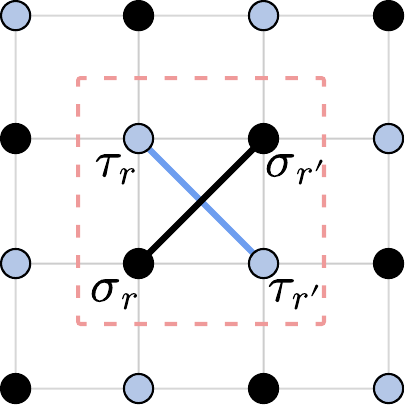}
\end{minipage}
\vspace{0.5ex}\linebreak
term couples spins within the same sublattice only, while
the plaquette product $s_1s_2s_3s_4$ on an elementary square contains two
$A$-spins and two $B$-spins. In this representation
Eq.~\eqref{eq:spinHam-Kform} becomes
\begin{equation} \label{eq:AT_from_QI_main}
	H_{\text{AT}}
	=
	-\sum_{\langle rr'\rangle}
	\Big[
	K_2 (\sigma_r\sigma_{r'}+\tau_r\tau_{r'})
	+
	K_3 (\sigma_r\sigma_{r'}) (\tau_r\tau_{r'})
	\Big],
\end{equation}
where the sums now run over nearest-neighbor bonds $\langle rr'\rangle$ of the
coarse-grained (diagonal) lattice. This Hamiltonian is known as the symmetric
Ashkin--Teller (AT) model, cf.~Refs.~\cite{AshkinTeller_1943,Aoun_AT_2023}. More
specifically, due to the equality of the couplings $K_2 = K_3=K$, the
Hamiltonian can -- up to an irrelevant additive constant -- be simplified
further to 
\begin{equation}
\begin{aligned}
    H_{\text{4P}}
    &=
    -K \sum_{\langle rr'\rangle}
    (1+\sigma_r\sigma_{r'})(1+\tau_r\tau_{r'})
    \\
    &=
    -\,4 K \sum_{\langle rr'\rangle}
    \delta_{(\sigma_r,\tau_r),(\sigma_{r'},\tau_{r'})},
\end{aligned}
\label{eq:4P_from_QI_main}
\end{equation}
where the Kronecker delta denotes projection onto pairwise equal configurations. This Hamiltonian is known as the four-state Potts (4P) model and `four-state' refers to the degrees of freedom, $(\sigma,\tau)$, which can take on four distinct states.

\begin{figure*}[t]
	\centering
	\includegraphics[width=\linewidth]{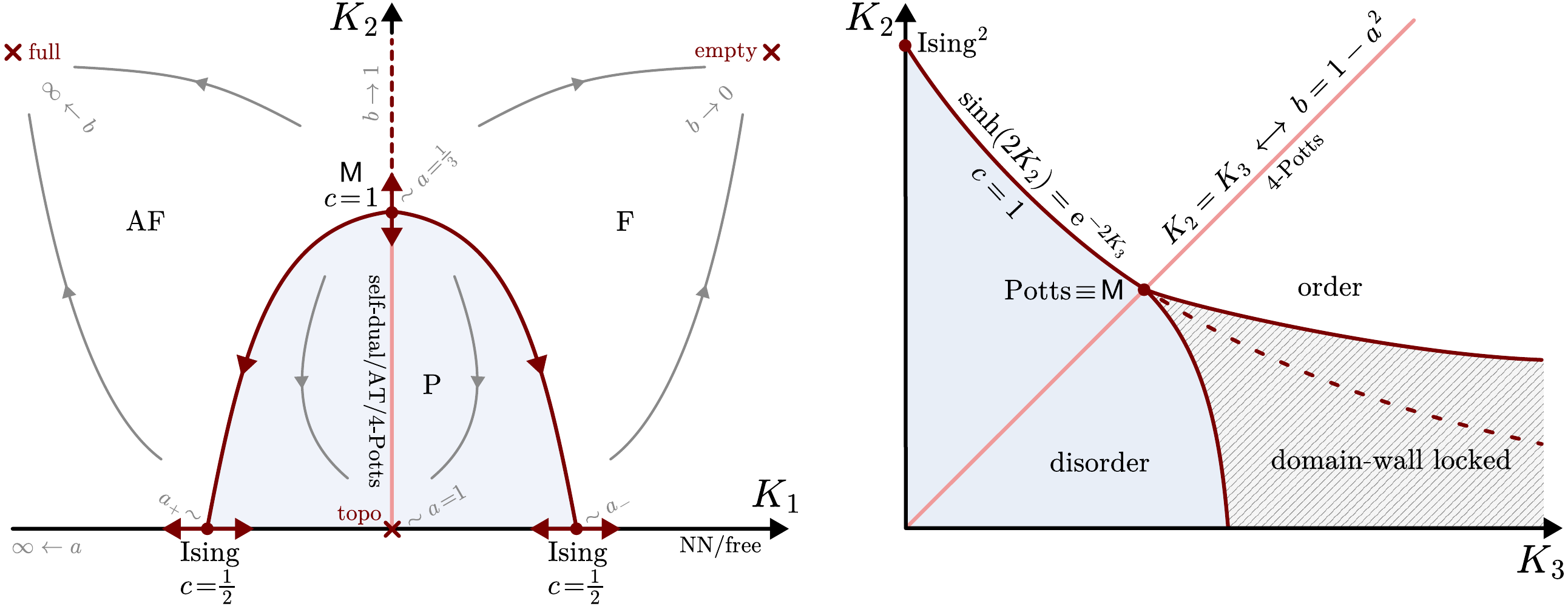}
  \caption{Two planar sections of the three-dimensional coupling space $(K_1,K_2,K_3)$. Left: the QI plane $K_2=K_3$, shown in coordinates $(K_1,K_2)$. The dashed line is the first-order separatrix between the ferromagnetic and antiferromagnetic basins; the solid boundaries are continuous transitions into the paramagnetic phase. The arrows are schematic projections of RG flows: generic perturbations flow to massive phases, while the critical separatrices run from the Potts ultraviolet fixed point at \textsf{M} toward Ising infrared fixed points, represented microscopically by $\textsf{A}_{\pm}$. Right: the symmetric AT plane $K_1=0$, shown in coordinates $(K_2,K_3)$~\cite{AshkinTeller_1943} and adapted from Ref.~\cite{Aoun_AT_2023}. The two planes intersect along the pink line $K_1=0$, $K_2=K_3$, which is the QI self-dual line and the four-state Potts line $J=U$ of the AT model.}
	\label{fig:phaseAT}
\end{figure*}

\paragraph{QI and AT as sections of a common coupling space.}
The comparison in Fig.~\ref{fig:phaseAT} is most naturally understood in the full three-dimensional coupling space $(K_1,K_2,K_3)$. The two-parameter QI model occupies the plane $\mathcal P_{\rm QI}: K_2=K_3$,
whereas the symmetric AT model occupies
$\mathcal P_{\rm AT}: K_1=0$.
The two planes intersect along the one-dimensional line $K_1=0$, $K_2=K_3$, which is simultaneously the QI self-dual line and the four-state Potts line $K_2=K_3$ of the AT model. Only the physics on this common line is exactly shared; away from it the two planes probe different microscopic perturbations and need not contain the same phases.

This distinction clarifies the phase correspondence. At $K_1=0$, the AT Hamiltonian has an independent $\mathbb Z_2$ spin-flip symmetry for each sublattice, and its phases may be characterized by
$m_\sigma=\avs{\sigma}$, $m_\tau=\avs{\tau}$, and $m_{\sigma\tau}=\avs{\sigma\tau}$.
A useful physical picture is obtained in terms of the bond, or domain-wall, variables $S_{rr'}=\sigma_r\sigma_{r'}$ and $T_{rr'}=\tau_r\tau_{r'}$.
The four-spin interaction is proportional to $-K_3S_{rr'}T_{rr'}$ and, for $K_3>0$, favors $S_{rr'}=T_{rr'}$: a domain wall in one Ising sector is therefore accompanied by a domain wall in the other. In the fully disordered phase, all three order parameters vanish and the domain walls of the two sectors fluctuate independently at long distances. For sufficiently strong positive sublattice coupling, both $\sigma$ and $\tau$ order ferromagnetically, so that $m_\sigma,m_\tau\neq0$ and domain walls are suppressed. However, since for AT $K_1=0$, the relative orientation of the two ordered sublattices is not selected. Upon returning to the QI plane, a nonzero $K_1$ lifts this degeneracy: $K_1>0$ favors equal sublattice orientations and hence the ferromagnetic QI phase, whereas $K_1<0$ favors opposite orientations and hence the antiferromagnetic QI phase. The common line inside the AT ordered regime therefore appears in the QI section as the first-order ferro--antiferro coexistence line.

Furthermore, the AT plane contains an intermediate product-ordered
(domain-wall-locked) phase~\cite{Wu_1977,Aoun_AT_2023}. There the individual Ising variables remain disordered, $m_\sigma=m_\tau=0$,
but their relative configuration is ordered,
$m_{\sigma\tau}\neq0$.
Equivalently, the two sublattices possess coincident domain-wall patterns even though neither sublattice develops magnetization by itself. This phase relies on the possibility of tuning $K_2$ and $K_3$ independently and is not encountered in the constrained QI plane $K_2=K_3$. On the weak-coupling side of the common line, both descriptions instead lie in their disordered regimes, corresponding to the paramagnetic/topological phase of the QI model. 

Despite the qualitative differences between the QI and AT phases, the analysis of the symmetric AT model can provide useful insights. Most importantly, it is known that along the 4P line (corresponding to the self-dual line of our model), the symmetric AT model has a multicritical point. Its location at $a=1/3$ ($K_2=\frac 1 4 \ln 3$) can be found by a Kramers-Wannier duality argument~\cite{FortuinKasteleyn1972,WuPottsReview1982} and its universality class can be derived from a simple argument that we review in the following.

\paragraph{Universality class.}
The AT embedding gives an intuitive argument for why the multicritical point \textsf{M} has central charge $c=1$. Consider first the limit of decoupled sublattices, $K_3=0$, with $K_2$ tuned to the critical nearest-neighbor Ising coupling. The model then consists of two independent critical Ising systems, so the continuum theory is the product CFT
\[
\mathrm{Ising}\times\mathrm{Ising},
\with
c=\frac12+\frac12=1.
\]
Turning on $K_3$ couples the two Ising copies through their bond-energy,
or equivalently domain-wall, variables. At the decoupled point this
generates the continuum perturbation
$\varepsilon_\sigma\varepsilon_\tau$, whose scaling dimension is
$\Delta_{\varepsilon_\sigma\varepsilon_\tau}
=\Delta_{\varepsilon_\sigma}+\Delta_{\varepsilon_\tau}
=1+1=2$.
The coupling is therefore marginal in two dimensions. After $K_2$ is
retuned so as to remain critical, the exact AT/eight-vertex solution
gives a critical line with continuously varying exponents connecting
the decoupled Ising$^2$ and four-state Potts points~\cite{BaxterExactly,Kadanoff1979}. Its continuum description is the
$c=1$ Gaussian, or equivalently $\mathbb Z_2$-orbifold, critical line of
a compact boson~\cite{Dijkgraaf1988,Dijkgraaf1989,Liu2026}.
This critical line intersects the Potts line at
$K_2=K_3=\frac14\ln3$, corresponding precisely to $a=1/3$ in the
QI model. We therefore identify \textsf{M} with the standard
four-state Potts critical point.

The Potts identification also organizes the leading perturbations around \textsf{M}. Motion along the common self-dual/Potts line changes only the Potts coupling $K=K_2=K_3$ and therefore acts as a temperature-like perturbation
$t\propto K_2-K_2^c\propto a-a_c$.
In the continuum, this deformation induces a coupling
$\lambda_\varepsilon$ to the Potts energy-density field
$\varepsilon(x)$. Switching on $K_1$ moves the QI model transversely
away from the AT plane and, under coarse graining, induces a coupling
$\lambda_P$ to the polarization field
$P(x)\sim\sigma(x)\tau(x)$. Thus the local two-parameter theory takes
the form
\begin{equation}
S_{\rm eff}
=
S_{4\rm P}^{(c=1)}
+\lambda_\varepsilon\int\dd^2x\,\varepsilon(x)
+\lambda_P\int\dd^2x\,P(x)
+\ldots,
\label{eq:EFT-M-main}
\end{equation}
where, to leading order near \textsf{M},
\begin{equation}
\lambda_\varepsilon \sim t+O(\delta K^2),
\qquad
\lambda_P \sim K_1+O(\delta K^2).
\label{eq:EFT-coupling-map-main}
\end{equation}

At the Potts point, the corresponding scaling dimensions are
$\Delta_\varepsilon=1/2$ and $\Delta_P=1/8$~\cite{Dijkgraaf1989,Liu2026}, so both perturbations are relevant. Generic displacements from \textsf{M} therefore flow into one of the surrounding massive phases. The two continuous phase boundaries correspond instead to tuned critical separatrices.

The thermal scaling dimension gives the RG eigenvalue
$y_t=d-\Delta_\varepsilon=3/2$ in $d=2$ dimensions~\cite{Cardy_1996}, and hence the correlation-length exponent
$\nu_{4\rm P}=y_t^{-1}=2/3$~\cite{WuPottsReview1982}.
Josephson hyperscaling then gives
$\alpha_{4\rm P}=2-d\nu_{4\rm P}=2/3$~\cite{Josephson1967,Cardy_1996}, so that the singular part of the free energy scales as
$f_{\rm sing}(t)\sim |t|^{2-\alpha}=|t|^{4/3}$.
The four-state Potts point also contains a marginally irrelevant field, which produces multiplicative logarithmic corrections. In particular,
\begin{equation}
\xi(t)
\sim
|t|^{-2/3}
\left[\log\frac1{|t|}\right]^{1/2},
\label{eq:Potts-xi-log-main}
\end{equation}
up to subleading logarithms~\cite{Salas_1997}. This correction is included in the finite-entanglement analysis of Appendix~\ref{app:numerics}.

Along either tuned separatrix, the RG flow remains massless: the four-state Potts theory governs the ultraviolet behavior close to \textsf{M}, while the infrared theory is the ordinary Ising CFT~\cite{Delfino1998,Fabrizio2000},
\begin{equation}
4\mathrm P_{\rm UV}
\longrightarrow
\mathrm{Ising}_{\rm IR},
\qquad
c_{\rm UV}=1
\longrightarrow
c_{\rm IR}=\frac12.
\label{eq:Potts-Ising-flow-main}
\end{equation}
In the microscopic QI phase diagram, these two trajectories connect \textsf{M} to the free-fermion Ising representatives $\textsf{A}_{-}$ and $\textsf{A}_{+}$. 

A fermionic interpretation provides an additional intuitive picture for this flow: at the multicritical point \textsf{M}, the two critical Ising sectors of the AT description combine into a single massless Dirac fermion, with their marginal coupling becoming the current--current interaction of the massless Thirring model~\cite{Thirring1958,Fabrizio2000},
\begin{equation}
S_{\rm Th}[\bar\psi,\psi]
=
\int\dd^2x\,
\left[
\bar\psi\,\gamma_\mu\partial_\mu\psi
+\frac{g_{\rm Th}}{2}
\bigl(\bar\psi\gamma_\mu\psi\bigr)^2
\right].
\label{eq:Thirring-main}
\end{equation}

Conversely, along either Ising separatrix, one of the two effective Majorana sectors becomes massive while the other remains critical, leaving a single-Majorana Ising CFT in the infrared~\cite{Delfino1998,Fabrizio2000,Ye2001}. In this sense, the Potts-to-Ising flow may be viewed as the reduction of a critical Dirac degree of freedom to one critical Majorana mode.
Note that this is a continuum mode-counting picture; the two Majorana sectors are those of the Ashkin--Teller representation and should not be identified directly with the two microscopic QI transition branches. The full Potts operator content and the corresponding orbifold qualification are discussed in Appendix~\ref{sec:eft-thirring}.


\subsection{Zero-field eight-vertex model and exact solvability}
\label{subsec:8V}
Our two-parameter model is a special instance of the eight-vertex model. The additional exchange symmetry along the self-dual line $b=1-a^2$ corresponds to the zero-field condition~\cite{BaxterExactly} and renders the model exactly solvable via the Bethe ansatz and commuting transfer matrices~\cite{BaxterExactly}. We briefly discuss this relation and its consequences.

General zero-field eight-vertex models are characterized by four vertex weights. Baxter's exact solution~\cite{BaxterExactly} shows that the free energy is analytic and the correlation length finite, except on the critical surfaces where one of the four weights equals the sum of the other three. 
For our case, with only two distinct weights ($1$ and $a>0$), the criticality
condition reduces to $1=3a$, yielding the multicritical point location at $a_c=1/3$. The exact eight-vertex result
therefore also confirms the assumption made in Sec.~\ref{subsec:loop-gas} that the self-dual line contains a single transition between the coexistence and topological regimes.

 \begin{figure}[t]
     \centering
     \includegraphics[width=\linewidth]{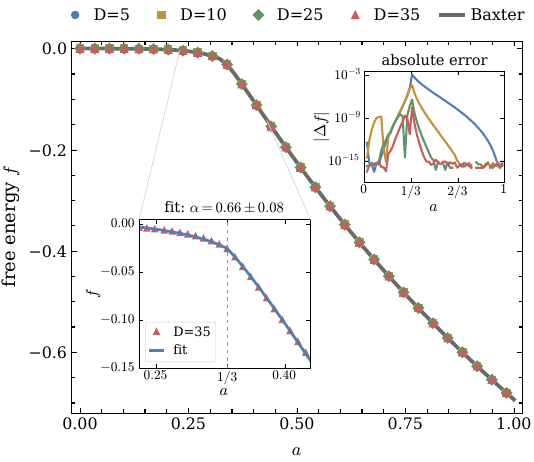}
	\caption{
	Free energy on the self-dual line calculated via Baxter's series solution compared with approximate TN contractions for different bond dimensions $D$. 
	The upper-right inset shows the absolute error $|f_{\tx{TN}}-f_{8\tx V}|$ on a logarithmic scale. 
	The lower-left inset shows the $D=35$ data fitted to Eq.~\eqref{eq:specific-heat-fit}, giving $\alpha=0.66(8)$.
}%
     \label{fig:fe-plots}
 \end{figure}

\paragraph{Free energy density from Baxter's solution.}
The exact solvability drastically reduces the computational cost. We demonstrate this by computing the free-energy density using expressions derived from the commuting-transfer-matrix ansatz~\cite{BaxterExactly} and comparing them with approximate tensor-network contractions.

Baxter's exact solution yields an expression for the 
dominant eigenvalue $\Lambda_{\max}$ of the row-to-row transfer matrix which determines the dimensionless free-energy density per vertex, $f_{8\tx V}=-\lim_{N\to\infty}\frac{1}{N}\ln \Lambda_{\max}$, where $N$ is the row length. 
The general solution is given by the series
\begin{equation}
\begin{split}
f_{8\tx V}
&=
-
2\sum_{n=1}^{\infty}
\frac{
\sinh^2\!\big[(\tau-\lambda)n\big]
\big[\!\cosh(n\lambda)-\cosh(n\zeta)\big]
}{
n\,\sinh(2n\tau)\cosh(n\lambda)
}
\\
&\quad\;
-\ln w_{8\tx V},
\end{split}
\label{eq:fe-general}
\end{equation}
where $\tau$, $\lambda$, and $\zeta$ are obtained from the elliptic parametrization of the four zero-field weights and $w_{8\tx V}$ is the empty-vertex configuration.

Baxter's free-energy density is thus obtained by evaluating the infinite series in Eq.~\eqref{eq:fe-general} after determining the corresponding elliptic parameters from the original weights. In general, this does not lead to a simple closed form. However, for our one-parameter specialization in terms of the single parameter $a$, it yields an efficient semi-analytic numerical procedure. Leaving the computational details to Appendix~\ref{app:fe}, we showcase the results in Fig.~\ref{fig:fe-plots}, benchmarked against the free energies computed via contraction of corner transfer matrices in the CTMRG procedure. 
The exact semi-analytic evaluation is orders of magnitude faster than the TN calculation. In our implementation, the TN contraction requires a wall time of order $10^2$\,s per data point, compared with $\sim10^{-2}$\,s for the semi-analytic evaluation.

We now test the specific-heat exponent against the four-state Potts value $\alpha_{4P}=2/3$. To this end, we fit the free-energy density near the multicritical point using an ansatz that isolates the leading singularity governing the specific heat,
$c_\tx{heat}=\partial_t^2 f \sim |t|^{-\alpha}$. Concretely, we use
\begin{equation}
\label{eq:specific-heat-fit}
f_{\mathrm{fit}}(a)
=
c_0
+
c_1t
+
c_2t^2
+
A |t|^{2-\alpha},
\end{equation}
where $t=3(a-a_c)$ is a temperature-like variable.\footnote{The factor 3 is included for consistency with the original literature.} 
This ansatz neglects the
multiplicative logarithmic corrections present at the four-state Potts point
and therefore captures only the leading algebraic singularity. The fit yields
$\alpha=0.66\pm0.08$, consistent with $\alpha=2/3$.

The fitted value depends appreciably on the choice of analytic background. A
more robust estimate follows from the Josephson hyperscaling
relation~\cite{Josephson1967,Cardy_1996} between $\alpha$ and the
correlation-length exponent $\nu$, using the log-corrected finite-entanglement
scaling result from the previous section. As shown in
Fig.~\ref{fig:data-collapse}(c), this gives
$\nu=0.67(2)\approx2/3$~\cite{WuPottsReview1982} and hence, from
$\alpha=2-2\nu$ in $d=2$,
$\alpha=0.66(4)\approx2/3$.
Since the scaling collapse uses the known logarithmic-correction exponent
$\hat\nu=1/2$ (cf.\ Eq.~\eqref{eq:delta-log-corrected-app}), it should be
viewed as a confirmation of four-state Potts criticality rather than an
independent determination.


\subsection{Fermion picture: band inversion}
\label{subsec:fermion-self}

In the fermionic picture, the self-dual line marks the loss of validity of one
low-energy theory (around $a_-$) and the beginning of another (around $a_+$).
This statement holds exactly at $b=0$, where the self-dual point $a=1$ marks
the closing of the higher-band gap, as shown in Fig.~\ref{fig:bands}. This band
crossing does not signal a phase transition, since it occurs away from $E=0$
and therefore does not coincide with the diverging length scales associated
with zero-energy eigenvalues of the Hamiltonian. Instead, it marks the point
where the two-band approximation breaks down because the low- and high-energy
bands are no longer separated. 

Similarly to the treatment of the critical lines
in Sec.~\ref{subsec:phase-diag}, we can track this breakdown into the finite-$b$
regime using an approximate effective quadratic theory. In contrast to the
critical-line analysis, which renormalizes the zero-energy mass gap, we here
renormalize the condition for the higher-band closing.

Concretely, for $b=0$, the band closing is given by
\begin{equation}
    \varepsilon_{2,+}(\pi,\pi)
    =
    \varepsilon_{1,+}(\pi,\pi)\;.
\end{equation}
For finite but small $b$, we consider an effective quadratic Hamiltonian
\begin{equation}
    H_{\mathrm{eff}}
    \approx
    H_2+b\,\delta H\;,
\end{equation}
where $H_2$ denotes the quadratic Gaussian Hamiltonian of
Sec.~\ref{subsec:recap}, retaining the full four-band structure and thus
distinct from the projected two-band Hamiltonian $H^{(2)}$. The correction
$\delta H$ is obtained by replacing pairs of fermionic modes in the quartic
term by their expectation value with respect to the Gaussian Hamiltonian
$H_2$. Symbolically,
\begin{equation}
    \eta_1\eta_2\eta_3\eta_4
    \approx
    \eta_1\eta_2
    \langle\eta_3\eta_4\rangle_{H_2}
    +\ldots\;,
\end{equation}
where the ellipsis represents the remaining choices of contracted pairs.

Assuming that the position of the band closing remains at
$k_x=k_y=\pi$, we calculate the shift of
$\varepsilon_{1,+}\equiv\varepsilon_{1,+}(\pi,\pi)$ from the contraction
\begin{equation*}
    \eta_{1,+}\eta_{1,-}
    \langle\eta_{2,+}\eta_{2,-}\rangle
    \propto
    \frac{1}{\varepsilon_{2,+}}
    \eta_{1,+}\eta_{1,-}\;,
\end{equation*}
and analogously for $\varepsilon_{2,+}$. To first order in $b$, this gives
\begin{equation}
    \varepsilon_{2,+}(a)
    +b\,\frac{c_1}{\varepsilon_{1,+}(a)}
    =
    \varepsilon_{1,+}(a)
    +b\,\frac{c_2}{\varepsilon_{2,+}(a)}\;,
    \label{eq:self_dual_analytic}
\end{equation}
where $c_1=1/\sqrt{2}$ and $c_2=-c_1$ are obtained from matrix elements of the
basis transformation that appears when expressing the quartic term
$b\,\theta_{1,x}\theta_{2,x}\theta_{3,x}\theta_{4,x}$ in the $H_2$
eigenbasis,
$\eta_{n,\pm}(k)=U_{n,\pm,i}(k)\theta_i(k)$.

Solving Eq.~\eqref{eq:self_dual_analytic} for $b(a)$, for $a_-<a<1$, yields
\begin{equation}
    b_{\text{self-dual}}(a)
    \approx
    1-a^2-\frac{(1-a)^2}{a}\;,
\end{equation}
which agrees to first order in $\Delta a=1-a$ with the exact result
$b=1-a^2$.


\section{Topological phase}
\label{sec:topo}


In free-fermion systems, topology is diagnosed from single-particle band
structure, e.g.\ through Berry curvature or Pfaffians at high-symmetry points.
For interacting Hamiltonian systems, such diagnostics cease to apply directly
and are instead replaced by many-body constructions based on flux insertion, or
equivalently twisted boundary conditions~\cite{NiuThoulessWu1985,AvronSeiler1985,HastingsMichalakis2015}.
We follow this approach here and construct a $\mathbb Z_2$ invariant from
fermionic partition functions for the four choices of $0$ or $\pi$ flux threaded
through the two cycles of a torus.

We then extend the duality between fermionic and bosonic tensor-network
contractions to the torus. This ultimately yields a bosonic formulation of the
topological invariant in terms of expectation values of winding-parity
observables, which admit an intuitive loop-gas interpretation and can be
evaluated numerically by finite-size tensor-network contraction.

\subsection{From Chern number to a twisted-partition-function invariant}
\label{subsec:fermionic-topo}

We now turn to the topological characterization of the loop-deconfined (paramagnetic) phase.
The starting point is the free-fermion line $b=0$, where the fermionic tensor network is
Gaussian. In this limit the Grassmann action is quadratic and defines a two-dimensional
Majorana/BdG problem in symmetry class D, $Z^{\tx{f}}=\operatorname{pf}(\imagunit H)$ with $H=-\imagunit(C+\bigoplus_x A)$.
A Fourier transformation in site space leads to a four-band Bloch Hamiltonian $H(k) = -\imagunit(C(k) + A)$. 

Along this free line, the total Chern number $\mathcal C$ of the occupied
BdG bands follows the sequence $\mathcal C: 0 \to 1 \to 0$
as $a$ crosses the gap-closing points
$a_\pm=\sqrt2\pm1$. The corresponding band-resolved Chern numbers and the
two-band descriptions of the two transitions are summarized in
App.~\ref{app:chern-numbers}. Thus, the intermediate region is distinguished
from the two surrounding phases by an odd Chern number.

For the interacting tensor network at $b>0$, however, the single-particle
band Chern number is no longer directly available as an exact diagnostic.
For the present model it is sufficient to retain only its parity: the free
phase sequence above distinguishes the topological region by
$(-1)^{\mathcal C}=-1$ and the two trivial regions by
$(-1)^{\mathcal C}=+1$. Our aim is therefore to reformulate this Chern-number
parity in a way that survives beyond the Gaussian limit.

On the free line, this can be done without evaluating Berry curvature.
At the high-symmetry momenta
($k_\star=-k_\star \!\!\mod 2\pi$) inside the Brillouin zone,
$k_\star\in\{(0,0),(\pi,0),(0,\pi),(\pi,\pi)\}$, define the four
$\mathbb Z_2$-valued quantities
\[
    \sigma_{k_\star}
    :=
    \sgn \pf (\imagunit H(k_\star)).
\]
Now note that a change of $\sigma_{k_\star}$ requires $\pf (\imagunit H(k_\star))=0$, i.e., a gap closing at the corresponding high-symmetry momentum.
In class~D, due to the symmetry of the spectrum around zero, gap closings generally come in $(k,-k)$ pairs; hence, for a gap closing away from the high-symmetry points $k_\star$, the Chern number $\mathcal C$ can only change by an even integer. An odd integer change, and thus a change in the parity of the Chern number, can therefore only occur at the unpaired momenta $k_\star$.
Consequently, the product
\begin{equation} \label{eq:s-topo-free}
	\sigma_{\tx{topo}} \coloneqq \prod_{k_\star}\sigma_{k_\star} \in \{\pm1\}
\end{equation} 
measures the parity of the free-fermion Chern number,
\begin{equation}
	\sigma_{\tx{topo}} \equiv (-1)^{\mathcal C},\quad \tx{(for }  b=0 \tx{)},
\end{equation}
in agreement with the class-D Pfaffian formulas of Refs.~\cite{Sato2010OddParity,KobayashiSato2024,Marra2024}.

This $\mathbb Z_2$ quantity moreover admits a formulation in terms of fermionic
partition functions with twisted boundary conditions. Put the fermionic system
on an $L_x\times L_y$ torus and let $\alpha_x,\alpha_y\in\{0,1\}$ label the four
possible choices, with $\alpha_\mu=0$ and $\alpha_\mu=1$ corresponding to
periodic and antiperiodic boundary conditions along the $\mu=x,y$ direction,
respectively. These four choices of fermionic boundary conditions are
equivalently referred to as the four spin structures of the
torus~\cite{CimasoniReshetikhin2007,Gaiotto_2016}. We denote the corresponding
fermionic partition functions (TN contractions) by $Z^{\tx
f}_{\alpha_x,\alpha_y}$. 

Concretely, an antiperiodic boundary condition is implemented by choosing a
boundary cut transverse to the relevant torus cycle and reversing the signs of
all fermionic connectivity terms crossing that cut, $C_{ij}\to-C_{ij}$ and
$C_{ji}\to-C_{ji}$; the resulting twisted kernel is denoted by $\imagunit
H_{\alpha_x,\alpha_y}=C_{\alpha_x,\alpha_y}+\bigoplus_xA$. In the Gaussian
limit, $Z^\tx{f}_{\alpha_x,\alpha_y}=\pf(\imagunit H_{\alpha_x,\alpha_y})$ are
Pfaffians of the respective twisted kernels; the sign of each twisted Pfaffian
is determined by the high-symmetry Pfaffian signs compatible with that boundary
condition.%
\footnote{On an odd-by-odd torus, every spin structure contains exactly one self-inverse momentum, so the four twisted Pfaffian signs correspond one-to-one to the four signs $\sigma_{k_\star}$, for fixed and consistent orientation conventions. For other lattice parities, an individual twisted Pfaffian sign may contain a product over several compatible self-inverse momenta, or none. Nevertheless, across the four spin structures each $k_\star$ occurs exactly once, so their total product still yields $(-1)^{\mathcal C}$.}
This motivates the expression
\begin{equation}
    \sigma_{\tx{topo}}
    \coloneqq
    \prod_{\alpha_x,\alpha_y\in\{0,1\}}
    \sgn Z^{\tx f}_{\alpha_x,\alpha_y}
    \in\{\pm1\}.
    \label{eq:topo-invariant}
\end{equation}
This quantity is defined whenever all four twisted partition functions are
nonzero; a zero of one of them marks a point at which its sign can change.
For $b=0$, the partition functions are Pfaffians, and
Eq.~\eqref{eq:topo-invariant} reduces to the parity of the class-D band Chern
number as above.

In condensed-matter theory, using boundary twists --- equivalently, flux insertions --- to extract topological information is standard~\cite{NiuThoulessWu1985}. A benefit of this approach is that it foregoes a momentum-space formulation. It thus remains applicable beyond the translationally
invariant free-fermion setting, e.g., in the presence of disorder~\cite{Altland2015TopoNonpert} or interactions~\cite{RaoSodemann2021}. 
In the present class-D problem, the four
discrete $0/\pi$ twists already recover the $\mathbb Z_2$ Chern parity on the
free line.

More specifically, Eq.~\eqref{eq:topo-invariant} closely parallels the
translation-resolved fermion-parity indicators of
Refs.~\cite{KouWen2010,RaoSodemann2021}. In those parity-preserving
$(2+1)$-dimensional Hamiltonian systems, the corresponding finite-torus signs
are fermion parities of many-body ground states. For each boundary twist, this
parity remains stable under parity-preserving interactions as long as the
ground state remains nondegenerate and the many-body gap does not close.%
\footnote{The pattern of individual twisted parities can contain additional
translation-resolved information, whose interpretation relies on translation
symmetry. Here we retain only their product, which gives the strong
Chern-parity combination and suffices to distinguish the phases realized in
the present model.}

The Niu--Thouless--Wu construction~\cite{NiuThoulessWu1985} uses the same
general idea of twisted boundary conditions but extracts a different
invariant. There, the integer many-body Chern number is obtained from the Berry
curvature of the many-body ground state over the continuous two-dimensional
torus of twist angles. Here, only the four discrete fermionic spin structures
enter, and we retain the $\mathbb Z_2$ product of their signs.

One relevant distinction here is that our $Z^{\tx f}_{\alpha_x,\alpha_y}$ are scalar $(2+0)$-dimensional Euclidean tensor-network contractions rather than ground-state parities. Away from the free-fermion line they are no longer Pfaffians, but they remain well-defined real contractions with the corresponding twisted boundary conditions. Equation~\eqref{eq:topo-invariant} therefore gives an algebraically well-defined interacting continuation of the free Chern parity. Its stability does not follow directly from the Hamiltonian constructions above and is established for the present tensor-network family below.

Moving a twist cut without changing its winding around the torus amounts only
to local sign redefinitions of the fermionic variables and therefore leaves
the physical boundary condition unchanged. We use the same mode-ordering and
cut conventions for all four contractions, as specified in
Sec.~\ref{subsec:JW_torus}; with these conventions,
$\sigma_{\tx{topo}}=+1$ at the trivial point $(a,b)=(0,0)$ provides a
consistency check.

In the fermionic language, the stability of this interacting continuation can be argued as follows. The perturbative fermionic reduction developed in Sec.~\ref{subsec:phase-diag} provides a controlled anchor close to the free-fermion line. In the vicinity of
the two free-fermion transition points, integrating out the gapped complementary pair of bands generates a self-energy for the two low-energy modes. To leading order, the mass coefficient is
renormalized according to
\begin{equation}
	h_{2,s}^{\mathrm{eff}}(0,a,b)
    =
    h_{2,s}(0,a)
    +
    b\mu_s
    +
    \Ocal(b^2),
    \qquad
    s=\pm.
\end{equation}
The remaining projected interaction is irrelevant at the Ising
critical point, as argued in Ref.~\cite{Wille_quarticTN_2025}. 
Thus, within the regime of perturbative control and away from a renormalized
mass closing, the interacting theory remains continuously connected to the
same massive Gaussian phase, and $\sigma_{\tx{topo}}$ retains its free value.
The interacting transition lines obtained in Sec.~\ref{subsec:phase-diag}
therefore provide the perturbative continuation of the changes in
Chern-number parity at $b=0$.

Beyond weak coupling, the stability of $\sigma_{\tx{topo}}$ is more naturally
understood in the bosonic formulation. As shown in the following subsections,
the four fermionic partition functions are fixed signed combinations of four
positive bosonic partition sums distinguished by their winding parities around
the two torus cycles. Within a massive phase, the corresponding winding-sector
pattern is stable under smooth changes of the local tensor. A change of
$\sigma_{\tx{topo}}$ therefore requires this pattern to change; for the
continuous phase boundaries of the present model, this occurs together with
the vanishing domain-wall tension and divergence of the correlation length, as
discussed in Sec.~\ref{subsec:winding-invariant}.

\begin{figure}[t]
    \centering
    \includegraphics[width=0.7\linewidth]{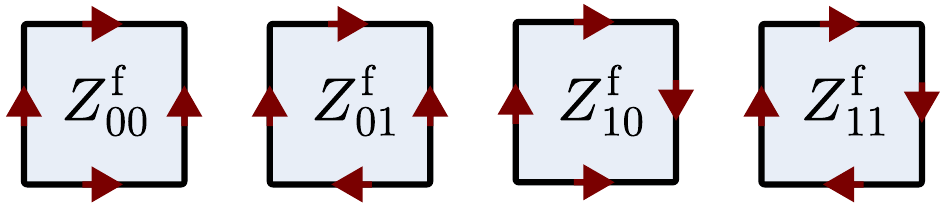}
    \caption{Partition functions on the torus with twisted boundary conditions. Aligned arrows indicate periodic boundary conditions, oppositely aligned --- antiperiodic ones.}
    \label{fig:torus-bc}
\end{figure}

To summarize, we view the topological quantity in
Eq.~\eqref{eq:topo-invariant} as a straightforward interacting continuation
of the free Chern-number parity. The perturbative fermionic calculation
grounds this identification near the free-fermion line, while finite-size TN
contractions reproduce the expected topological island, as shown in
Fig.~\ref{fig:invariant-num}. The bosonic TN formulation provides both an
efficient way to evaluate these contractions and a winding-parity
interpretation that relates the invariant directly to the thermodynamic
phases of the model. We therefore next turn to the mapping from the fermionic
to the bosonic TN on a torus.


\subsection{Jordan--Wigner transformation on a torus}
\label{subsec:JW_torus}

We have defined the topological invariant through the four twisted fermionic
contractions. Now, we translate these quantities to the bosonic TN by extending
the Jordan--Wigner map to the torus. This leads to
Eq.~\eqref{eq:JW-torus-sector-map}: each fermionic spin structure is a fixed
signed combination of the four bosonic winding sectors.

In Sec.~\ref{subsec:recap}, we introduced the (parity-preserving) tensor-network model in two flavors: the fermionic and the bosonic one. On the level of a local tensor, each bosonic leg $k$ with index value $i_k=0,1$ corresponds to a fermionic leg occupied by a Grassmann mode $\theta^{i_k}=1,\theta_k$, where $k=1,2,\dots,4$ labels the legs clockwise (cf.\ Eq.~\eqref{eq:fermionic-tensor-def}). Labeling order as well as bond contraction order is significant in a fermionic TN, as permutations of Grassmann variables lead to sign factors~\cite{BarthelPinedaEisert2009,Wille_2024}.
For instance, a cyclic permutation of the fermionic modes of a parity-even tensor, i.e.\ $i_1 + i_2 + i_3 + i_4 = 0 \pmod 2$, yields a sign factor, represented as a red dot below:
\vspace{-0.5ex}
\begin{center}
\includegraphics[width=0.98\linewidth]{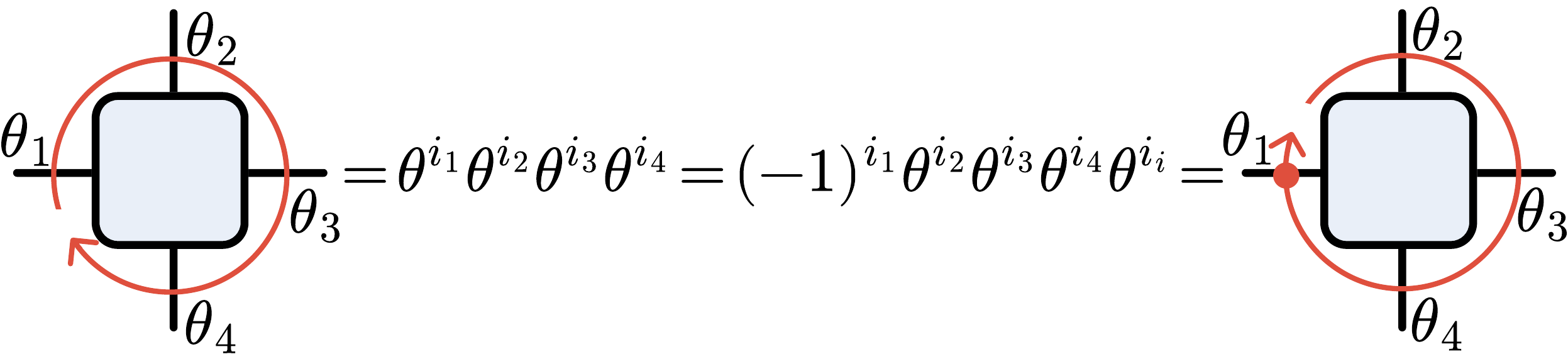}
\end{center}
\vspace{-1ex}
As a second example, contracting the indices $B_b$ and $A_a$ of two fermionic tensors defines the contraction identity
\vspace{-0.5ex}
\begin{center}
\includegraphics[width=0.98\linewidth]{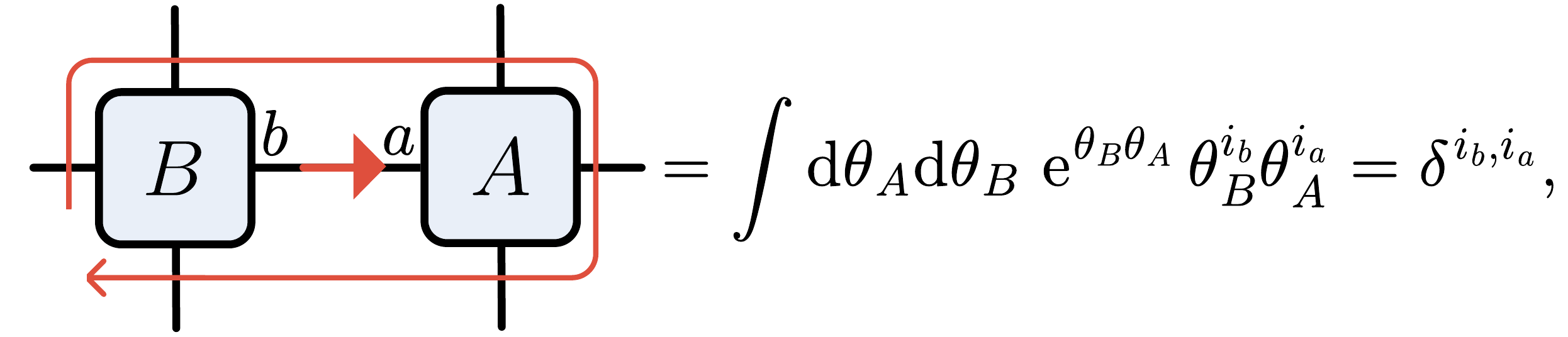}
\end{center}
\vspace{-1ex}
where an orientation choice of a contracted bond is needed for a fixed sign. 

A fixed choice of bond orientations and mode orderings is therefore required to define the bosonic correspondence consistently. We fix the orientation of all fermionic bonds to be left-to-right for horizontal links and top-to-bottom for vertical links. The mode ordering of each local tensor is chosen to be clockwise, with the starting point at the topmost index along the left edge. This convention is referred to as the standard ordering. 

\paragraph{Parity sectors.}
For a planar network with open boundaries, the chosen bond orientations and local mode orderings are sufficient to make the fermionic and bosonic contractions coincide, $Z^\tx{f} = Z^\tx{b}$, cf.\ Sec.~\ref{subsec:recap}.
(Here and throughout Sec.~\ref{sec:topo},
$Z^{\tx b}$ denotes the ordinary bosonic TN contraction, denoted simply by $Z$ in Sec.~\ref{subsec:recap}.)
On a torus, however, the remaining boundary bonds must be closed around two independent directions. Bringing the corresponding Grassmann variables into the contraction order requires global reorderings and produces an additional sign that depends only on the winding parities of the configuration. Equivalently, the bosonic contraction decomposes into four disjoint sectors of fixed winding parity (whose existence follows from the parity-preserving property of the TNs considered here). 

The same homology structure underlies the logical sectors of the toric
code~\cite{Kitaev2003,Dennis_2002}. There, the two noncontractible
$\mathbb Z_2$ loop parities label four ground-state sectors, equivalently two
logical qubits. Here, the same homology classes (winding parities) are used to decompose a statistical partition sum.

We denote these parities by $p_x,p_y\in\{0,1\}$, where $p_\mu$
records whether the occupied-edge configuration has even or odd net winding
around the $\mu$-direction of the torus. Equivalently, $p_\mu$ is the parity
of the occupied bonds crossing a cut transverse to the $\mu$-cycle. The pair
$(p_x,p_y)$ therefore labels one of four winding-parity sectors.
These winding parities are global topological labels: local rearrangements can
deform loops or create and remove contractible loops without changing
$(p_x,p_y)$. Changing either parity requires changing the noncontractible loop
content. The four sectors are therefore the four $\mathbb Z_2$ homology
classes of loop configurations on the torus.

We then define $Z_{p_x,p_y}$ as the bosonic tensor-network contraction restricted to the winding sector $(p_x,p_y)$. For the nonnegative tensor class considered throughout this work, these sector partition sums satisfy $Z_{p_x,p_y}\geq0$. The complete torus bosonic partition function thus has the sector-decomposed form
\begin{equation}
	Z^{\tx b}
    =
    \sum_{p_x,p_y\in\{0,1\}}
    Z_{p_x,p_y},
    \label{eq:bosonic-sum-sectors}
\end{equation}

\paragraph{Fermion--boson map in standard ordering.}
For the standard ordering, closing the fermionic network around the $\mu$-cycle produces the factor $(-1)^{p_\mu}$; when both closures are present, their fermionic reorderings additionally contribute the mixed factor $(-1)^{p_xp_y}$. The total reordering sign is therefore given by
\begin{equation}
\begin{aligned}
    &(-1)^{q_0(p_x,p_y)},
    \qquad \tx{with}
    \\[1ex]
    &q_0(p_x,p_y)
    \coloneqq
    p_x + p_y + p_x p_y \pmod 2.
\end{aligned}
    \label{eq:JW-closure-sign}
\end{equation}
That is, on a torus, recovering the fermionic contraction from the
bosonic one requires keeping track of the two winding parities together with the sign generated when both noncontractible directions are closed. A fermionic partition sum on a torus maps to a weighted sum of bosonic partition sums
\begin{equation}
	Z^\text{f}=\sum_{p_x,p_y} (-1)^{q_0(p_x,p_y)} Z_{p_x,p_y}\;.
\end{equation}

\begin{figure*}[htbp]
    \centering
    \includegraphics[width=0.98\linewidth]{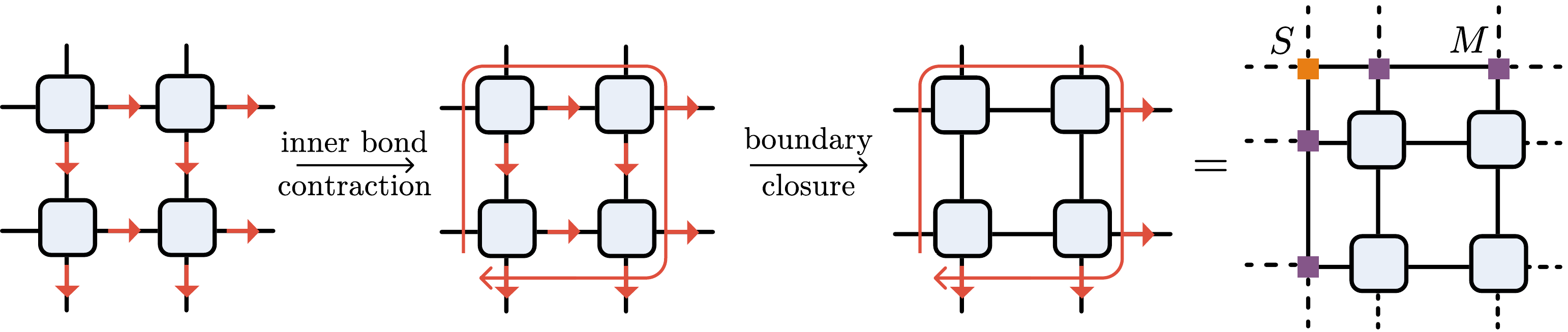}
    \caption{Contraction of fermionic tensors on a torus. The inner bonds are first
    contracted according to the standard ordering, indicated by the red curve.
    The boundary bonds are then closed following the directions of the red arrows.
    Auxiliary parity-tracking MPOs $M$ (purple) record the winding parities
    $p_x$ and $p_y$ of the configuration. Their crossing is contracted with the
    tensor $S$ (yellow), which combines these parities into the fermionic sign
    $(-1)^{p_x+p_y+p_xp_y}$.}
    \label{fig:jw-torus}
\end{figure*}

\paragraph{MPO implementation.}
This bookkeeping can be implemented directly in the tensor network by placing
an auxiliary two-state matrix-product operator (MPO) along each noncontractible cycle, as shown in
Fig.~\ref{fig:jw-torus}. Each occupied bond crossing the cycle flips the
auxiliary state, while an unoccupied bond leaves it unchanged. The corresponding
local MPO tensor is
\begin{equation}
    M_{00}=\mathbb 1,
    \qquad
    M_{11}=\sigma_x,
    \qquad
    M_{01}=M_{10}=0.
    \label{eq:parity-tracking-MPO}
\end{equation}
After the MPO is closed, its auxiliary state distinguishes even from odd
winding around that cycle.

The two parity-tracking MPOs intersect once on the torus. Since the fermionic
sign in Eq.~\eqref{eq:JW-closure-sign} contains the mixed term $p_xp_y$,
the intersection must also account for configurations that wind around both
cycles. This is implemented by the crossing tensor
\begin{equation}
\begin{aligned}
    S_{ijkl}
    &=
    \frac{1}{4}s_{i+k,j+l},
    \\[1ex]
    s_{0,0}&=1,\quad
    s_{0,1}=s_{1,0}=s_{1,1}=-1,
\end{aligned}
    \label{eq:JW-crossing-tensor}
\end{equation}
where the index addition is understood modulo two. With the normalization of
the complete MPO contraction shown in Fig.~\ref{fig:jw-torus}, contracting
$S$ for fixed winding parities $(p_x,p_y)$ gives
\begin{equation}
    s_{p_x,p_y}
    =
    (-1)^{p_x+p_y+p_xp_y},
\end{equation}
and reproduces the fermionic sign in
Eq.~\eqref{eq:JW-closure-sign}.%
\footnote{Mathematically, the function $q_0(p_x,p_y)=p_x+p_y+p_xp_y$ is known as the quadratic refinement of the mod-two intersection rule for the two torus cycles; changing the fermionic boundary conditions adds linear terms to this function. This reordering structure appears more generally in Pfaffian formulations on surface graphs and in fermionic state-sum constructions~\cite{Bravyi_2008,CimasoniReshetikhin2007,Gaiotto_2016}.}

\paragraph{Twisted boundary conditions and flux insertions.}
The signs above arise solely from the fermionic reordering required to close the network on the torus and are fixed by the standard ordering convention.
The twisted boundary conditions introduced in
Sec.~\ref{subsec:fermionic-topo} give a separate sign: an antiperiodic
boundary condition in the $\mu$-direction multiplies a configuration by $(-1)$ for every occupied bond crossing the corresponding cut, and therefore contributes $(-1)^{p_\mu}$.
For boundary-condition labels $\alpha_x,\alpha_y\in\{0,1\}$, the total sign multiplying a configuration in the winding sector $(p_x,p_y)$ can therefore be written as
\begin{equation}
\begin{split}
    &(-1)^{
        q_{\alpha_x,\alpha_y}(p_x,p_y)
    }, \qquad \tx{with}
    \\
    &q_{\alpha_x,\alpha_y}(p_x,p_y)
    \coloneqq
    p_x+p_y+p_xp_y
    \\
    &\hspace{29.5mm}+
    \alpha_xp_x+\alpha_yp_y
    \pmod 2.
    \label{eq:JW-twisted-sign}
\end{split}
\end{equation}
Thus, the terms $p_x+p_y+p_xp_y$ encode the fixed fermionic reordering sign, whereas $\alpha_xp_x+\alpha_yp_y$ encodes the chosen boundary twists.
Collecting all sign factors, we obtain the torus Jordan--Wigner relation between the twisted fermionic contractions $Z^{\tx f}_{\alpha_x,\alpha_y}$ and the bosonic parity-sector sums $Z_{p_x,p_y}$. This is a torus realization of the standard relation between fermionic spin structures and bosonic winding sectors~\cite{CimasoniReshetikhin2007,Gaiotto_2016}:
\begin{equation}
    Z^{\tx f}_{\alpha_x,\alpha_y}
    =
    \sum_{p_x,p_y\in\{0,1\}}
    (-1)^{ q_{\alpha_x,\alpha_y}(p_x,p_y) }
    Z_{p_x,p_y}.
    \label{eq:JW-torus-sector-map}
\end{equation}
This equation represents a linear transformation between the four bosonic winding sectors and the four fermionic spin structures. Its
matrix form and inverse are given in App.~\ref{app:winding-algebra}. Inverting Eq.~\eqref{eq:JW-torus-sector-map} and summing the four bosonic sectors yields
\begin{equation}
Z^{\tx b}
    =
    \frac{1}{2}
    \left(
        -Z^{\tx f}_{00}
        +Z^{\tx f}_{10}
        +Z^{\tx f}_{01}
        +Z^{\tx f}_{11}
    \right).
    \label{eq:bosonic-from-fermionic-torus}
\end{equation}
That is, on a torus, the ordinary bosonic contraction corresponds to this fixed signed combination of all four boundary conditions. In the following subsection, the bosonic sector sums $Z_{p_x,p_y}$ acquire
their interpretation as loop-gas winding sectors.


\subsection{Loop-winding sectors and the topological invariant}
\label{subsec:winding-invariant}

\begin{figure*}[t]
	\centering
	\includegraphics[width=0.98\linewidth]{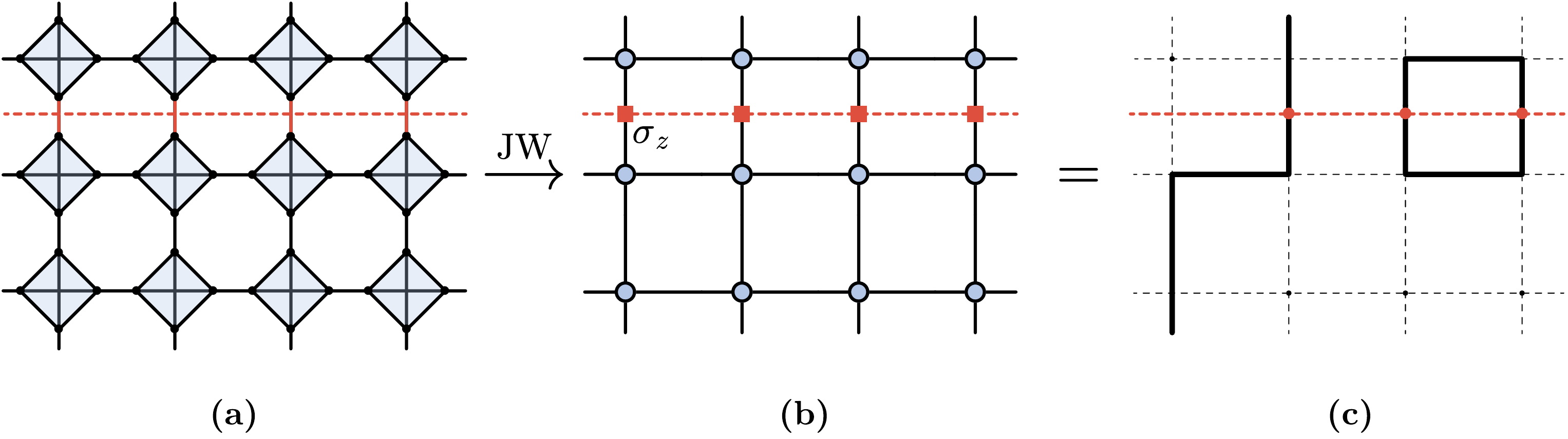}
	\caption{Boundary twists and winding parity in the three representations. 
	(a) In the fermionic TN, the antiperiodic boundary condition (here, in the vertical $y$-direction) is implemented by a transverse cut (here, in the horizontal $x$-direction, red dotted line), along which the connectivity matrix entries are flipped, $C_{ij} \to -C_{ij}$ (red bonds). 
	(b) In the bosonic TN, this is equivalent to a string of Pauli $\sigma_z$ insertions (red boxes) on the legs crossing this cut. The fixed fermionic reordering factor $(-1)^{p_x+p_y+p_xp_y}$ is not displayed.
	(c) In the loop gas, the winding observable $\mathcal W_y=(-1)^{p_y}$ counts the parity of the number of intersections of loops (dark red) with the cut (red dotted line). Only noncontractible loops contribute, since contractible loops cross the cut an even number of times.
	}
	\label{fig:twists-windings}
\end{figure*}

We now employ the Jordan--Wigner map above to reformulate the topological invariant in the loop gas model and obtain a physical interpretation in terms of loop observables. 

The fermionic topological invariant is expressed through the four partition functions with twisted fermionic boundary conditions. The torus Jordan--Wigner map expresses each of them as a linear combination of four bosonic winding-sector sums $Z_{p_x,p_y}$, distinguished in the loop-gas picture by the even or odd winding of loops around the two torus cycles. 
This suggests a connection to winding-parity observables. 
Indeed, together with the total partition sum, the winding-parity expectation values determine the four winding-sector partition sums. This sequence of reformulations --- from fermionic boundary twists to bosonic winding sectors, and then to winding-parity expectation values --- yields a compact expression with a transparent physical interpretation and convenient numerical access.

In the loop gas interpretation introduced in Sec.~\ref{subsec:recap}, the tensor entry at each vertex corresponds exactly to a loop-gas weight; thus, the unrestricted bosonic TN contraction $Z^\tx{b} = Z^\tx{loop}$ equals the statistical-mechanical partition sum of our loop gas model. For the sector-restricted contractions, it likewise holds
\begin{equation}
    Z_{p_x,p_y}
    =
    Z^{\tx{loop}}_{p_x,p_y}
    =
    \sum_{\substack{\ell : \tx{loops}\\
                    p_x(\ell)=p_x,\;
                    p_y(\ell)=p_y}}
    w(\ell),
    \label{eq:TN-loop-sector-identification}
\end{equation}
where $w(\ell)$ denotes the loop-gas weight in Eq.~\eqref{eq:loop-gas-Z}, and
$p_\mu(\ell)$ is the winding parity of the given loop configuration w.r.t.\ a cut transverse to the $\mu$-direction (i.e.\ the parity of the number of occupied edges crossing such a cut). The full loop gas partition sum is given by the sum over winding-sector sums as in Eq.~\eqref{eq:bosonic-sum-sectors}.


The winding parities are measured by noncontractible string observables in the bosonic tensor network. Inserting $\sigma_z=\operatorname{diag}(1,-1)$ on the virtual bonds crossed by a cut $\gamma_\mu$ contributes a factor of $-1$ for each occupied bond and therefore assigns a loop configuration $\ell$ the sign
\begin{equation}
    \mathcal W_\mu(\ell)
    \coloneqq
    (-1)^{p_\mu(\ell)}.
    \label{eq:winding-parity-operators}
\end{equation}
Thus, $\mathcal W_\mu$ measures the parity of loops winding around the $\mu$-cycle (cf.\ Fig.~\ref{fig:twists-windings}).

%
%
%

We define the expectation values of the winding-parity operators as
\begin{equation}
\begin{aligned}
    m_x
    &\coloneqq
    \av{\mathcal W_x}_{\tx{loop}},
    \qquad
    m_y
    \coloneqq
    \av{\mathcal W_y}_{\tx{loop}},
    \\[1ex]
    m_{xy}
    &\coloneqq
    \av{\mathcal W_x\mathcal W_y}_{\tx{loop}},
\end{aligned}
    \label{eq:winding-parity-expectations}
\end{equation}
where the loop-gas average is defined by $\av{\mathcal O}_{\tx{loop}} \coloneqq \frac{1}{Z^{\tx b}}\sum_{\ell} \Ocal(\ell)\,w(\ell)$.
Concretely, these expectation values can be computed as
\begin{equation}
\begin{aligned}
    m_\mu
    &=
    \frac{1}{Z^{\tx b}}
    \sum_{p_x,p_y}
    (-1)^{p_\mu}Z_{p_x,p_y},
    \qquad
    \mu=x,y,
    \\
    m_{xy}
    &=
    \frac{1}{Z^{\tx b}}
    \sum_{p_x,p_y}
    (-1)^{p_x+p_y}Z_{p_x,p_y}.
\end{aligned}
    \label{eq:winding-characters-main}
\end{equation}

We refer to $m_x$, $m_y$, and $m_{xy}$ as winding characters. In particular, $m_\mu$ measures the imbalance between even and odd $\mu$-winding sectors:
$m_\mu=1$ if only even sectors contribute, $m_\mu=-1$ if only odd sectors contribute, and $m_\mu=0$ when the two carry equal total weight. 
Notably, the winding-operator insertions supply the factors $(-1)^{\alpha_xp_x+\alpha_yp_y}$ associated with the fermionic boundary twists, distinct from the reordering factor $(-1)^{p_x+p_y+p_xp_y}$ appearing in the JW map via the auxiliary MPO construction (cf.\ Eq.~\eqref{eq:JW-closure-sign}). 
The winding operator in the fermionic TN, bosonic TN, and loop gas is visualized in Fig.~\ref{fig:twists-windings}.



Together with normalization, Eqs.~\eqref{eq:winding-characters-main}
determine the four sector partition sums $Z_{p_x,p_y}$.
Substituting these into the torus Jordan--Wigner relation
Eq.~\eqref{eq:JW-torus-sector-map} and using the definition
Eq.~\eqref{eq:topo-invariant} yields the compact loop-gas expression
\begin{equation}
\begin{aligned}
    \sigma_{\tx{topo}}
    &=
    \sgn\!
    \left\{
        \left[
            (m_x+m_y)^2-(1-m_{xy})^2
        \right]
        \right.
    \\
    &\hspace{3.2em}\left.
        {}\times
        \left[
            (1+m_{xy})^2-(m_x-m_y)^2
        \right]
    \right\}.
\end{aligned}
    \label{eq:topological-sign-winding}
\end{equation}
The corresponding linear algebra is collected in
App.~\ref{app:winding-algebra}.
%

The physical meaning of Eq.~\eqref{eq:topological-sign-winding} becomes transparent in two limiting forms of the winding-sector
distribution. 

In the loop-deconfined phase, loops that wind around the torus remain present
with appreciable weight as the system size increases. Even and odd winding
therefore occur with comparable weight. Consequently, on a torus of fixed aspect ratio, the relative differences between the four winding-sector weights are exponentially suppressed with the linear system size (this statement is made explicit in the spin representation below). Hence,
\begin{equation}
    Z_{00}
    \simeq
    Z_{10}
    \simeq
    Z_{01}
    \simeq
    Z_{11},
    \qquad
    m_x,m_y,m_{xy}\to0.
    \label{eq:equal-winding-sectors}
\end{equation}
Inserting these winding characters into the topological sign formula \eqref{eq:topological-sign-winding} leads to $\sigma_\tx{topo} = -1$, signaling a topological phase.

In the two loop-confined (ordered) phases, by contrast, a single winding sector $(p_x^\star,p_y^\star)$ dominates in the thermodynamic limit. In this case, the winding characters become
\begin{equation}
\begin{split}
    m_x
    &\to
    (-1)^{p_x^\star},
    \qquad
    m_y
    \to
    (-1)^{p_y^\star},
    \\
    m_{xy}
    &\to
    (-1)^{p_x^\star+p_y^\star}.
\end{split}
\end{equation}
Deep in the empty phase the dominant sector is $(0,0)$, while in the
full phase it is
$(L_y\bmod2,L_x\bmod2)$ for the present cut convention. In either case
Eq.~\eqref{eq:topological-sign-winding} gives
$\sigma_{\tx{topo}}=+1$; the corresponding winding characters are listed in
Tab.~\ref{tab:winding-characters}.

Away from a phase transition, these characteristic winding-sector patterns are
stable under smooth changes of the tensor parameters. The corresponding
twisted fermionic partition functions therefore keep the same signs, and
$\sigma_{\tx{topo}}$ cannot change within either phase.

Changing $\sigma_{\tx{topo}}$ requires the winding-sector weights themselves
to reorganize. At the continuous boundaries of the loop-deconfined phase, the
system crosses between a regime dominated by a single winding sector and one
in which all four sectors have comparable weight. The spin representation
below gives a complementary description of this change in terms of domain
walls and their free-energy cost.

\paragraph{Spin interpretation.}

We now translate the winding sectors into the domain-wall spin language.
As explained in Sec.~\ref{subsec:recap}, an occupied loop edge separates
opposite Ising spins on the dual lattice. Starting from one reference spin,
crossing the loop edges reconstructs the complete spin configuration.

A domain wall winding around the $x$-cycle changes the spin boundary condition
along the transverse $y$-cycle, and vice versa. Thus $p_x=1$ corresponds to an
antiperiodic spin boundary condition along $y$, while $p_y=1$ corresponds to
one along $x$. The four winding-sector sums are therefore, up to the global
factor of two from the spin-flip redundancy, the four quartic-Ising partition
functions with periodic or antiperiodic boundary conditions in the two
directions~\cite{WuHu2002}.

The limiting cases above are then interpreted as follows. 
In the paramagnetic phase, changing a periodic boundary
condition to an antiperiodic one costs vanishing free energy in the large-system
limit. The four boundary-condition partition functions therefore become equal,
matching the equal-weight winding sectors of the loop gas.
In an ordered phase, by contrast, an antiperiodic boundary condition forces a
domain wall across the system. Its free-energy cost grows with the transverse
system size, so sectors containing such a wall are exponentially suppressed
and a single winding sector dominates.%
\footnote{The noncontractible $\sigma_z$ strings used here measure the winding parity of domain walls already present in the configuration. They should not be confused with Kadanoff--Ceva disorder lines~\cite{KadanoffCeva1971}, which flip the Ising couplings along a branch cut. In the Gaussian case, a Hadamard
mapping~\cite{Wille_2024} relates the loop gas to a different Ising model in which the same $\sigma_z$ strings do become Kadanoff--Ceva disorder lines. Extending this mapping to the non-Gaussian case is beyond the scope of this work.}

On the Gaussian line, this suppression can be computed explicitly from the exact domain-wall tension of the square-lattice Ising model. Consider the ferromagnetic case. The Ising coupling is $J=-\frac12\ln a$ and the dual coupling $J^\star$ is defined through the Kramers--Wannier duality~\cite{KramersWannier1941} by $\exp(-2J^\star)=\tanh J$. 
The free-energy cost per unit length of an axis-aligned domain wall is given by the string tension~\cite{Onsager1944,GallavottiMartinLof1972,
AbrahamGallavottiMartinLof1973}
\begin{equation}
    \tau(J)
    =
    2(J-J^\star)
    =
    2 J+\ln\tanh J.
\end{equation}
At the same time, $\tau(J)$ sets the correlation length of the dual Ising excitation, $\xi = \tau(J)^{-1}$~\cite{BaxterExactly,McCoyWu1973}.

Let $Z_{p_x^\star,p_y^\star}$ denote the winding sector selected by the ordered phase; for the ferromagnetic phase,
$(p_x^\star,p_y^\star)=(0,0)$. Changing the winding parity forces a domain
wall to span the torus. In the ordered phase, such a wall has a free-energy
cost proportional to its length. For a single parity change, where the wall is
axis-aligned and has length $L_\perp$, this gives
\begin{equation}
    \frac{Z_{p_x,p_y}}{Z_{p_x^\star,p_y^\star}}
    \sim
    \exp\!\left[
        -\tau(J)L_\perp+o(L_\perp)
    \right].
    \label{eq:Ising-sector-suppression}
\end{equation}
Sectors differing in both winding parities are likewise exponentially
suppressed with system size, although the precise domain-wall cost depends on
the geometry of the corresponding domain walls.

At the self-dual point $J=J^\star$, the domain-wall tension vanishes and $\xi=\tau^{-1}$ diverges, marking the transition between the ferromagnetic and paramagnetic phases. Indeed, with $a=\e^{-2J}$, this occurs at $a=a_-=\sqrt2-1$.
The antiferromagnetic side follows by a sublattice spin flip, which maps $J<0$ to $|J|$ and sends $a\to a^{-1}$. The corresponding transition is therefore at $a^{-1}=a_-$, or $a=a_+=\sqrt2+1$.

Away from $b=0$, the exact formula for $\tau(J)$ no longer applies, but the same qualitative picture holds. Along the continuous phase boundaries (and away from the multicritical point), the
transition remains in the two-dimensional Ising universality class. The domain-wall tension therefore scales with the inverse correlation length, $\tau\sim\xi^{-1}$~\cite{Stauffer1972}, and vanishes as the
critical line is approached from the ordered side.


\subsection{Numerical evaluation of the topological invariant}
\label{subsec:topo-numerics}

We now evaluate the twisted-contraction invariant directly on finite tori and compare its sign structure with the thermodynamic phase diagram.

\begin{figure*}[t]
    \begin{minipage}{0.49\textwidth}\centering
        \includegraphics[width=\textwidth]{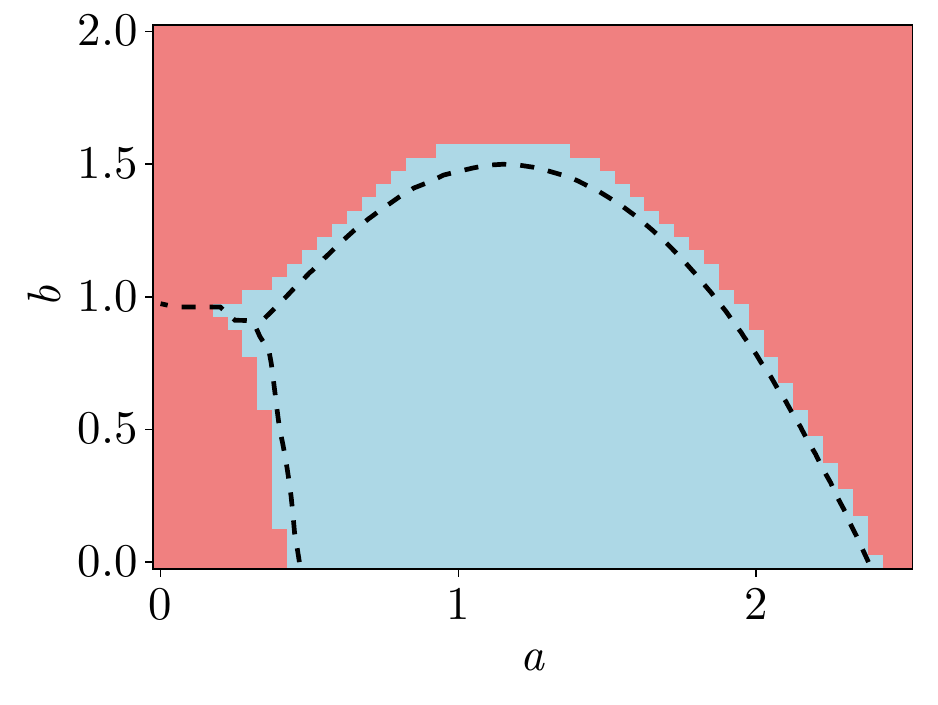}
        \par\smallskip\textbf{(a)} $3\times2$
    \end{minipage}%
    \begin{minipage}{0.49\textwidth}\centering
        \includegraphics[width=\textwidth]{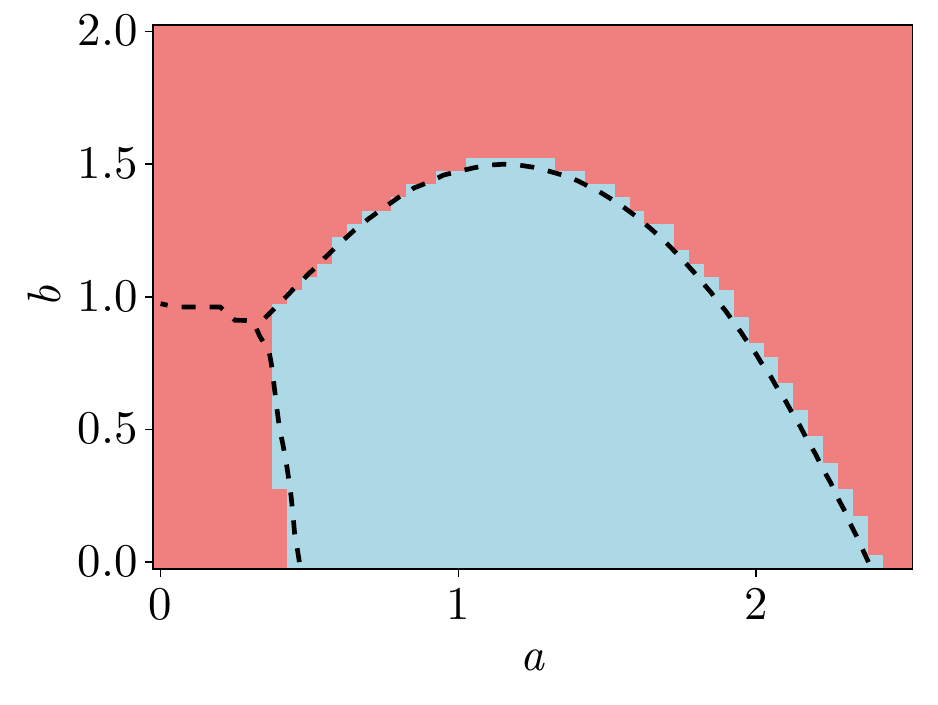}
        \par\smallskip\textbf{(b)} $6\times4$
    \end{minipage}
    \par\medskip
    \begin{minipage}{0.49\textwidth}\centering
        \includegraphics[width=\textwidth]{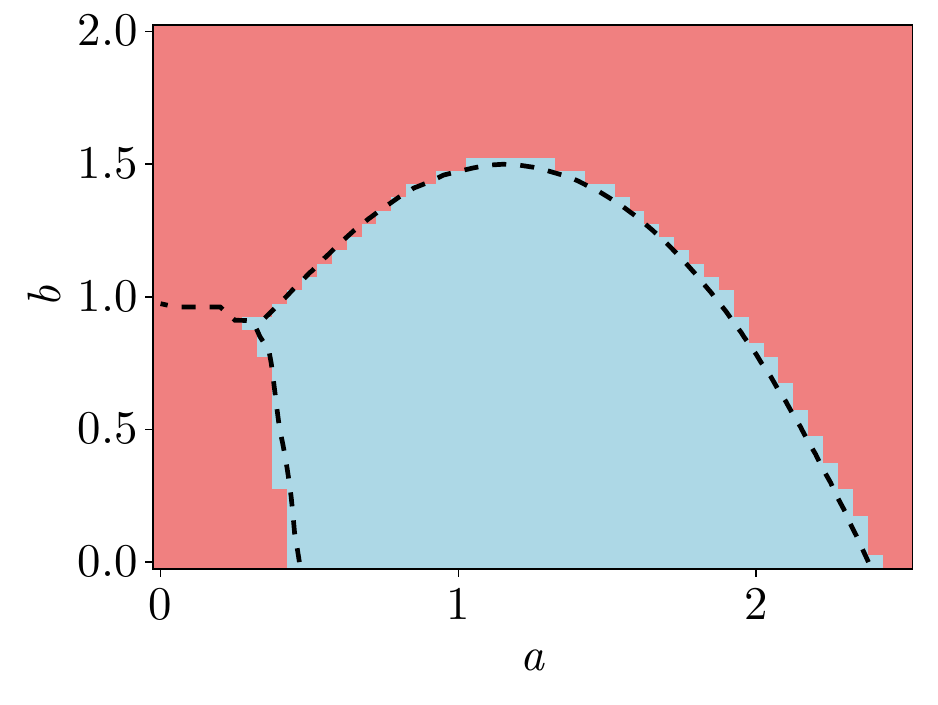}
        \par\smallskip\textbf{(c)} $10\times5$
    \end{minipage}%
    \begin{minipage}{0.49\textwidth}\centering
        \includegraphics[width=\textwidth]{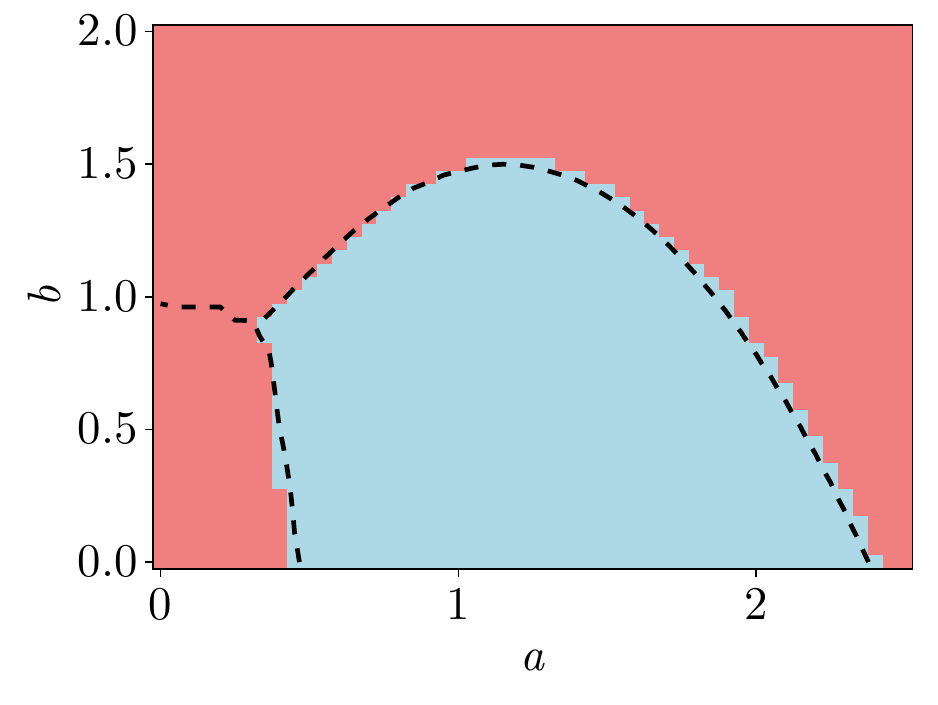}
        \par\smallskip\textbf{(d)} $9\times6$
    \end{minipage}
    \caption{%
    The $\mathbb Z_2$ invariant $\sigma_{\tx{topo}}$ computed by exact
    TN contraction on tori of dimensions
    $(L_x,L_y)=$ (a) $3\times2$, (b) $6\times4$, (c) $10\times5$, and
    (d) $9\times6$.
    Pink and blue correspond to $\sigma_{\tx{topo}}=+1$ (trivial) and
    $\sigma_{\tx{topo}}=-1$ (topological), respectively; these phase labels refer
    to the thermodynamic limit.
    The black dashed curves show the thermodynamic-limit phase boundaries.}  
    \label{fig:invariant-num}
\end{figure*}

The numerical calculation requires only bosonic tensor-network contractions.
Although the invariant is defined through the four fermionic quantities $Z^{\tx f}_{\alpha_x,\alpha_y}$,
$(\alpha_x,\alpha_y)\in\{0,1\}^2$, the torus Jordan--Wigner map of
Sec.~\ref{subsec:JW_torus} represents the four fermionic spin structures by the auxiliary parity-tracking MPOs introduced above. The required signs can therefore be incorporated directly into ordinary bosonic contractions.

For the numerical implementation, an antiperiodic twist is absorbed directly
into the parity-tracking MPO rather than inserted as a separate
$\sigma_z$ string. Writing $\phi_\mu=\pi\alpha_\mu$, we attach the twist phase
to the unoccupied MPO channel,
\begin{equation}
\begin{aligned}
    &M_{00}(\phi_\mu)
    =
    \e^{\imagunit\phi_\mu}\mathbb 1,
    \qquad
    M_{11}(\phi_\mu)
    =
    \sigma_x,
    \\[1ex]
    &M_{01}
    =M_{10}=0.
\end{aligned}
    \label{eq:numerical-twist-MPO}
\end{equation}
Thus $\phi_\mu=0$ gives periodic boundary conditions, while
$\phi_\mu=\pi$ implements the corresponding antiperiodic twist.

For $\phi_\mu=\pi$, this convention places the minus sign on unoccupied bonds
crossed by the cut, whereas the
$\sigma_z=\operatorname{diag}(1,-1)$ string of
Sec.~\ref{subsec:winding-invariant} places it on occupied bonds. Since every
crossed bond is either unoccupied or occupied, the two prescriptions differ only
by the configuration-independent factor
$(-1)^{L_{\gamma_\mu}}$, where $L_{\gamma_\mu}$ is the number of bonds crossed
by the cut. These fixed seam signs cancel in the product over the four boundary
conditions and therefore do not affect $\sigma_{\tx{topo}}$.

Denoting the contractions obtained with this numerical convention by
$\widetilde Z^{\tx f}_{\alpha_x,\alpha_y}$, the cancellation of the seam
signs (cf.\ Appendix~\ref{app:winding-algebra}) yields
\begin{equation}
    \sigma_{\tx{topo}}
    =
    \prod_{\alpha_x,\alpha_y}
    \sgn \widetilde Z^{\tx f}_{\alpha_x,\alpha_y}
    =
    \prod_{\alpha_x,\alpha_y}
    \sgn Z^{\tx f}_{\alpha_x,\alpha_y}.
\end{equation}
In the thermodynamic limit, the winding-sector argument of
Sec.~\ref{subsec:winding-invariant} fixes this sign throughout each pure massive
phase. On a finite torus, several winding sectors retain finite relative
weight, so their signed combinations can pass through zero away from the
thermodynamic phase boundary. The finite systems in
Fig.~\ref{fig:invariant-num} should therefore be viewed as finite-size
estimates, with these shifts decreasing as the system size grows.

Figure~\ref{fig:invariant-num} shows clear convergence toward the
thermodynamic phase diagram. Already for $(L_x,L_y)=(10,5)$, the region with
$\sigma_{\tx{topo}}=-1$ closely reproduces the expected topological island,
while both the empty and full phases have
$\sigma_{\tx{topo}}=+1$. The empty--full first-order line does not separate
different values of the invariant, as expected.


\section{Conclusion and outlook}
\label{sec:con}

In this work, we used a minimal two-parameter ppTN as a case study of how physical insight can be gained from a non-Gaussian tensor-network contraction. 
The core message is that the physical content of such a contraction goes far beyond its direct numerical evaluation: its complementary fermionic and
statistical-mechanical formulations reveal and interpret the phase structure of the TN parameter plane, while providing languages suited to different questions. 
A coherent picture emerges by combining these descriptions, with direct numerical TN contractions providing a common benchmark.

Apart from conceptual insight, our results provide a degree of analytical control. 
Beyond the exactly tractable free-fermion limit, perturbative fermionic methods 
yield a mean-field-inspired effective quadratic theory that tracks the continuous phase boundaries.
Statistical-mechanical mappings identify the multicritical theory and reveal an additional exactly solvable line within the non-Gaussian regime. 
Topology offers perhaps the clearest example of the relation between the different pictures: the same parameter region --- the topological island --- is characterized by an interacting $\Zbb_2$ indicator constructed from fermionic boundary twists as well as by the winding-sector structure of the domain-wall configurations.
These dualities thus relate not only model parameters and partition functions, but also observables, boundary conditions, and notions of locality.

More broadly, our analysis exemplifies a cross-disciplinary use of many-body methods to study tensor-network contraction problems. 
Here, the bridge is concrete: parity preservation and planarity give rise to an interacting fermionic formulation, while nonnegative tensor entries permit a statistical-mechanical interpretation. 
Other tensor families will require different tools, but the underlying strategy --- using the algebraic structure of the local tensors to identify effective physical descriptions --- extends beyond the present model.

\paragraph{Outlook.}
A concrete generalization of our present approach is to move beyond the statistical mechanics setting by allowing complex tensor entries. This leads to a much broader class of tensor networks, encompassing unitary quantum circuits as well as more general non-equilibrium and non-unitary processes. The fermionic correspondence and in particular the distinction between Gaussian and interacting structures continue to apply. Conceptually, this distinction provides a common yardstick for departures from the tractable Gaussian limit. Practically, it provides access to the toolbox of many-body theory, with individual methods requiring suitable adaptations and extensions.

One concrete application is provided by parity-preserving brickwork circuits, a direction we are currently pursuing~\cite{CircuitsInPrep}. At the local level, a parity-preserving two-site gate defines a four-legged tensor: matchgate circuits yield Gaussian fermionic networks, while departures from the matchgate limit introduce local interaction terms. Unlike the static $(2+0)$-dimensional setting considered here, the input--output structure of a circuit selects one lattice direction as physical time and turns the tensor network into a causal $(1+1)$-dimensional evolution. Then, the relation between physical unitarity and the auxiliary fermionic kernel governing the tensor-network contraction becomes important. While a circuit composed of unitary gates defines unitary physical time evolution, the Gaussian kernel of the associated two-dimensional contraction need not be Hermitian. This raises the question of how the spectral and topological structure of the generally non-Hermitian kernel is reflected in physical properties of the underlying unitary circuit.

In the circuit setting, the distinction between Gaussian and interacting structures acquires a direct computational meaning. Matchgate circuits are efficiently classically simulable, whereas suitable non-Gaussian extensions supply the resources for universal quantum computation~\cite{Brod2011,Oszmaniec2017,HebenstreitJosza2019}. Non-Gaussianity therefore marks both the onset of fermionic interactions and the transition from efficient classical simulability toward universal quantum computation. Perturbative treatments of weak non-Gaussianity already lead to improved classical simulation algorithms~\cite{Dias_2024,ReardonSmith2024,WilleStrelchuk2025}. A broader objective is to determine whether many-body approximations can identify new regimes of classical simulability.

Unitary real-time circuits constitute only one distinguished sector of the broader complex-ppTN framework. Beyond this sector, complex ppTNs describe hybrid unitary--non-unitary dynamics. Many such settings are intrinsically stochastic, making disorder a central part of the problem rather than a separate extension. Random matchgate tensor-network ensembles with Gaussian-distributed disorder already admit a many-body field-theoretic description~\cite{Usoltcev2026}. Quantum error correction supplies a concrete application for the discrete counterpart: stochastic Pauli errors correspond to discrete disorder variables in otherwise Gaussian fermionic tensor networks. Coherent errors take this problem beyond the Gaussian limit, giving rise to random-bond Ising models with complex couplings and four-spin interactions~\cite{BehrendsBeri2025}. 
Quantum error correction thus provides a natural application of the present framework, in which disorder and departures from the free-fermion limit arise together in a physically motivated setting. 

Beyond these examples, hybrid and disordered circuit dynamics support a rich nonequilibrium phenomenology, including measurement-induced transitions in entanglement, mutual information, purification and quantum-information retention, learnability, magic, and classical simulability. These questions are currently the focus of intense activity at the interface of nonequilibrium many-body physics and quantum information~\cite{SkinnerRuhmanNahum2019,LiChenFisher2019,ChoiBaoQiAltman2020,GullansHuse2020,IppolitiKhemani2021,IppolitiKhemani2024,BejanMcLauchlanBeri2024,Jian_2022}. The ppTN formulation may provide a common fermionic perspective on these phenomena.

%

\begin{acknowledgments}
M.U. thanks Konstantin Weisenberger for insightful discussions concerning the CFT content
of the model. N.H.D. thanks Erik Weerda and Niklas Tausendpfund for helpful discussions
concerning finite-entanglement scaling.

This work was supported by the Deutsche Forschungsgemeinschaft (DFG, German
Research Foundation) under Germany's Excellence Strategy, Cluster of Excellence
Matter and Light for Quantum Computing (ML4Q), EXC 2004/1 (grant 390534769);
the EU Quantum Flagship PASQuanS2.1 (grant 101113690); CRC TRR 183
(grant 277101999; subproject B04); and the EPSRC UK Quantum Technologies
Programme (grant EP/Z53318X/1). The authors also acknowledge
FZ Jülich for JURECA~\cite{JURECA} (Institute Project No.~PGI-8) at the Jülich
Supercomputing Centre (Grant No.~NeTeNeSyQuMa).

\emph{AI use.}
The authors used OpenAI GPT-5.5 and GPT-5.6 Sol for language polishing,
consistency checks, and assistance with manuscript revision.
OpenAI Codex 5.6 Sol was used to assist with code verification, repository
organization, and code development. All AI-assisted content and code changes
were reviewed and verified by the authors.
\end{acknowledgments}

\section*{Data Availability}
The numerical data, analysis code, and figure-reproduction workflows are
available in the accompanying Zenodo repository~\cite{repo}.

\appendix

\section{Numerical methods and finite-entanglement scaling}
\label{app:numerics}

In this appendix we summarize the tensor-network methods used to obtain the phase diagram and the universality data shown in Sec.~\ref{subsec:phase-diag}. Our goal is twofold: to specify the numerical contraction procedure in sufficient detail for reproducibility, and to explain the finite-entanglement-scaling analysis used to extract critical points, correlation-length exponents, and central charges. The corresponding numerical data and analysis workflows are provided in the accompanying Zenodo repository~\cite{repo}.

\paragraph{Boundary-MPS contraction in the thermodynamic limit.}
To contract the bosonic tensor network in the thermodynamic limit, we approximate the fixed points of the row-to-row transfer matrix $\mathcal T$ by uniform matrix product states (MPS). We use two closely related approaches: Corner Transfer Matrix Renormalization Group (CTMRG)~\cite{Naumann_2024} and Variational Uniform Matrix Product States (VUMPS)~\cite{Zauner2018}. Both methods represent the effective environment of the infinite two-dimensional tensor network by boundary tensors of finite bond dimension $D$.

In the CTMRG approach, the environment is obtained iteratively by growing the effective corner and edge tensors and truncating back to bond dimension $D$ after each step. 
In the VUMPS approach, one instead solves directly for the dominant fixed point of $\mathcal T$ 
by optimizing within the tangent space of the uniform-MPS manifold. 
In practice, we find that VUMPS, formulated as self-consistent effective eigenvalue equations for the fixed point, 
converges more rapidly and robustly than power-method-based CTMRG near criticality of statistical models, in line with the observations of Ref.~\cite{Fishman_2018}.

Expectation values of local observables are evaluated in the standard impurity-tensor formulation: one replaces a single bulk tensor by the corresponding impurity tensor and contracts the resulting network with the effective boundary environments. In particular, we use the local loop density as a diagnostic of local loop occupation across the phase diagram.

\paragraph{Transfer-matrix spectrum and correlation length.}
Let $\lambda_\alpha$ denote the leading eigenvalues of the transfer matrix $\mathcal T$, ordered by decreasing magnitude. We define
\begin{equation}
\varepsilon_\alpha = -\log |\lambda_\alpha/\lambda_1| ,
\end{equation}
so that $\varepsilon_1=0$. The MPS correlation length at bond dimension $D$ is then
\begin{equation}
\xi_D = \frac{1}{\varepsilon_2}.
\end{equation}
Tracking $\xi_D$ as a function of $(a,b)$ provides a first diagnostic of criticality: peaks in $\xi_D$ that grow with increasing $D$ identify the continuous transition lines in Fig.~\ref{fig:pd-num}.

\paragraph{Finite-entanglement scaling of the correlation length.}
To extract critical points and exponents quantitatively, we employ the finite-entanglement scaling hypothesis of Ref.~\cite{Vanhecke_2019}. At a critical point, a finite MPS bond dimension $D$ imposes an effective infrared cutoff, so that correlation functions decay exponentially even though the exact theory is gapless. Equivalently, finite $D$ acts as a relevant perturbation of the underlying conformal field theory (CFT), and the true critical behavior is recovered only in the limit $D\to\infty$.

Following Ref.~\cite{Vanhecke_2019}, we characterize this induced infrared scale by the discreteness of the low-lying transfer-matrix spectrum. In our analysis we define
\begin{equation}
\delta_D = \varepsilon_3-\varepsilon_2.
\end{equation}
The quantity $\delta_D$ measures the discreteness of the transfer matrix spectrum induced by finite $D$ and $\delta_D^{-1}$ 
defines the effective infrared length scale generated by the MPS approximation. The scaling ansatz is
\begin{equation}
\label{eq:delta-based-app}
\delta_D\,\xi_D(t)=f\!\left(\delta_D^{-1/\nu} t\right),
\end{equation}
where $t$ is the reduced distance to criticality, $\nu$ is the correlation-length exponent, and $f$ is a scaling function.

Compared to the more conventional finite-entanglement ansatz written directly in terms of the bond dimension~\cite{Tagliacozzo2008},
\begin{equation}
\xi_D D^{-\kappa} = f\!\left(D^{\kappa/\nu} t\right),
\end{equation}
the $\delta_D$-based formulation has two advantages. First, $\delta_D^{-1}$ directly quantifies the artificial length scale, whereas $D$ is only an indirect proxy for it, assuming the relation $\xi_D\sim D^\kappa$. Second, 
it avoids introducing the additional exponent $\kappa$ which must be fitted together with the critical point and $\nu$,
thereby enlarging the parameter space. This generally makes the collapse more stable away from the asymptotic limit of large $D$.

At the multicritical point $\mathsf M$, the expected universality class is that of the two-dimensional four-state Potts model, equivalently the endpoint of the Ashkin--Teller critical line. The leading algebraic correlation-length exponent is therefore $\nu_{\mathrm{Potts}}=2/3$~\cite{WuPottsReview1982}. Importantly, the four-state Potts point contains a marginally irrelevant perturbation~\cite{Salas_1997}, leading to multiplicative logarithmic corrections to the correlation-length power law,
\begin{equation}
\xi(t)
\sim
|t|^{-\nu} \times
\left[\log\frac{1}{|t|}\right]^{\hat{\nu}},
\end{equation}
up to subleading logarithmic factors. Thus, Eq.~\eqref{eq:delta-based-app} should be interpreted as a collapse of the leading algebraic scaling behavior. Very close to the Potts point, a log-corrected scaling variable may be used,
\begin{equation}
\label{eq:delta-log-corrected-app}
x_{\log}
=
\frac{\delta_D^{-1/\nu}t}
{\left[\log\left(1/\delta_D\right)\right]^{\hat{\nu}/\nu}},
\end{equation}
with $\nu=2/3$ and $\hat{\nu}=1/2$~\cite{Salas_1997}.
(The logarithmic factor leads to slowly drifting effective exponents in finite-$D$ data if one fits to a pure power law.)

A data collapse based on Eq.~\eqref{eq:delta-based-app} yields estimates for the critical point and the
correlation length exponent $\nu$. We perform this analysis for representative horizontal and vertical cuts
in the phase diagram, as shown in Fig.~\ref{fig:data-collapse}. In particular, the extracted value $\nu\simeq 1$ 
is consistent with Ising criticality along both continuous transition lines. At the multicritical point, the expected leading exponent is instead $\nu=2/3$, with the logarithmic corrections discussed above characteristic of the four-state Potts universality class.

\paragraph{Central charge from entanglement-entropy scaling.}
For a critical system described by a CFT, the bipartite entanglement entropy of the boundary MPS obeys the universal scaling relation~\cite{Pollmann_2009}
\begin{equation}
\label{eq:central-charge-app}
S_D=\frac{c}{6}\ln \xi_D+\mathrm{const.},
\end{equation}
where $c$ is the central charge of the underlying CFT and $\xi_D$ is the finite-entanglement correlation length defined above. 
By extracting $S_D$ and $\xi_D$ at the critical points identified from the correlation-length analysis, we obtain 
a direct estimate of $c$. The results are shown in Fig.~\ref{fig:data-collapse}. 
At the free-fermion critical points $(a,b)=(a_\pm,0)$ we obtain values
consistent with $c=1/2$, as expected for the Ising universality class. The same value is obtained, within numerical accuracy,
at two critical points along the two continuous lines extending into the interacting regime. At the multicritical point \textsf{M}, 
we find $c\simeq 1$, in agreement with the Ashkin--Teller / four-state Potts universality class discussed in Sec.~\ref{subsec:spins-AT}.

\paragraph{Free energy per site.}
The converged CTMRG corner and edge tensors (shown in orange below) determine the thermodynamic-limit
partition function per site. The blue tensor in the contraction below is the
local bosonic weight tensor $T(a,b)$ introduced in Sec.~\ref{subsec:recap},
see also Fig.~\ref{fig:weights}(b), with
$T_{0000}=1$, the six two-leg configurations weighted by $a$, and
$T_{1111}=a^2+b$. This provides the CTMRG estimate of the free-energy density used in
Fig.~\ref{fig:fe-plots} to benchmark the tensor-network contraction
against Baxter's exact result and to analyze the singular behavior near
$\mathsf M$. 

\begin{center}
    \includegraphics[width=0.98\linewidth]{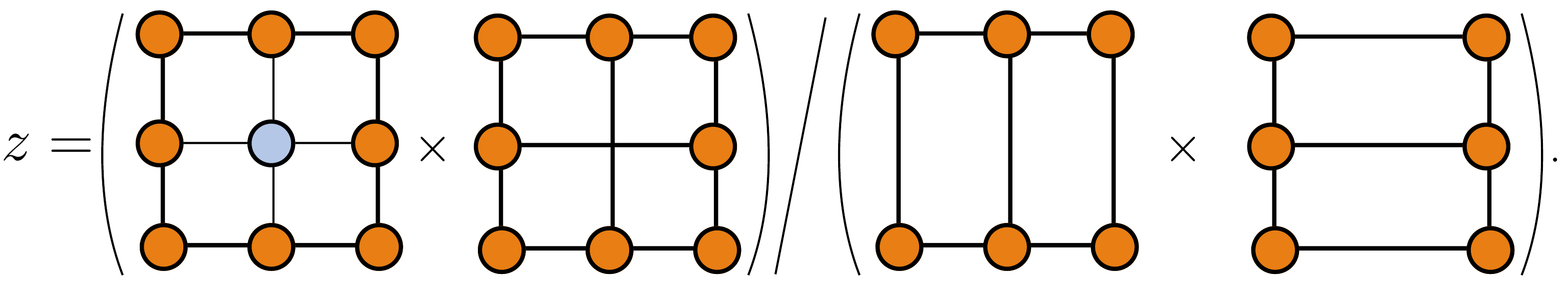}
\end{center}

The ratio of contractions removes the arbitrary normalization of the CTMRG
environment and yields the multiplicative partition-function contribution
$z$ per lattice site. The corresponding dimensionless free-energy density is
\begin{equation}
    f=-\ln z.
\end{equation}
%
%

\section{Perturbative treatment of critical lines at weak nonlinearity}
\label{app:tadpole-slopes}

This appendix summarizes the perturbative fermionic calculation used
in Sec.~\ref{subsec:phase-diag} to estimate the initial slopes of the
two continuous phase boundaries emanating from
$a_-=\sqrt2-1$ and $a_+=\sqrt2+1$, cf.\
Fig.~\ref{fig:pd-slopes}. We label the two branches by $s=\pm$.
In addition to the first-order mass shifts at $b=0$, we define the two
finite-$b$ continuations shown in Fig.~\ref{fig:pd-slopes}(a): the
exact result within the single-momentum problem and the continuation
obtained by reevaluating the bare high-band covariance at the four
self-inverse momenta. These constructions retain selected finite-$a$
dependence of the Gaussian problem, but do not constitute a systematic
higher-order expansion in $b$.

At $b=0$, the fermionic tensor network is Gaussian. With the conventions of
Sec.~\ref{subsec:recap}, we write
\begin{equation}
    \begin{aligned}
        Z^{\tx f}
        &=
        \int \dd\theta\,\e^{-S[\theta]},
        \\
        -S[\theta]
        &=
        \frac12\theta^\T K\theta
        +
        b\sum_x
        \theta_{1,x}\theta_{2,x}\theta_{3,x}\theta_{4,x},
        \\
        K&=A+C=\imagunit H.
    \end{aligned}
    \label{eq:app-gaussian-kernel}
\end{equation}
Thus $H=-\imagunit K$ is the Hermitian single-particle Hamiltonian of the Gaussian problem. 
In momentum space, $H(k)$ has gap closings at
\[
    (a_-,k_-)=(\sqrt2-1,(\pi,\pi)),
    \quad
    (a_+,k_+)=(\sqrt2+1,(0,0)).
\]
Near either point we write $k=k_s+q$, with $s=\pm$, and project
$H(k_s+q,a)$ onto the corresponding two-dimensional low-energy
subspace,
\begin{equation}
    H_s^{(2)}(q,a)
    =
    \sum_{\alpha=1}^3
    h_{\alpha,s}(q,a)\sigma_\alpha.
\end{equation}
The transition is controlled by the mass coefficient $h_{2,s}$. We
use a right-oriented mass convention, in which
\begin{equation}
    h_{2,-}(0,a)=-\frac{a-a_-}{a_-},
    \qquad
    h_{2,+}(0,a)=+\frac{a-a_+}{a_+}.
\end{equation}
Equivalently,
\begin{equation}
    \begin{aligned}
        \kappa_s
        &:={}
        \left.
        \partial_a h_{2,s}(0,a)
        \right|_{a=a_s},
        \\
        \tx{hence}\quad
        \kappa_-&=-\frac1{a_-}=-(\sqrt2+1),
        \\
        \kappa_+&=+\frac1{a_+}=\sqrt2-1.
    \end{aligned}
    \label{eq:app-kappa-values}
\end{equation}
Since the microscopic kernel is linear in $a$ and the projection
basis is kept fixed at the corresponding free critical point, the
projected mass is
\begin{equation}
    h_{2,s}(0,a)=\kappa_s(a-a_s)
\end{equation}
exactly within this fixed-basis projection.

We now include the quartic term to first order in $b$. At the critical
point $(a_s,k_s)$, decompose the local Grassmann variables into low-
and high-band components,
\begin{equation}
    \theta_i=\theta_{\ell,i}^{(s)}+\theta_{h,i}^{(s)},
    \qquad
    \theta_{\ell,i}^{(s)}
    =
    P_{s,i+}\eta_+^{(s)}
    +
    P_{s,i-}\eta_-^{(s)}.
    \label{eq:app-low-high-decomposition}
\end{equation}

Here $s=\pm$ labels the critical branch, while the explicit signs on
$\eta_\pm^{(s)}$ distinguish the two Grassmann modes within its
low-energy pair.

Expanding one quartic contribution according to the number of low- and
high-band fields gives five sectors. The terms containing one or three
high-band fields vanish under the centered high-band Gaussian average
at first order in $b$. The four-high-band term generates only a field-independent constant that can be discarded from the low-energy effective action. The four-low-band term vanishes at zeroth
order in momentum and first appears with two gradients, as derived in
Ref.~\cite{Wille_quarticTN_2025}. It therefore does not contribute to
the zero-momentum mass considered here. The induced quadratic kernel
comes from the sector containing two low- and two high-band fields.
At zero external momentum, the low-mode bilinears are
\begin{equation}
\begin{split}
    \theta_{\ell,i}^{(s)}
    \theta_{\ell,j}^{(s)}
    &=
    p_{ij}^{(s)}
    \eta_+^{(s)}\eta_-^{(s)},
    \\[0.5ex]
    p_{ij}^{(s)}
    &=
    P_{s,i+}P_{s,j-}
    -
    P_{s,i-}P_{s,j+},
    \label{eq:app-pij-def}
\end{split}
\end{equation}
where $P_{s,i,\pm}$ are the zero-momentum versions of the generally momentum dependent expansion coefficients defined in Eq.~\eqref{eq:app-low-high-decomposition}. 
In the following local formulas we work at fixed branch $s$ and
suppress this label on $\theta_{\ell}$, $\theta_h$, and $\eta_\pm$
where no ambiguity can arise.

The perturbative step is to contract two legs of the quartic term in the high-band sector,
leaving an effective quadratic term for the low modes. With
\begin{equation}
    \Gamma_{h,s}^{ij}
    :=
    \langle\theta_{h,i}\theta_{h,j}\rangle_h ,
\end{equation}
the part of one quartic vertex quadratic in the low fields is
\begin{align}
    \bigl\langle
        \theta_1\theta_2\theta_3\theta_4
    \bigr\rangle_h^{(2)}
    &=
    \Gamma_{h,s}^{12}\theta_{\ell,3}\theta_{\ell,4}
    -
    \Gamma_{h,s}^{13}\theta_{\ell,2}\theta_{\ell,4}
    \notag\\[0.5ex]
    &\quad
    +
    \Gamma_{h,s}^{14}\theta_{\ell,2}\theta_{\ell,3}
    +
    \Gamma_{h,s}^{23}\theta_{\ell,1}\theta_{\ell,4}
    \notag\\[0.5ex]
    &\quad
    -
    \Gamma_{h,s}^{24}\theta_{\ell,1}\theta_{\ell,3}
    +
    \Gamma_{h,s}^{34}\theta_{\ell,1}\theta_{\ell,2}.
    \label{eq:app-tadpole-wick}
\end{align}
The signs are the Grassmann Wick signs of the ordered product
$\theta_1\theta_2\theta_3\theta_4$. Using Eq.~\eqref{eq:app-pij-def}, this becomes
\begin{equation}
    b\,
    \bigl\langle
        \theta_1\theta_2\theta_3\theta_4
    \bigr\rangle_h^{(2)}
    =
    b\,\mu_s\,\eta_+\eta_- ,
\end{equation}
where
\begin{align}
    \mu_s
    &=
    \Gamma_{h,s}^{12}p_{34}^{(s)}
    -
    \Gamma_{h,s}^{13}p_{24}^{(s)}
    +
    \Gamma_{h,s}^{14}p_{23}^{(s)}
    +
    \Gamma_{h,s}^{23}p_{14}^{(s)}
    \notag\\
    &\quad
    -
    \Gamma_{h,s}^{24}p_{13}^{(s)}
    +
    \Gamma_{h,s}^{34}p_{12}^{(s)}.
    \label{eq:app-mu-def}
\end{align}
For completeness, the momentum-dependent form used in
Eq.~\eqref{eq:main-text-tadpole} is obtained by retaining the external
relative momentum $q$ in the low-band eigenvectors while integrating
over the internal high-band momentum $p$. Define
\begin{align}
    \overline\Gamma_{h,s}^{ij}
    &=
    \int_{\rm BZ}\frac{\dd^2p}{(2\pi)^2}\,
    \Gamma_{h,s}^{ij}(p),
    \\[1ex]
    p_{ij}^{(s)}(q)
    &=
    P_{s,i+}(-q)P_{s,j-}(q)
    \notag\\
    &\quad
    -
    P_{s,i-}(-q)P_{s,j+}(q).
\end{align}
The induced mass coefficient can then be written compactly as
\begin{equation}
    \mu_s(q)
    =
    \frac14\,
    \epsilon_{ijkl}\,
    \overline\Gamma_{h,s}^{ij}\,
    p_{kl}^{(s)}(q).
    \label{eq:app-mu-momentum-dependent}
\end{equation}
Equivalently, expanding the antisymmetric tensors reproduces the six
terms in Eq.~\eqref{eq:app-mu-def}. The lattice symmetries imply that
the mass channel is even in the external momentum, and hence
\begin{equation}
    \mu_s(q)=\mu_s(0)+\mathcal O(q^2).
\end{equation}
The slope calculation below requires only
$\mu_s\equiv\mu_s(0)$.

Since the low-energy Gaussian exponent in the mass channel is
\begin{equation}
    -S_\ell
    =
    \frac{i}{2}\eta^\T h_{2,s}\sigma_2\eta
    =
    h_{2,s}\eta_+\eta_-,
\end{equation}
the quartic term shifts the mass as $h_{2,s}\mapsto h_{2,s}+b\mu_s$. The
first-order criticality condition is therefore
\begin{equation}
    h_{2,s}(0,a)+b\mu_s=0.
\end{equation}

Let $a_{c,s}(b)$ denote the critical value of $a$ on branch $s$, with
$a_{c,s}(0)=a_s$. Linearizing the criticality condition near
$a=a_s$ gives
\begin{equation}
    \left.
    \frac{\dd a_{c,s}}{\dd b}
    \right|_{b=0}
    =
    -\frac{\mu_s}{\kappa_s},
    \qquad
    \Rightarrow
    \qquad
    \left.
    \frac{\dd b_s}{\dd a}
    \right|_{a=a_s}
    =
    -\frac{\kappa_s}{\mu_s}.
    \label{eq:app-slope-general}
\end{equation}
The second form is the one plotted in Fig.~\ref{fig:pd-slopes},
where $b$ is the vertical axis.

The only remaining input is the high-band contraction $\Gamma_{h,s}$. We use three successive approximations.

\paragraph{Crude single-momentum estimate.}
The simplest estimate evaluates the contraction only at the critical momentum. Since
$K_s^\star=K(k_s,a_s)$ has two zero modes, we invert only on the high-band
subspace,
\begin{equation}
    \Gamma_{h,s}^{ij}
    =
    \bigl[(K_s^\star)^+\bigr]_{ji},
\end{equation}
where $(K_s^\star)^+$ denotes the high-subspace pseudoinverse. This yields
\begin{equation}
    \mu_-=-\frac{2+\sqrt2}{4},
    \qquad
    \mu_+=+\frac{2-\sqrt2}{4},
\end{equation}
and therefore
\begin{equation}
    \begin{aligned}
        \left.
        \frac{\dd a_{c,-}}{\dd b}
        \right|_{b=0}
        &=
        \left.
        \frac{\dd a_{c,+}}{\dd b}
        \right|_{b=0}
        =
        -\frac{\sqrt2}{4},
        \\
        &\Rightarrow
        \left.
        \frac{\dd b_s}{\dd a}
        \right|_{a=a_s}
        =
        -2\sqrt2.
    \end{aligned}
\end{equation}
This captures the direction of the shift, but artificially predicts equal slopes at the two transitions.

The $q=0$ two-pair truncation can also be evaluated without linearizing in $a-a_s$. At the corresponding critical momentum, the reduced problem contains one low-energy pair and one high-energy pair. Integrating out the latter exactly within this four-mode truncation gives the criticality condition
\begin{equation}
    b_s^{(0)}(a)
    =
    -E_{\ell,s}(a)E_{h,s}(a),
    \label{eq:app-single-point-general}
\end{equation}
where $E_{\ell,s}$ and $E_{h,s}$ are the two Gaussian pair energies in the corresponding oriented basis.
The signed Gaussian pair energies entering
Eq.~\eqref{eq:app-single-point-general}, evaluated at the corresponding
critical momentum $k_s$, are
\begin{equation}
    \begin{aligned}
        E_{\ell,-}(a)
        &=
        -\frac{a-a_-}{a_-},
        &\qquad
        E_{h,-}(a)
        &=
        -a_-(a+a_+),
        \\
        E_{\ell,+}(a)
        &=
        +\frac{a-a_+}{a_+},
        &
        E_{h,+}(a)
        &=
        +a_+(a+a_-).
    \end{aligned}
    \label{eq:app-single-point-pair-energies}
\end{equation}
Their signs are fixed by the oriented basis used for each branch. In
particular, $E_{h,-}(a)<0$ and $E_{h,+}(a)>0$ for $a\geq0$, while
$E_{\ell,s}(a)$ changes sign at the corresponding free transition
$a=a_s$. Substituting Eq.~\eqref{eq:app-single-point-pair-energies} into
Eq.~\eqref{eq:app-single-point-general} gives
\begin{equation}
    b_-^{(0)}(a)=1-2a-a^2,
    \qquad
    b_+^{(0)}(a)=1+2a-a^2.
    \label{eq:app-single-point-domes}
\end{equation}
Their derivatives at $a=a_\pm$ reproduce the crude slopes
$\dd b/\dd a=-2\sqrt2$, while the two branches meet at $(a,b)=(0,1)$.
Thus, the crude truncation already predicts a bounded, dome-shaped topological region.
Equation~\eqref{eq:app-single-point-domes} is exact within the single-momentum problem, but its global accuracy is limited by the discarded momentum sectors.

\paragraph{TRIM average.}
A simple diagnostic improvement is to replace the critical-momentum contraction by an
average over the four self-inverse momenta,
\begin{equation}
    \Gamma_{h,s}^{ij}
    \longrightarrow
    \overline\Gamma_{h,s}^{{\rm TRIM},ij}
    :=
    \frac14
    \sum_{q_\star\in\{(0,0),(\pi,0),(0,\pi),(\pi,\pi)\}}
    \Gamma_{h,s}^{ij}(q_\star).
    \label{eq:app-trim-average}
\end{equation}
Here ``TRIM'' refers only to the property $q_\star=-q_\star$ modulo reciprocal lattice vectors.
This four-point estimate respects the elementary high-symmetry structure of the Brillouin
zone, but it is not controlled: symmetry fixes the allowed matrix form of $\overline\Gamma_{h,s}$, not its numerical entries.

For the finite-$b$ continuation shown in
Fig.~\ref{fig:pd-slopes}(a), we keep the low-mode coefficients
$p_{ij}^{(s)}$ fixed at the corresponding free critical point, but
reevaluate the bare high-band covariance at general $a$:
\begin{equation}
    \overline\Gamma_{h,s}^{{\rm TRIM},ij}(a)
    \coloneqq
    \frac14
    \sum_{p_\star\in
    \{(0,0),(\pi,0),(0,\pi),(\pi,\pi)\}}
    \bigl[K_{h,s}^{-1}(p_\star;a)\bigr]_{ji}.
    \label{eq:app-running-trim-gamma}
\end{equation}
Here $p_\star$ denotes the absolute lattice momentum, while the label
$s$ specifies which band pair is retained by continuation from the
free critical point $(a_s,k_s)$. Inserting
Eq.~\eqref{eq:app-running-trim-gamma} into
Eq.~\eqref{eq:app-mu-def} defines
$\mu_s^{\rm TRIM}(a)$, and the corresponding phase-boundary estimate is
\begin{equation}
    b_s^{\rm TRIM}(a)
    =
    -\frac{h_{2,s}(0,a)}
    {\mu_s^{\rm TRIM}(a)}.
    \label{eq:app-running-trim-line}
\end{equation}
Since $h_{2,s}(0,a_s)=0$, its derivative at $a=a_s$ reproduces the
TRIM slope quoted in the main text.

The continuation remains well-defined only while the Gaussian low- and high-band pairs are separated. For the branch emanating from $\mathsf A_+$, the two positive-energy pairs become degenerate at
\begin{equation}
    a=1,
    \qquad
    k=(\pi,\pi),
    \qquad
    \varepsilon_{1,+}(\pi,\pi)
    =
    \varepsilon_{2,+}(\pi,\pi).
    \label{eq:app-running-trim-breakdown}
\end{equation}
Accordingly, the denominator
$\varepsilon_{2,+}^2-\varepsilon_{1,+}^2$ in
Eq.~\eqref{eq:app-eigenvector-free-inverse} vanishes and the restricted high-band inverse
is no longer uniquely defined. The endpoint of the running TRIM curve therefore marks the
breakdown of the bare low/high-band closure, rather than a physical termination of the
critical line. (The crude $\mathsf A_+$ continuation remains smooth since it retains only the critical momentum $k_+=(0,0)$ and therefore does not resolve the inversion at the opposite TRIM.)

\paragraph{Full Brillouin-zone contraction.}
The most reliable first-order lattice estimate replaces the local contraction by the full
Brillouin-zone average
\begin{equation}
    \begin{aligned}
        \Gamma_{h,s}^{ij}
        &\longrightarrow
        \overline\Gamma_{h,s}^{ij}
        =
        \int_{\rm BZ}
        \frac{d^2p}{(2\pi)^2}
        \Gamma_{h,s}^{ij}(p),
        \\
        \Gamma_{h,s}^{ij}(p)
        &=
        \bigl[K_{h,s}^{-1}(p)\bigr]_{ji},
    \end{aligned}
    \label{eq:app-full-bz-gamma}
\end{equation}
with $K_s(q):=K(k_s+q,a_s)$ and $K_{h,s}^{-1}(q)$ the inverse restricted to the
high-band pair. This restricted inverse can be evaluated without fixing eigenvector phases.
If $\varepsilon_{1,s}(q)$ and $\varepsilon_{2,s}(q)$ are the positive energy pairs, with
$n=1$ connected to the critical low modes, then
\begin{equation}
    K_{h,s}^{-1}(q)
    =
    \frac{
        K_s(q)
        \left[
            K_s(q)^2+\varepsilon_{1,s}(q)^2\mathbb{1}
        \right]
    }{
        \varepsilon_{2,s}(q)^2
        \left[
            \varepsilon_{2,s}(q)^2-\varepsilon_{1,s}(q)^2
        \right]
    }.
    \label{eq:app-eigenvector-free-inverse}
\end{equation}
Indeed, the factor in square brackets annihilates the low-band pair and is proportional to the projector onto the
high-band pair. The BZ integral in Eq.~\eqref{eq:app-full-bz-gamma} is then evaluated by a
uniform midpoint quadrature.

By square-lattice rotation symmetry the averaged contraction takes the form
\begin{equation}
    \overline\Gamma_{h,s}
    =
    \begin{pmatrix}
        0 & \alpha_s & \gamma_s & \alpha_s\\
        -\alpha_s & 0 & \alpha_s & \gamma_s\\
        -\gamma_s & -\alpha_s & 0 & \alpha_s\\
        -\alpha_s & -\gamma_s & -\alpha_s & 0
    \end{pmatrix},
    \label{eq:app-gamma-alpha-gamma}
\end{equation}
which reduces Eq.~\eqref{eq:app-mu-def} to
\begin{equation}
    \mu_-=\gamma_- -\sqrt2\,\alpha_-,
    \qquad
    \mu_+=\gamma_+ +\sqrt2\,\alpha_+.
\end{equation}
The resulting slopes $\dd b/\dd a$ are
\begin{center}
    \renewcommand{\arraystretch}{1.2}
    \setlength{\tabcolsep}{7pt}
    \begin{tabular}{c|cc}
        & $a_-=\sqrt2-1$ & $a_+=\sqrt2+1$
        \\
        \hline
        crude $q=0$ estimate & $-2.83$ & $-2.83$ \\
        TRIM average         & $-17.91$ & $-2.43$ \\
        full BZ estimate     & $-18.06$ & $-2.49$
    \end{tabular}
\end{center}

The full-BZ estimate explains the main qualitative feature seen in the numerical phase boundary: the lower transition line has a much steeper tangent than the upper one. 
It is the most reliable first-order lattice estimate and is used here only to determine the local slopes at $b=0$.

The `crude' single-momentum and TRIM curves retain additional finite-$a$
dependence and provide qualitative continuations of the same
fermionic mass equation. The former yields a complete dome within the
single-momentum problem, while the latter tracks the numerical phase
boundaries more closely over its domain of validity and exposes the
higher-band interchange that obstructs its continuation toward the
multicritical region.

Neither finite-$b$ construction is a controlled higher-order
approximation to the full lattice theory. The calculation renormalizes
only the low-band mass while keeping the remaining Gaussian band
structure and the low-mode projectors fixed. Higher-order
contractions, the momentum dependence of the self-energy beyond the
mass channel, projector corrections, and velocity renormalizations
can be incorporated in a self-consistent treatment but lie beyond the
scope of the present work.

\section{Semianalytic evaluation of Baxter's free energy}
\label{app:fe}

This appendix summarizes the numerical evaluation of Baxter's exact
zero-field eight-vertex free energy used in Sec.~\ref{subsec:8V}. The
complete implementation, including convergence diagnostics and the
comparison with tensor-network data, is provided in the Zenodo repository~\cite{repo}. 

\paragraph{Elliptic parametrization.}
It is sufficient to evaluate the free energy in the fundamental domain
$0<a<1/3$. In the zero-field eight-vertex labeling used here, the four
weights are chosen as
\begin{equation}
\bigl(a_{8\tx V},b_{8\tx V},w_{8\tx V},d_{8\tx V}\bigr)
=
(a,a,1,a).
\label{eq:8V-weights-app}
\end{equation}
Baxter's elliptic parametrization may be written as
\begin{equation}
\bigl(a_{8\tx V},b_{8\tx V},w_{8\tx V},d_{8\tx V}\bigr)
=
\frac{1}{C}(A,B,C,D),
\label{eq:baxter-param-app}
\end{equation}
where
\begin{align}
A&=\operatorname{snh}(\eta-u),\quad
B=\operatorname{snh}(\eta+u),\quad
C=\operatorname{snh}(2\eta),
\nonumber\\[1ex]
D&=
k\,\operatorname{snh}(2\eta)
\operatorname{snh}(\eta-u)
\operatorname{snh}(\eta+u),
\label{eq:baxter-ABCD-app}
\end{align}
and
\begin{equation}
\operatorname{snh}(x|k)
\equiv
-\imagunit\,\operatorname{sn}(\imagunit x|k).
\end{equation}
Here $k$ is the elliptic modulus and $\operatorname{sn}$ is the Jacobi
elliptic function. Since $a_{8\tx V}=b_{8\tx V}$, we choose $u=0$.
Matching Eq.~\eqref{eq:baxter-param-app} to
Eq.~\eqref{eq:8V-weights-app} then gives
\begin{equation}
a
=
\frac{\operatorname{snh}(\eta)}
     {\operatorname{snh}(2\eta)}
=
k\,\operatorname{snh}^{2}(\eta).
\label{eq:baxter-matching-app}
\end{equation}

Defining $y\equiv\operatorname{snh}^{2}(\eta)$ and using the Jacobi
duplication formula reduces these two transcendental matching
conditions to
\begin{equation}
4a^{2}y^{2}
+
\bigl(3a^{4}+6a^{2}-1\bigr)y
+
4a^{4}
=
0.
\end{equation}
The branch appropriate to $0<a<1/3$ is
\begin{equation}
    \begin{aligned}
        y(a)
        &=
        \frac{1}{8a^{2}}
        \Big[
            1-6a^{2}-3a^{4}
            \\
        &\qquad\quad
            +\sqrt{(1-a)^{3}(1+a)^{3}(1-3a)(1+3a)}
        \Big],
        \\
        k(a)
        &=
        \frac{a}{y(a)}.
    \end{aligned}
    \label{eq:baxter-y-app}
\end{equation}
Thus the only remaining inversion is the scalar equation
\begin{equation}
\operatorname{snh}\bigl(\eta|k(a)\bigr)=\sqrt{y(a)},
\label{eq:baxter-eta-app}
\end{equation}
which is solved numerically for $\eta(a)$.

Let
\begin{equation}
K=K(k),
\qquad
K'=K\!\left(\sqrt{1-k^{2}}\right)
\end{equation}
denote the complete elliptic integrals of the first kind. The parameters
appearing in Eq.~\eqref{eq:fe-general} are then
\begin{equation}
\tau=\frac{\pi K'}{2K},
\qquad
\lambda=\frac{\pi\eta}{K},
\qquad
\zeta=0,
\label{eq:baxter-series-parameters-app}
\end{equation}
where the last equality follows from $u=0$. Since
$w_{8\tx V}=1$, Baxter's series specializes to
\begin{equation}
    \begin{aligned}
        f_{<}(a)
        &=-2\sum_{n=1}^{\infty}
        \frac{
        \sinh^{2}\bigl((\tau-\lambda)n\bigr)
        \bigl[\cosh(n\lambda)-1\bigr]
        }{
        n\,\sinh(2n\tau)\cosh(n\lambda)
        },
        \\
        0&<a<1/3.
    \end{aligned}
    \label{eq:baxter-special-series-app}
\end{equation}

\paragraph{Duality reconstruction and numerical convergence.}
The complementary part of the self-dual line is obtained from the
Hadamard duality map derived in Sec.~\ref{subsec:loop-gas},
\begin{equation}
a^\star=\frac{1-a}{1+3a}.
\end{equation}
Restoring the normalization of the local vertex weights produces an
additive contribution to the free-energy density. The result on the
full physical interval $0\leq a\leq1$ is therefore
\begin{equation}
f_{8\tx V}(a)
=
\begin{cases}
f_{<}(a),
&
0\leq a<\dfrac13,
\\[2mm]
\displaystyle
\lim_{x\to1/3^-}f_{<}(x),
&
a=\dfrac13,
\\[2mm]
\displaystyle
f_{<}(a^\star)
-
\ln\!\left(\frac{1+3a}{2}\right),
&
\dfrac13<a\leq1.
\end{cases}
\label{eq:baxter-full-line-app}
\end{equation}
In particular, the endpoint limits give $f_{8\tx V}(0)=0$ and
$f_{8\tx V}(1)=-\ln2$. The formulas for $y(a)$ and the elliptic
parameters are evaluated only on the open interval $0<a<1/3$; the
endpoints and the Potts point are obtained by continuity.

Numerically, Eq.~\eqref{eq:baxter-eta-app} is solved at arbitrary
precision and the two matching conditions in
Eq.~\eqref{eq:baxter-matching-app} are checked explicitly. The series
in Eq.~\eqref{eq:baxter-special-series-app} is evaluated adaptively:
starting from a truncation $N$, the cutoff is doubled until
\begin{equation}
\left|S_{2N}(a)-S_N(a)\right|<\epsilon,
\end{equation}
where $S_N$ denotes the partial sum through order $N$. The public
notebook uses high working precision and records the final cutoff,
matching residuals, and doubling difference as internal error
diagnostics. This procedure converges rapidly throughout the
fundamental domain, with the slowest convergence occurring close to
$a=1/3$, and yields the semianalytic benchmark shown in
Fig.~\ref{fig:fe-plots}.

\section{Effective field theory and Potts--Ising flows near \textsf{M}}
\label{sec:eft-qi}

Section~\ref{subsec:spins-AT} established the exact lattice identification of the multicritical point \textsf{M} with the four-state Potts point of the Ashkin--Teller embedding. Here we determine how the two independent microscopic deformations of the QI model project onto continuum scaling fields. We then explain the massless flows from the Potts ultraviolet fixed point to the Ising infrared fixed point and give the equivalent emergent Thirring description. The lattice mapping and the Potts critical coupling are not repeated.

\subsection{Coarse graining and microscopic operator matching}
\label{sec:eft-setup}

In the QI plane $K_2=K_3$ (cf.\ Fig.~\ref{fig:phaseAT}), a neighborhood of \textsf{M} is parametrized by
\begin{equation}
K_1=\delta K_1,
\qquad
K_2=K_2^c+\delta K_2,
\qquad
K_2^c=\frac14\ln3.
\end{equation}
Using Eq.~\eqref{eq:spinHam-Kform}, the corresponding deformation of the lattice Hamiltonian can be written as
\begin{equation}
H_{\rm QI}
=
H_{\textsf M}
+\delta K_1\,\mathcal O_P
+\delta K_2\,\mathcal O_t,
\label{eq:QI-deformation-M}
\end{equation}
with lattice operators
\begin{align}
    \mathcal O_P
    &=-\sum_{\langle ij\rangle}s_i s_j,
    \\
    \mathcal O_t
    &=-\sum_{\langle\!\langle ij\rangle\!\rangle}s_i s_j
      -\sum_{\square}s_1s_2s_3s_4.
    \label{eq:ops-lattice}
\end{align}
The labels anticipate their continuum projections: $\mathcal O_t$ is tangent to the Potts line, whereas $\mathcal O_P$ moves transversely away from the $K_1=0$ AT plane.

To implement the long-wavelength limit explicitly, choose a block scale $\ell$ with $1\ll \ell\ll \xi$
in lattice units and define block fields on the rotated AT lattice,
\begin{equation}
\sigma_\ell(x)
=\frac{1}{|B_x|}\sum_{r\in B_x}\sigma_r,
\qquad
\tau_\ell(x)
=\frac{1}{|B_x|}\sum_{r\in B_x}\tau_r.
\label{eq:block-spin-AT}
\end{equation}
At scales much larger than $\ell$, these become the continuum fields $\sigma(x)$ and $\tau(x)$. Coarse graining a local lattice operator means expanding all shifted block fields about a common continuum point and retaining the lowest-dimension field compatible with its symmetries.

For the nearest-neighbor perturbation, every original-lattice bond joins opposite checkerboard sublattices. With the unit-cell convention shown in the
sublattice diagram of Sec.~\ref{subsec:spins-AT}, one obtains the exact lattice stencil
\begin{equation}
\sum_{\langle ij\rangle}s_i s_j
=
\sum_r\sigma_r
\Big(
\tau_r+\tau_{r-\vec e_x}+\tau_{r-\vec e_y}
+\tau_{r-\vec e_x-\vec e_y}
\Big),
\label{eq:K1_stencil_merged}
\end{equation}
where $\vec e_x$ and $\vec e_y$ are primitive vectors of the rotated lattice. Expanding the four $\tau$ fields about the center of the corresponding cell gives a nonzero local term proportional to $\sigma(x)\tau(x)$; all remaining terms contain derivatives. Thus
\begin{equation}
    \begin{aligned}
        \mathcal O_P
        &\longrightarrow
        c_P\int\dd^2x\,P(x)
        +\text{derivative-suppressed terms},
        \\
        P(x)&\equiv\sigma(x)\tau(x),
    \end{aligned}
    \label{eq:O1_to_P}
\end{equation}
with a nonuniversal coefficient $c_P\neq0$. The operator $P$ is the AT polarization field; it is odd under either separate sublattice flip $\sigma\mapsto-\sigma$ or $\tau\mapsto-\tau$, but even under the diagonal spin flip inherited from the original QI model. Accordingly, $K_1$ breaks the enhanced $\mathbb Z_2^{(\sigma)}\times\mathbb Z_2^{(\tau)}$ symmetry at $K_1=0$ down to its diagonal subgroup and selects the relative orientation of the two sublattice magnetizations.

Meanwhile, the tangent operator $\mathcal O_t$ varies the common AT coupling $K_2=K_3$ while remaining on the four-state Potts line. Microscopically, it is the lattice energy operator conjugate to this common coupling. To define its continuum counterpart, one may distribute the bond and plaquette terms in $\mathcal O_t$ over local cells, subtract the constant expectation value at \textsf{M}, and block-average the resulting local energy density. The leading $\mathbb Z_2$-even scalar obtained in this way is the Potts thermal field $\varepsilon(x)$; alternative local definitions differ only by the identity and RG-subleading operators. Thus,
\begin{equation}
\mathcal O_t
\longrightarrow
c_\varepsilon\int\dd^2x\,\varepsilon(x)
+\text{RG-subleading even operators},
\label{eq:O2_to_eps}
\end{equation}
where $c_\varepsilon\neq0$ and $\Delta_\varepsilon=1/2$ at \textsf{M}. The enhanced sublattice symmetry distinguishes the two projections: $\mathcal O_t$ is even under both independent flips, whereas $\mathcal O_P$ is odd under each. They therefore cannot mix at linear order.

We reserve $t\propto\delta K_2$ for the microscopic temperature-like
coordinate along the Potts line, while denoting the corresponding
continuum couplings by $\lambda_\varepsilon$ and $\lambda_P$. To leading
order near \textsf{M},
\begin{equation}
\lambda_\varepsilon
\sim t+\Ocal(\delta K^2),
\qquad
\lambda_P
\sim \delta K_1+\Ocal(\delta K^2),
\label{eq:tg-linear-adapted}
\end{equation}
where $\sim$ denotes equality up to finite nonuniversal multiplicative
normalizations.
The diagonal QI spin-flip symmetry forbids terms linear in the individual sublattice spin fields $\sigma$ and $\tau$; the remaining allowed Potts order-parameter component is precisely $P=\sigma\tau$, already included above. Within the two-dimensional microscopic deformation space accessible to the QI model, $\varepsilon$ and $P$ are therefore the leading relevant scalar perturbations.

\subsection{Minimal EFT and massless Potts--Ising flows}
\label{sec:eft-flow}

The operator matching above gives the long-distance action
\begin{equation}
    \begin{split}
        S_{\rm eff}
        =
        S_{4\rm P}^{(c=1)}
        &+\lambda_\varepsilon\int\dd^2x\,\varepsilon(x)
        +\lambda_P\int\dd^2x\,P(x)
        \\
        &
        +\lambda_{\rm marg}\int\dd^2x\,\mathcal O_{\rm marg}(x)
        +\ldots.
    \end{split}
    \label{eq:minimal_EFT_M}
\end{equation}
Here $\mathcal O_{\rm marg}$ is the marginally irrelevant scalar responsible for the logarithmic corrections at the lattice four-state Potts point, $\lambda_{\rm marg}$ is its coupling, and the ellipsis denotes less relevant operators. The two relevant fields have
\begin{equation}
    \begin{aligned}
        \Delta_\varepsilon&=\frac12,
        &
        y_t&=2-\Delta_\varepsilon=\frac32,
        \\
        \Delta_P&=\frac18,
        &
        y_P&=2-\Delta_P=\frac{15}{8}.
    \end{aligned}
    \label{eq:Potts-relevant-dimensions}
\end{equation}
Here $y_t$ is the thermal RG eigenvalue associated with the microscopic temperature-like variable $t$. 
The value $\Delta_P=1/8$ follows from the enhanced symmetry of the
four-state Potts point: the three lattice fields $\sigma$, $\tau$, and
$P=\sigma\tau$ form the three equivalent components of the Potts order
parameter and therefore have the same scaling dimension. In the
orbifold description, $\sigma$ and $\tau$ are represented by twist
fields with conformal weights $h=\bar h=1/16$, while $P$ is a distinct
primary whose scaling dimension varies along the generic
Ashkin--Teller critical line. At the four-state Potts radius it becomes
degenerate with the two twist fields, giving
$\Delta_P=1/8$~\cite{Dijkgraaf1988,Dijkgraaf1989}, which is
consistent with the Ashkin--Teller scaling relation
$\Delta_P=\Delta_\varepsilon/4$~\cite{Kadanoff1979,Giuliani2020}.

Since both $\lambda_\varepsilon$ and $\lambda_P$ are relevant, a generic perturbation of \textsf{M} flows into one of the surrounding massive phases. Criticality survives only after one relation between the two couplings is imposed. The resulting codimension-one separatrices are the two continuous phase boundaries seen in the QI plane. Along either boundary the flow is massless: short-distance observables are governed by the four-state Potts fixed point, while at long distances the polarization perturbation locks the two Ising sectors so that only one critical Ising mode remains. The endpoint theories are therefore
\begin{equation}
4\mathrm P_{\rm UV}
\longrightarrow
\mathrm{Ising}_{\rm IR},
\qquad
c_{\rm UV}=1
\longrightarrow
c_{\rm IR}=\frac12.
\label{eq:cflow_Potts_Ising}
\end{equation}
This describes the standard Ising critical line of the generalized Ashkin--Teller or double-frequency sine-Gordon deformation~\cite{Delfino1998,Fabrizio2000,Ye2001}. The theory is not conformal at intermediate scales; instead, Zamolodchikov's running $c$-function decreases monotonically between the two fixed-point values~\cite{Zamolodchikov1986}.

In the microscopic QI section, the two massless trajectories separate the paramagnetic basin from the ferromagnetic and antiferromagnetic basins, respectively. Their intersections with the free line $b=0$ are the points $\textsf A_-$ and $\textsf A_+$. These are distinct lattice realizations of the same Ising infrared CFT: they involve different microscopic ordering channels and different gap-closing momenta, but share $c=1/2$ and the ordinary Ising critical exponents. Accordingly, the arrows in Fig.~\ref{fig:phaseAT} should be read as schematic projections of the full RG trajectories onto the two-dimensional QI coupling plane, since coarse graining generally generates additional irrelevant couplings outside that plane.

Close to \textsf{M}, a finite observation scale can lie between the ultraviolet and infrared crossover scales. Numerical observables may then display effective Potts behavior before crossing over to their asymptotic Ising values. 
In addition, the marginal coupling $\lambda_{\rm marg}$ in Eq.~\eqref{eq:minimal_EFT_M} runs only logarithmically, producing the multiplicative correction quoted in Eq.~\eqref{eq:Potts-xi-log-main}.
These two effects explain why finite-entanglement estimates close to \textsf{M} can drift slowly even though the extended transition lines are asymptotically Ising critical.

\subsection{Relation to the Thirring description}
\label{sec:eft-thirring}

The fermionic interpretation stated in the main text can be made more
precise by starting from the decoupled Ising$\times$Ising point of the
Ashkin--Teller critical line. Each critical Ising sector admits a
massless Majorana representation, and the two Majorana fields can be
combined into a single Dirac fermion. The coupling between the two
Ising energy densities then becomes the current--current interaction
of the massless Thirring model\footnote{For standard derivations of the Ashkin--Teller/Thirring correspondence, see e.g.\ Ref.~\cite{Fabrizio2000}.} in
Eq.~\eqref{eq:Thirring-main}~\cite{Thirring1958,Fabrizio2000}.
After the thermal perturbation is tuned to criticality, the Thirring
coupling $g_{\rm Th}$ is exactly marginal and parametrizes the
Ashkin--Teller critical manifold. In this refermionized description,
the $c=1$ theory at \textsf{M} may therefore be viewed as two
Majorana sectors packaged into one interacting massless Dirac degree
of freedom.

The relevant perturbations in Eq.~\eqref{eq:minimal_EFT_M} become
competing vertex operators after bosonization, giving the
double-frequency sine-Gordon description of the Ising separatrices.
Equivalently, in its refermionized formulation, one effective Majorana
mass remains tuned to zero along the separatrix while the complementary
Majorana sector becomes massive. The infrared theory consequently
contains one massless Majorana mode and is the ordinary Ising CFT with
$c=1/2$~\cite{Delfino1998,Fabrizio2000,Ye2001}. This makes precise the
mode-counting picture used in the main text. The two Majorana sectors
are those of the Ashkin--Teller continuum representation and should
not be identified separately with the two microscopic QI transition
branches.

Under Abelian bosonization, the Thirring model becomes a compact
Gaussian boson whose stiffness, or equivalently compactification
radius, varies continuously along the critical line~\cite{Fabrizio2000}. A qualification is essential: this Gaussian
theory captures the local fermionic dynamics and the continuously
varying scaling dimensions, but not by itself the complete operator
content of the Ashkin--Teller spin model. The latter is described by
the corresponding $\mathbb Z_2$ orbifold under
$\phi\mapsto-\phi$, including twisted sectors that represent spin and
disorder fields which are local in the lattice model but nonlocal in
the fermionic variables. The four-state Potts CFT is realized at a
distinguished compactification radius of this orbifold line~\cite{Dijkgraaf1988,Dijkgraaf1989}.
Thus, the Thirring model describes the local fermionic dynamics of the $c=1$ critical
manifold, while the orbifold specification is required to recover the
full Ashkin--Teller/Potts theory.

Finally, the Dirac field in Eq.~\eqref{eq:Thirring-main} is an
emergent continuum variable obtained from the Ashkin--Teller
description. It is distinct from the microscopic four-component
Grassmann field of the fermionic tensor-network representation in
Sec.~\ref{subsec:recap}. The two fermionic constructions encode the
same universal long-distance physics near \textsf{M}, but arise from
different microscopic mappings.

\section{Chern numbers of the clean fermionic model}
\label{app:chern-numbers}

For $b=0$, the fermionic tensor network is Gaussian and defines the
four-band Bloch Hamiltonian $H(k)$ introduced in
Sec.~\ref{subsec:recap}. The topological invariant of a
two-dimensional class-D band structure is the total Chern number of the
occupied BdG bands~\cite{AltlandZirnbauer,QiZhang2011review}.
Writing
\begin{equation}
    \mathcal C
    =
    \sum_{n:\,\tx{occ}} c_n,
\end{equation}
the Chern number $c_n$ of an individual Bloch band is obtained from its Berry
curvature
\begin{equation}
    \Omega_n(k)
    \coloneqq
    \imagunit\epsilon_{\mu\nu}
    \avs{\del_{k_\mu}n|\del_{k_\nu}n},
\end{equation}
where $\ket{n}$ denotes the momentum-dependent band eigenvector and
$k_\mu=k_1,k_2$ are the components of the crystal momentum. The corresponding
band Chern numbers are~\cite{TKNN1982}
\begin{equation}
    c_n
    \coloneqq
    \frac{1}{2\pi}
    \int_{\tx{BZ}}
    \dd^2k\,
    \Omega_n(k)
    .
\end{equation}
For the four-band Hamiltonian $H(k)$, the resulting band-resolved Chern
numbers are listed in Table~\ref{tab:chern-numbers}
(cf.\ Ref.~\cite{Wille_2024}).

\begin{table}[t]
    \centering
    \renewcommand{\arraystretch}{1.25}
    \setlength{\tabcolsep}{7pt}
    \begin{tabular}{c|c|c|c|c}
        $a \in$
        & $(0,a_-)$
        & $(a_-,1)$
        & $(1,a_+)$
        & $(a_+,\infty)$
        \\
        \hline
        $c_4$ & $-1$ & $-1$ & $0$  & $0$ \\
        $c_3$ & $1$  & $0$  & $-1$ & $0$ \\
        $c_2$ & $-1$ & $0$  & $1$  & $0$ \\
        $c_1$ & $1$  & $1$  & $0$  & $0$ \\
        \hline
        $\mathcal C$ & $0$ & $1$ & $1$ & $0$
    \end{tabular}
    \caption{
        Chern numbers carried by the individual bands of the four-band
        Hamiltonian $H(k)$, ordered by increasing energy
        $\eps_1(k)<\eps_2(k)<\eps_3(k)<\eps_4(k)$.
        The total occupied-band Chern number
        $\mathcal C=c_1+c_2$ changes at
        $a_\pm=\sqrt2\pm1$, signaling the two topological transitions.
        At $a=1$, individual band Chern numbers rearrange due to a band
        touching away from the Fermi level, while the total occupied-subspace
        Chern number remains unchanged.
    }
    \label{tab:chern-numbers}
\end{table}

Thus, the free-fermion line contains a topological region of odd Chern number,
\begin{equation}
    \mathcal C
    =
    \begin{cases}
        0, & 0<a<a_-,\\
        1, & a_-<a<a_+,\\
        0, & a>a_+,
    \end{cases}
\end{equation}
while the redistribution of Chern numbers between individual bands at $a=1$
does not change the topology of the occupied subspace.

Near either transition $\mathsf A_\pm$, one particle--hole-related pair of
bands approaches zero energy while the complementary pair remains gapped.
Projecting onto this low-energy pair gives a two-band Hamiltonian of the form
\begin{equation}
    H_\pm^{(2)}(q)
    =
    \sum_{a=1}^3
    h_{a,\pm}(q)\sigma_a,
    \qquad
    q=k-k_\pm,
\end{equation}
with the explicit lattice form given in
Ref.~\cite{Usoltcev2026}. For a globally defined two-band Hamiltonian
$H^{(2)}(q)=h(q)\cdot\sigma$, the Chern number of the negative-energy band can
be written in terms of the normalized vector
\begin{equation}
    n(q)
    \coloneqq
    \frac{h(q)}{|h(q)|}
\end{equation}
as~\cite{QiWuZhang_2006,Moreno2023}
\begin{equation}
    \mathcal C
    =
    \frac{1}{4\pi}
    \int_{\tx{BZ}}
    \dd^2q\,
    n\cdot
    \left(
        \del_{q_1}n
        \times
        \del_{q_2}n
    \right).
    \label{eq:two-band-chern}
\end{equation}

The projected Hamiltonians $H_\pm^{(2)}$ have the structure of
Qi--Wu--Zhang-type two-band lattice models~\cite{QiWuZhang_2006} and provide
convenient representatives of the topology change at the two critical points.
Their use as approximations to the original four-band problem is, however,
controlled only in the vicinity of the corresponding gap closing:
$H_-^{(2)}$ describes the transition at $\mathsf A_-$, while
$H_+^{(2)}$ describes the transition at $\mathsf A_+$. We therefore use
Eq.~\eqref{eq:two-band-chern} to characterize the unit change of Chern
number across each transition, rather than regarding either projected model as
a globally controlled two-band description of the entire $b=0$ line. The
absolute Chern numbers of the three gapped regions are fixed by the full
four-band calculation summarized in Table~\ref{tab:chern-numbers}.

\section{Winding-sector algebra}
\label{app:winding-algebra}

In this appendix, we collect the linear relations between the bosonic
winding-sector contractions $Z_{p_x,p_y}$, the fermionic twisted partition functions $Z^{\tx f}_{\alpha_x,\alpha_y}$, and the normalized winding characters. We then derive the loop-gas expression
\eqref{eq:topological-sign-winding} for the topological invariant and record its limiting values in the thermodynamic phases.

We use the ordering $(00,10,01,11)$ for both the bosonic winding sectors and the fermionic boundary conditions. The torus Jordan--Wigner relation derived in Sec.~\ref{subsec:JW_torus} then takes the matrix form
\begin{equation}
    \begin{pmatrix}
        Z^{\tx f}_{00}\\
        Z^{\tx f}_{10}\\
        Z^{\tx f}_{01}\\
        Z^{\tx f}_{11}
    \end{pmatrix}
    =
    \underbrace{
    \begin{pmatrix}
        1 & -1 & -1 & -1\\
        1 &  1 & -1 &  1\\
        1 & -1 &  1 &  1\\
        1 &  1 &  1 & -1
    \end{pmatrix}
    }_{M_{\tx b\to\tx f}}
    \begin{pmatrix}
        Z_{00}\\
        Z_{10}\\
        Z_{01}\\
        Z_{11}
    \end{pmatrix}.
    \label{eq:JW-sector-matrix}
\end{equation}

The matrix is orthogonal up to normalization,
\begin{equation}
    M_{\tx b\to\tx f}
    M_{\tx b\to\tx f}^{\T}
    =
    4\id,
    \qquad
    M_{\tx b\to\tx f}^{-1}
    =
    \frac14
    M_{\tx b\to\tx f}^{\T}.
\end{equation}

In particular, summing the four bosonic sectors gives
\begin{equation}
    Z^{\tx b}
    =
    \sum_{p_x,p_y}Z_{p_x,p_y}
    =
    \frac12
    \left(
        -Z^{\tx f}_{00}
        +Z^{\tx f}_{10}
        +Z^{\tx f}_{01}
        +Z^{\tx f}_{11}
    \right),
    \label{eq:bosonic-from-fermionic-appendix}
\end{equation}
in agreement with Eq.~\eqref{eq:bosonic-from-fermionic-torus}.

Define the normalized winding-sector weights by
\begin{equation}
    P_{p_x,p_y}
    \coloneqq
    \frac{Z_{p_x,p_y}}{Z^{\tx b}}.
\end{equation}

The contractions with noncontractible $\sigma_z$ strings yield the
normalized winding-character relation
\begin{equation}
    \begin{pmatrix}
        1\\
        m_x\\
        m_y\\
        m_{xy}
    \end{pmatrix}
    =
    \underbrace{
    \begin{pmatrix}
        1 &  1 &  1 &  1\\
        1 & -1 &  1 & -1\\
        1 &  1 & -1 & -1\\
        1 & -1 & -1 &  1
    \end{pmatrix}
    }_{M_{\tx b\to m}}
    \begin{pmatrix}
        P_{00}\\
        P_{10}\\
        P_{01}\\
        P_{11}
    \end{pmatrix},
    \label{eq:winding-character-matrix}
\end{equation}
where the normalized characters are
\begin{equation}
    \begin{aligned}
        m_x
        &=
        \frac{Z^{\tx b}[\sigma_z(\gamma_x)]}{Z^{\tx b}},
        \\
        m_y
        &=
        \frac{Z^{\tx b}[\sigma_z(\gamma_y)]}{Z^{\tx b}},
        \\
        m_{xy}
        &=
        \frac{Z^{\tx b}[\sigma_z(\gamma_x)\sigma_z(\gamma_y)]}{Z^{\tx b}}.
    \end{aligned}
    \label{eq:normalized-winding-characters}
\end{equation}

Equivalently,
\begin{equation}
    m_x
    =
    \sum_{p_x,p_y}
    (-1)^{p_x}P_{p_x,p_y},
    \qquad
    m_y
    =
    \sum_{p_x,p_y}
    (-1)^{p_y}P_{p_x,p_y},
\end{equation}
and
\begin{equation}
    m_{xy}
    =
    \sum_{p_x,p_y}
    (-1)^{p_x+p_y}P_{p_x,p_y}.
\end{equation}

Since for the inversion it holds
\begin{equation}
    M_{\tx b\to m}^{-1}
    =
    \frac14
    M_{\tx b\to m},
\end{equation}
the sector weights can conversely be expressed as
\begin{equation}
    \begin{aligned}
        P_{00}
        &=
        \frac14
        \left(
            1+m_x+m_y+m_{xy}
        \right),
        \\
        P_{10}
        &=
        \frac14
        \left(
            1-m_x+m_y-m_{xy}
        \right),
        \\
        P_{01}
        &=
        \frac14
        \left(
            1+m_x-m_y-m_{xy}
        \right),
        \\
        P_{11}
        &=
        \frac14
        \left(
            1-m_x-m_y+m_{xy}
        \right).
    \end{aligned}
    \label{eq:sector-probabilities-from-winding}
\end{equation}

Thus, the three winding characters completely determine the probability distribution over the four winding sectors.

Eliminating the sector sums between Eqs.~\eqref{eq:JW-sector-matrix} and \eqref{eq:winding-character-matrix} yields
\begin{equation}
    \begin{aligned}
        \frac{2Z^{\tx f}_{00}}{Z^{\tx b}}
        &=
        -1+m_x+m_y+m_{xy},
        \\
        \frac{2Z^{\tx f}_{10}}{Z^{\tx b}}
        &=
        1-m_x+m_y+m_{xy},
        \\
        \frac{2Z^{\tx f}_{01}}{Z^{\tx b}}
        &=
        1+m_x-m_y+m_{xy},
        \\
        \frac{2Z^{\tx f}_{11}}{Z^{\tx b}}
        &=
        1+m_x+m_y-m_{xy}.
    \end{aligned}
    \label{eq:fermionic-Z-winding}
\end{equation}

Since the bosonic partition sum is positive, $Z^{\tx b}>0$, this common normalization does not affect the signs of the fermionic partition functions. Pairing the first and fourth expressions in Eq.~\eqref{eq:fermionic-Z-winding} leads to
\begin{equation}
    \left(
        \frac{2Z^{\tx f}_{00}}{Z^{\tx b}}
    \right)
    \left(
        \frac{2Z^{\tx f}_{11}}{Z^{\tx b}}
    \right)
    =
    (m_x+m_y)^2-(1-m_{xy})^2,
    \label{eq:paired-fermionic-factors-1}
\end{equation}
while pairing the second and third yields
\begin{equation}
    \left(
        \frac{2Z^{\tx f}_{10}}{Z^{\tx b}}
    \right)
    \left(
        \frac{2Z^{\tx f}_{01}}{Z^{\tx b}}
    \right)
    =
    (1+m_{xy})^2-(m_x-m_y)^2.
    \label{eq:paired-fermionic-factors-2}
\end{equation}

Taking the sign of their product thus produces the winding-character formula of the topological invariant in Eq.~\eqref{eq:topological-sign-winding},
\begin{equation}
    \begin{split}
        \sigma_{\tx{topo}}
        &=
        \prod_{\alpha_x,\alpha_y\in\{0,1\}}
        \sgn Z^{\tx f}_{\alpha_x,\alpha_y}
        \\[0.5ex]
        &=
        \sgn\Bigl\{
            \left[(m_x+m_y)^2-(1-m_{xy})^2\right]
            {}
            \\
        &\hspace{3.2em}{}
            \times
            \left[(1+m_{xy})^2-(m_x-m_y)^2\right]
        \Bigr\}.
    \end{split}
    \label{eq:topological-sign-winding-appendix}
\end{equation}

For a square torus and a rotation-invariant local tensor, rotational symmetry
enforces
\[
    m_x=m_y\equiv m.
\]
If, in addition, the two binary winding parities are statistically independent,
their joint character factorizes,
\[
    m_{xy}=m_xm_y=m^2.
\]
Under this additional assumption,
Eq.~\eqref{eq:sector-probabilities-from-winding} reduces to
\begin{equation}
    \begin{aligned}
        P_{00}
        &=
        \frac{(1+m)^2}{4},
        \\
        P_{10}=P_{01}
        &=
        \frac{1-m^2}{4},
        \\
        P_{11}
        &=
        \frac{(1-m)^2}{4},
    \end{aligned}
    \label{eq:factorized-sector-probabilities}
\end{equation}
and the invariant becomes a function of the single winding parameter $m$ (as mentioned in Sec.~\ref{subsec:winding-invariant}),
\begin{equation}
    \sigma_{\tx{topo}}
    =
    \sgn\!
    \left\{
        \left[
            4m^2-(1-m^2)^2
        \right]
        (1+m^2)^2
    \right\}.
    \label{eq:topological-sign-winding-m}
\end{equation}

Note that statistical independence is not a
consequence of rotational symmetry; it is exact in the equal-sector and single-sector limiting distributions, but need not hold near criticality.

For equal sector weights,
\begin{equation}
    P_{00}
    =
    P_{10}
    =
    P_{01}
    =
    P_{11}
    =
    \frac14,
\end{equation}
the winding characters and the topological sign read
\begin{equation}
    m_x=m_y=m_{xy}=0,
    \qquad
    \sigma_{\tx{topo}}=-1,
\end{equation}
producing the thermodynamic winding-sector pattern of the loop-deconfined phase.

When a single winding sector $(p_x^\star,p_y^\star)$ dominates,
\begin{equation}
\begin{split}
    m_x&=(-1)^{p_x^\star},
    \qquad
    m_y=(-1)^{p_y^\star},
    \\[0.5ex]
    m_{xy}&=(-1)^{p_x^\star+p_y^\star},
    \label{eq:single-sector-winding-characters}
\end{split}
\end{equation}
and Eq.~\eqref{eq:topological-sign-winding-appendix} gives $\sigma_{\tx{topo}}=+1$ for all four possible sectors. The corresponding values are summarized in Tab.~\ref{tab:winding-characters}.

With the cut convention of Sec.~\ref{subsec:JW_torus}, the empty configuration always belongs to the sector $(0,0)$. In the fully occupied configuration, a cut transverse to the $x$-cycle intersects $L_y$ occupied edges, while a cut transverse to the $y$-cycle intersects $L_x$ occupied edges. Hence
\begin{equation}
    (p_x^\star,p_y^\star)_{\tx{full}}
    =
    (L_y\bmod2,L_x\bmod2).
    \label{eq:full-sector-parities}
\end{equation}

\begin{table*}[t]
    \centering
    \renewcommand{\arraystretch}{1.25}
    \setlength{\tabcolsep}{7pt}
    \begin{tabular}{c|cccc| l}
        $(p_x^\star,p_y^\star)$
        &
        $m_x$
        &
        $m_y$
        &
        $m_{xy}$
        &
        $\sigma_{\tx{topo}}$
        &
        Realized deep in
        \\
        \midrule
        $(0,0)$
        &
        $+1$
        &
        $+1$
        &
        $+1$
        &
        $+1$
        &
        `empty' \& `full' for $L_x,L_y$ even
        \\
        $(1,0)$
        &
        $-1$
        &
        $+1$
        &
        $-1$
        &
        $+1$
        &
        `full' for $L_x$ even, $L_y$ odd
        \\
        $(0,1)$
        &
        $+1$
        &
        $-1$
        &
        $-1$
        &
        $+1$
        &
        `full' for $L_x$ odd, $L_y$ even
        \\
        $(1,1)$
        &
        $-1$
        &
        $-1$
        &
        $+1$
        &
        $+1$
        &
        `full' for $L_x,L_y$ odd
        \\
    \end{tabular}
    \caption{
        Winding characters and topological sign when a single winding sector
        $(p_x^\star,p_y^\star)$ dominates. The last column identifies the
        corresponding limiting configurations in the present model. The empty
        configuration always lies in $(0,0)$, whereas the fully occupied
        configuration lies in $(L_y\bmod2,L_x\bmod2)$ because the $x$- and
        $y$-transverse cuts intersect $L_y$ and $L_x$ occupied edges,
        respectively. In every single-sector limit, substitution into
        Eq.~\eqref{eq:topological-sign-winding} gives
        $\sigma_{\tx{topo}}=+1$.
    }
    \label{tab:winding-characters}
\end{table*}

Finally, the numerical twist convention of Sec.~\ref{subsec:topo-numerics} attaches the factor $-1$ to the unoccupied, rather than the occupied, channel along an antiperiodic seam. For a cut
$\gamma_\mu$ containing $L_{\gamma_\mu}$ bonds,
\begin{equation}
    (-1)^{N_{0,\mu}}
    =
    (-1)^{L_{\gamma_\mu}}
    (-1)^{N_{1,\mu}},
\end{equation}
where $N_{0,\mu}$ and $N_{1,\mu}$ are the numbers of unoccupied and occupied indices on the cut. Consequently, the raw numerical contractions $\widetilde Z^{\tx f}_{\alpha_x,\alpha_y}$ and the convention used in Eqs.~\eqref{eq:JW-sector-matrix}--\eqref{eq:fermionic-Z-winding} are related by
\begin{equation}
    \widetilde Z^{\tx f}_{\alpha_x,\alpha_y}
    =
    (-1)^{
        \alpha_x L_y+\alpha_y L_x
    }
    Z^{\tx f}_{\alpha_x,\alpha_y}.
    \label{eq:numerical-analytical-seam-sign}
\end{equation}

The fixed seam factors cancel from the product over all four boundary conditions,
\begin{equation}
    \prod_{\alpha_x,\alpha_y\in\{0,1\}}
    (-1)^{
        \alpha_x L_y+\alpha_y L_x
    }
    =
    1,
\end{equation}
so the numerical convention gives the same value of $\sigma_{\tx{topo}}$.

\bibliography{refs}

\end{document}